\documentclass[aps,prl,twocolumn,superscriptaddress,floatfix]{revtex4-2}

\usepackage{graphicx}
\usepackage{dcolumn}

\usepackage{bm}
\usepackage{amsmath}
\usepackage{amssymb}
\usepackage{microtype}
\usepackage{xfrac}
\usepackage{physics}
\usepackage{array}
\usepackage[dvipsnames]{xcolor}
\usepackage[colorlinks,plainpages=false,linkcolor=blue,urlcolor=blue,citecolor=blue,pdfpagemode=UseNone,pdfstartview=FitBH]{hyperref}
\usepackage{nicematrix}

\newcommand{\angstrom}{\text{\normalfont\AA}}

\begin{document}

\title{Two-Peak Heat Capacity Accounts for \textit{R}ln(2) Entropy\\ and Ground State Access in the Dipole-Octupole Pyrochlore  \texorpdfstring{Ce$_2$Hf$_2$O$_7$}~}

\author{E.~M.~Smith}
\affiliation{Department of Physics and Astronomy, McMaster University, Hamilton, Ontario L8S 4M1, Canada}
\affiliation{Brockhouse Institute for Materials Research, McMaster University, Hamilton, Ontario L8S 4M1, Canada}

\author{A.~Fitterman}
\affiliation{D\'epartement de Physique, Universit\'e de Montr\'eal, Montr\'eal, Quebec H2V 0B3, Canada}
\affiliation{Regroupement Qu\'eb\'ecois sur les Mat\'eriaux de Pointe (RQMP), Quebec H3T 3J7, Canada}

\author{R.~Sch\"{a}fer}
\affiliation{Department of Physics, Boston University, Boston, Massachusetts 02215, USA}

\author{B.~Placke}
\affiliation{Max Planck Institute for the Physics of Complex Systems, N\"{o}thnitzer Stra{\ss}e 38, Dresden 01187, Germany}
\affiliation{Rudolf Peierls Centre for Theoretical Physics, University of Oxford, Oxford OX1 3PU, United Kingdom}

\author{A.~Woods}
\affiliation{Los Alamos National Laboratory, Los Alamos, New Mexico 87545, USA}

\author{S.~Lee}
\affiliation{Los Alamos National Laboratory, Los Alamos, New Mexico 87545, USA}

\author{S.~H.-Y.~Huang}
\affiliation{Department of Physics and Astronomy, McMaster University, Hamilton, Ontario L8S 4M1, Canada}

\author{J.~Beare}
\affiliation{Department of Physics and Astronomy, McMaster University, Hamilton, Ontario L8S 4M1, Canada}
\affiliation{Neutron Scattering Division, Oak Ridge National Laboratory, Oak Ridge, Tennessee 37831, USA}

\author{S.~Sharma}
\affiliation{Department of Physics and Astronomy, McMaster University, Hamilton, Ontario L8S 4M1, Canada}

\author{D.~Chatterjee}
\affiliation{Universit\'e Paris-Saclay, CNRS, Laboratoire de Physique des Solides, 91405 Orsay, France}

\author{C.~Balz}
\affiliation{Neutron Scattering Division, Oak Ridge National Laboratory, Oak Ridge, Tennessee 37831, USA}
\affiliation{ISIS Neutron and Muon Source, STFC Rutherford Appleton Laboratory, Didcot OX11 0QX, United Kingdom}

\author{M.~B.~Stone}
\affiliation{Neutron Scattering Division, Oak Ridge National Laboratory, Oak Ridge, Tennessee 37831, USA}

\author{A.~I.~Kolesnikov}
\affiliation{Neutron Scattering Division, Oak Ridge National Laboratory, Oak Ridge, Tennessee 37831, USA}

\author{A.~R.~Wildes}
\affiliation{Institut Laue-Langevin, 71 Avenue des Martyrs CS 20156, 38042 Grenoble Cedex 9, France}

\author{E.~Kermarrec}
\affiliation{Universit\'e Paris-Saclay, CNRS, Laboratoire de Physique des Solides, 91405 Orsay, France}

\author{G.~M.~Luke}
\affiliation{Department of Physics and Astronomy, McMaster University, Hamilton, Ontario L8S 4M1, Canada}
\affiliation{Brockhouse Institute for Materials Research, McMaster University, Hamilton, Ontario L8S 4M1, Canada}

\author{O.~Benton}
\affiliation{Max Planck Institute for the Physics of Complex Systems, N\"{o}thnitzer Stra{\ss}e 38, Dresden 01187, Germany}
\affiliation{School of Physical and Chemical Sciences, Queen Mary University of London, London, E1 4NS, United Kingdom}

\author{R.~Moessner}
\affiliation{Max Planck Institute for the Physics of Complex Systems, N\"{o}thnitzer Stra{\ss}e 38, Dresden 01187, Germany}

\author{R.~Movshovich}
\affiliation{Los Alamos National Laboratory, Los Alamos, New Mexico 87545, USA}

\author{A.~D.~Bianchi}
\affiliation{D\'epartement de Physique, Universit\'e de Montr\'eal, Montr\'eal, Quebec H2V 0B3, Canada}
\affiliation{Regroupement Qu\'eb\'ecois sur les Mat\'eriaux de Pointe (RQMP), Quebec H3T 3J7, Canada}

\author{B.~D.~Gaulin}
\affiliation{Department of Physics and Astronomy, McMaster University, Hamilton, Ontario L8S 4M1, Canada}
\affiliation{Brockhouse Institute for Materials Research, McMaster University, Hamilton, Ontario L8S 4M1, Canada}
\affiliation{Canadian Institute for Advanced Research, 661 University Avenue, Toronto, Ontario M5G 1M1, Canada.}

\date{July 16, 2025}

\begin{abstract} 
Magnetic heat capacity measurements of a high-quality single crystal of the dipole-octupole pyrochlore Ce$_2$Hf$_2$O$_7$ down to a temperature of $T = 0.02$~K are reported. These show a two-peaked structure, with a Schottky-like peak at $T_1 \sim 0.065$~K, similar to what is observed in its sister Ce-pyrochlores Ce$_2$Zr$_2$O$_7$ and Ce$_2$Sn$_2$O$_7$. However, a second sharper peak is observed at $T_2 \sim 0.025$~K, signifying the entrance to the ground state. The ground state appears to have gapped excitations, as even the most abrupt extrapolation to $C_P=0$ at $T = 0$~K fully accounts for the $R\ln(2)$ entropy associated with the pseudospin-1/2 doublet for Ce$^{3+}$ in this environment. The ground state could be conventionally ordered, although theory predicts a much larger anomaly in $C_P$ at much higher temperatures than the measured $T_2$ for expectations from an all-in all-out ground state of the XYZ Hamiltonian for Ce$_2$Hf$_2$O$_7$. The sharp low-temperature peak could also signify a cross-over from a classical spin liquid to a quantum spin liquid (QSL). For both scenarios, comparison of the measured $C_P$ with NLC calculations suggests that weak interactions beyond the nearest-neighbor XYZ Hamiltonian become relevant below  $T \sim 0.25$~K. The diffuse magnetic neutron scattering observed from Ce$_2$Hf$_2$O$_7$ at low temperatures between $T_2$ and $T_1$ resembles that observed from Ce$_2$Zr$_2$O$_7$, which is well established as a $\pi$-flux quantum spin ice (QSI). Together with the peak in the heat capacity at $T_2$, this diffuse scattering from Ce$_2$Hf$_2$O$_7$ is suggestive of a classical spin liquid regime above $T_2$ that is distinct from the zero-entropy quantum ground state below $T_2$.
\end{abstract}
\maketitle

Cerium-based pyrochlore insulators have recently attracted attention as the best candidates to display QSI ground states. QSIs are a specific form of QSL, and ones which map on to an emergent quantum electrodynamics with exotic elementary excitations corresponding to magnetic and electric monopoles as well as emergent photons~\cite{Hermele2004, Banerjee2008, Lee2012, Benton2012, Savary2012b, GingrasReview2014}. For these pyrochlores, with Ce$^{3+}$ ions decorating networks of corner-sharing tetrahedra, crystal electric field (CEF) effects break the $J=5/2$ Hund's rule ground state into three well separated doublets~\cite{Sibille2015, Gaudet2019, Gao2019, Sibille2020, Poree2022}, and yield a quantum pseudospin-$1/2$ degree of freedom at low temperature~\cite{Curnoe2007, Onada2011, Huang2014, RauReview2019}. The wavefunctions associated with the CEF ground state doublet correspond to a $z$-component of pseudospin with a dipole moment, while the $x$ and $y$ components carry octupole moments and transform different from each other under time-reversal symmetry and the point group symmetry at the Ce-site. Such materials are known as dipole-octupole pyrochlores and their QSI ground states can have either a dipolar or octupolar character~\cite{Huang2014, Benton2020, Patri2020, Huang2020}.

Experimental work on cerium-based pyrochlores is most advanced on single crystal Ce$_2$Zr$_2$O$_7$~\cite{Gao2019, Gaudet2019, Smith2022, Changlani2022, Gao2022, Smith2023, Beare2023, Smith2024, Gao2024, Smith2025}, where detailed cases have been made for Ce$_2$Zr$_2$O$_7$ displaying a $\pi$-flux QSI ground state. Prior studies have focused on estimating the interaction parameters in the symmetry-allowed XYZ Hamiltonian for Ce$_2$Zr$_2$O$_7$ through comparison of measurements to relevant theory~\cite{Smith2022,Changlani2022,Smith2023}. Changlani \emph{et al.}~\cite{Changlani2022} concluded that the magnetic ground state in Ce$_2$Zr$_2$O$_7$ is an octupolar QSI while Smith \emph{et al.}~\cite{Smith2022,Smith2023} concluded that the magnetic ground state in Ce$_2$Zr$_2$O$_7$ is a QSI near the boundary between the dipolar and octupolar regimes. While the estimated interaction parameters and corresponding magnetic ground states vary slightly between Ref.~\cite{Changlani2022} and Refs.~\cite{Smith2022, Smith2023}, both works suggest a rare QSI in Ce$_2$Zr$_2$O$_7$, consistent with the original reports of QSI behavior~\cite{Gaudet2019, Gao2019}. More recent gauge mean field theory calculations from Desrochers and Kim~\cite{Desrochers2024a, Desrochers2024b} for a $\pi$-flux QSI ground state of the XYZ Hamiltonian can account for both the low-temperature structure factor and the zone boundary non-spin-flip scattering from Ce$_2$Zr$_2$O$_7$~\cite{Gaudet2019, Gao2019, Smith2022}. Together these provide a strong case for Ce$_2$Zr$_2$O$_7$ displaying a $\pi$-flux QSI ground state.

Ce$_2$Sn$_2$O$_7$, another dipole-octupole pyrochlore, has also been examined. However, synthesis difficulties have prevented the study of large single crystals to date. Early experiments on powder samples of Ce$_2$Sn$_2$O$_7$ were interpreted in terms of an octupole-based QSI phase~\cite{Sibille2015, Sibille2020, Poree2023}. However, new results on hydrothermally-grown powder and small single crystal samples of Ce$_2$Sn$_2$O$_7$ suggest that the magnetic ground state in Ce$_2$Sn$_2$O$_7$ may be an ``all-in all-out'' non-colinear N\'eel state that is proximate to a QSI phase with dynamics that persist down to very low temperature~\cite{Yahne2024}. 

A third member of this dipole-octupole Ce-pyrochlore family, Ce$_2$Hf$_2$O$_7$, has recently been studied and the resulting analysis of these experiments is consistent with a QSI ground state~\cite{Poree2022, Poree2023b, Bhardwaj2024}. However this material has also presented synthesis challenges. While large single crystals can be grown, the ones grown previously are black in color and opaque, which is not typical of insulators and implies structural disorder.  

An important issue relevant to studies on Ce-based pyrochlores is that the energy scale of the underlying XYZ Hamiltonian is on order of 1~K or less~\cite{Sibille2020, Smith2022, Changlani2022, Smith2023, Poree2023b, Yahne2024}, and it is difficult to perform equilibrium measurements below $T \sim 0.1$~K. To date, no heat capacity measurements below $\sim 0.06$~K have been published for any of the Ce-based pyrochlores. Therefore existing measurements do not allow full coverage of the Schottky-like peak at low temperature in any of these three systems, let alone allow for study of the true quantum ground state regime.

In this letter, we report heat capacity measurements on a high-quality single crystal of Ce$_2$Hf$_2$O$_7$, which extend a factor of $\sim$ 3 lower in temperature than those previously-reported on any Ce-based pyrochlore. These measurements access the ground state and recover virtually all of the $R\ln(2)$ entropy associated with the Ce$^{3+}$ CEF ground state doublet.


\begin{figure}[t]
\linespread{1}
\par
\includegraphics[width=3.4in]{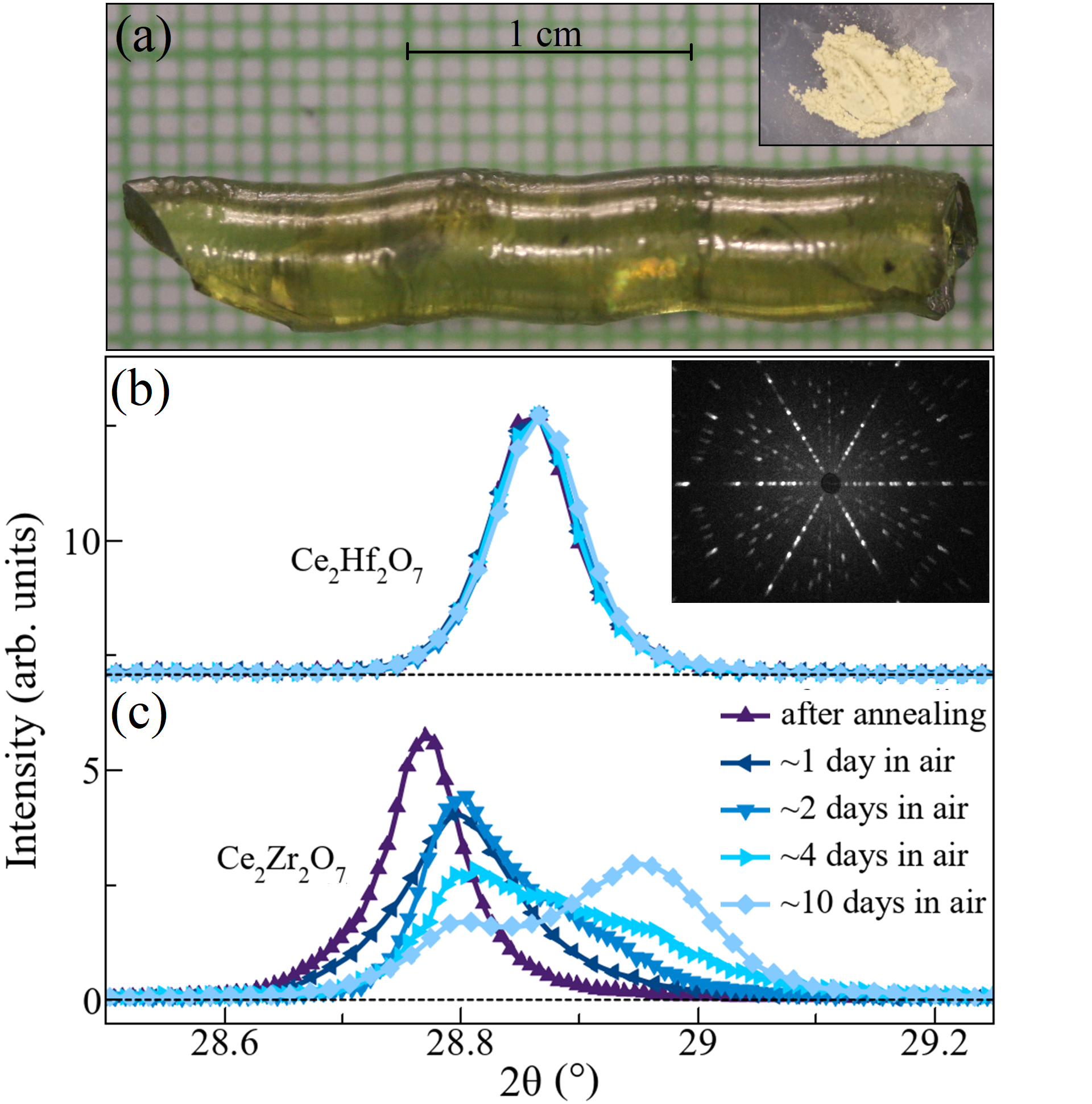}
\par
\caption{(a) The single crystal of Ce$_2$Hf$_2$O$_7$ used in this work. The inset to (a) shows the bright yellow color of a powder of the single crystal. The inset to (b) shows a neutron Laue diffraction pattern measured from our single crystal with the incident neutron beam along $(1,1,1)$. (b) shows x-ray diffraction measurements of the $\mathbf{Q} = (2,2,0)$ Bragg peak from a typical powder sample of Ce$_2$Hf$_2$O$_7$ while (c) shows the same for a typical powder sample of Ce$_2$Zr$_2$O$_7$~\cite{Gaudet2019}, and their stability after exposure to air at ambient conditions for up to 10 days after being annealed in hydrogen.} 
\label{Figure1}
\end{figure}



\begin{figure*}[t]
\linespread{1}
\par
\includegraphics[width=7.2in]{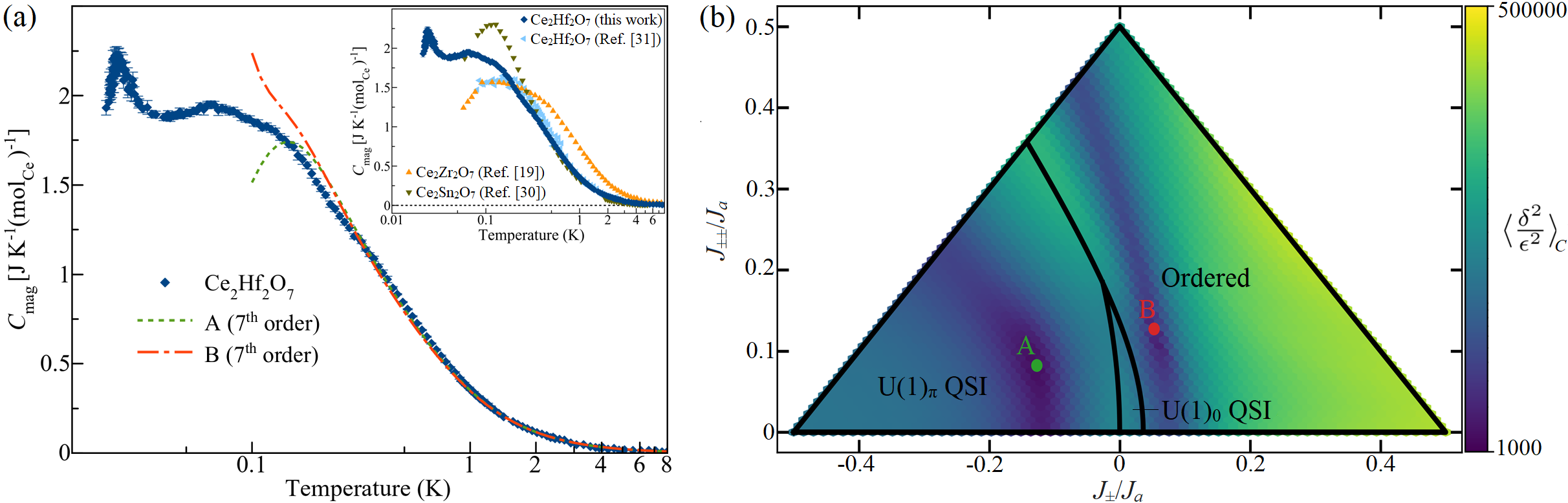}
\par
\caption{(a) The $C_\mathrm{mag}$ measured from single crystal Ce$_2$Hf$_2$O$_7$ in this work. The inset shows a comparison of the $C_\mathrm{mag}$ measured from single crystal Ce$_2$Hf$_2$O$_7$ in this work with that from Ref.~\cite{Poree2023b}, as well as the $C_\mathrm{mag}$ measured from single crystal Ce$_2$Zr$_2$O$_7$~\cite{Smith2022} and Ce$_2$Sn$_2$O$_7$~\cite{Yahne2024}. The curves in (a) show $C_\mathrm{mag}$ calculated via seventh-order NLC using the best fitting parameters obtained from our fitting to the experimental $C_\mathrm{mag}$, $(J_a,J_b,J_c) = (0.050, 0.021, 0.004)$~meV (labeled as A) and $(0.051, 0.008, -0.018)$~meV (labeled as B). (b) The goodness-of-fit parameter $\langle \delta^2/\epsilon^2 \rangle_C$ for our sixth-order NLC fits to the measured $C_\mathrm{mag}$ of Ce$_2$Hf$_2$O$_7$, shown on a logarithmic scale. We also show the phase boundaries and corresponding phases in the ground state phase diagram predicted at the nearest-neighbor level for dipole-octupole pyrochlores~\cite{Benton2020}.} 
\label{Figure2}
\end{figure*}

As shown in Fig.~\ref{Figure1}(a), our high-quality single crystal of Ce$_2$Hf$_2$O$_7$ is light greenish-yellow as expected for a high Ce$^{3+}$ to Ce$^{4+}$ ratio~\cite{Gaudet2019}, and semi-transparent as expected for a magnetic insulator with little structural disorder. In contrast to the earlier-studied Ce$_2$Zr$_2$O$_7$~\cite{Gaudet2019}, it is stable in air, even in powder form. Fig.~\ref{Figure1}(b) and ~\ref{Figure1}(c) show powder x-ray diffraction from Ce$_2$Hf$_2$O$_7$ and Ce$_2$Zr$_2$O$_7$, respectively, as a function of time in air over 10 days of exposure. Ce$_2$Hf$_2$O$_7$ is stable while Ce$_2$Zr$_2$O$_7$ clearly oxidizes, adding Ce$^{4+}$ over time within the Ce$^{3+}_{2-2\delta}$Ce$^{4+}_{2\delta}$Zr$_{2}$O$_{7+\delta}$ structure. Our Ce$_2$Hf$_2$O$_7$ sample is also of high single crystalline quality as evidenced by its neutron Laue pattern [inset to Fig.~\ref{Figure1}(b)].

Our main experimental result is shown in Fig.~\ref{Figure2}(a), which is the magnetic contribution to the heat capacity ($C_\mathrm{mag}$) of our Ce$_2$Hf$_2$O$_7$ single crystal as a function of temperature on a logarithmic temperature scale, down to $T \sim 0.02$~K. This low base-temperature is possible due to the relatively large sample mass (57.7~mg) used for the measurements, allowing for a strong thermal linkage between the sample and the dilution refrigerator. Furthermore, the large relaxation time-constant of Ce$_2$Hf$_2$O$_7$ at low temperature allowed for careful equilibration protocols with relatively slow heat pulses and a long averaging-time for thermometer readings, leading to high precision and equilibrated measurements at very low temperatures. 

Our $C_\mathrm{mag}$ measurements are plotted from 0.02~K to 8~K in Fig.~\ref{Figure2}(a), but the heat capacity ($C_P$) measurements themselves extend to $T \sim 125$~K, at which point $C_P$ is dominated by phonon contributions [see Supplemental Material (SM)~\cite{SM}]. Our $C_\mathrm{mag}$ measurements on Ce$_2$Hf$_2$O$_7$ are overplotted with measurements on Ce$_2$Zr$_2$O$_7$, Ce$_2$Sn$_2$O$_7$, and earlier measurements from a different single crystal of Ce$_2$Hf$_2$O$_7$ in the inset to Fig.~\ref{Figure2}(a)~\cite{Smith2022, Poree2023b, Yahne2024}. The earlier measurements on a different single crystal sample of Ce$_2$Hf$_2$O$_7$ agree with the new measurements above $T=0.5$~K. However, below $\sim$ 0.5~K the two sets of $C_\mathrm{mag}$ measurements diverge from each other with the measurements of this work yielding much higher $C_\mathrm{mag}$ values for $T<0.25$~K.

Most importantly, the low-temperature $C_{\mathrm{mag}}$ from our single crystal of Ce$_2$Hf$_2$O$_7$ shows a sharp peak at $\sim$0.025~K, a qualitatively new feature for these dipole-octupole QSI-candidate pyrochlores. At higher temperatures, above the sharp peak, the measured $C_{\mathrm{mag}}$ broadly resembles that measured on other cerium-based dipole-octupole pyrochlores, Ce$_2$Zr$_2$O$_7$ and Ce$_2$Sn$_2$O$_7$~\cite{Gao2019, Sibille2020, Smith2022, Yahne2024}, and from the different Ce$_2$Hf$_2$O$_7$ sample in Refs.~\cite{Poree2022, Poree2023b}. Each of these display a hump in $C_{\mathrm{mag}}$ with maximum just above $T\sim0.1$~K [inset to Fig.~\ref{Figure2}(a)]. This is above the temperature of the maximum in the $C_{\mathrm{mag}}$ hump measured from Ce$_2$Hf$_2$O$_7$ in this work, $T_1 \sim 0.065$~K. 

The low-temperature peak in $C_{\mathrm{mag}}$ at $T_2\sim0.025$~K could signify a phase transition to an ordered state in Ce$_2$Hf$_2$O$_7$. However, we argue below that this peak is {\it much} smaller than that expected from a transition to an all-in all-out ordered ground state of the symmetry-allowed nearest-neighbor XYZ Hamiltonian, and it also occurs too low in temperature. Accordingly, if the peak in $C_{\mathrm{mag}}$ does indicate an ordering transition, then the ordered state likely possesses significant interactions not included in the XYZ Hamiltonian, such as dipole-dipole interactions beyond nearest neighbors for example.

To understand our data quantitatively, we follow earlier work on Ce$_2$Zr$_2$O$_7$~\cite{Smith2022} and Ce$_2$Sn$_2$O$_7$~\cite{Yahne2024} and compare the $C_{\mathrm{mag}}$ measured from our single crystal of Ce$_2$Hf$_2$O$_7$ with numerical linked cluster (NLC) calculations~\cite{Schafer2020, schaefer_magnetic_2022}, which allows estimates for the nearest-neighbor exchange parameters in the XYZ Hamiltonian for Ce$_2$Hf$_2$O$_7$. The calculations are performed using a permutation of the XYZ Hamiltonian's exchange parameters which are conventionally called $J_a$, $J_b$, and $J_c$, where $|J_a|\geq|J_b|,|J_c|$ and $J_b \geq J_c$. This allows a unique fit to $C_\mathrm{mag}$ but does not specify the permutation relating $(J_{a}, J_{b}, J_{c})$ to the XYZ Hamiltonian parameters $(J_{\tilde{x}}, J_{\tilde{y}}, J_{\tilde{z}})$. This procedure determines whether the corresponding ground state of the XYZ Hamiltonian is an ordered phase or a QSI phase but does not distinguish between the octupolar or dipolar nature of the ground state.

Following earlier work~\cite{Smith2022}, this nearest-neighbor Hamiltonian can be written as:
\begin{equation}
\label{eq:1}
\begin{split}
    \mathcal{H}_\mathrm{ABC} & = \sum_{\langle ij \rangle}[J_{a}{S_i}^{a}{S_j}^{a} - J_{\pm}({S_i}^{+}{S_j}^{-} + {S_i}^{-}{S_j}^{+}) \\
    & + J_{\pm\pm}({S_i}^{+}{S_j}^{+} + {S_i}^{-}{S_j}^{-})] \\
\end{split}
\end{equation}

\noindent in zero field, where $J_\pm = -\frac{1}{4}(J_b + J_c)$, $J_{\pm\pm} = \frac{1}{4}(J_b - J_c)$. Here ${S_{j}}^{a}$ is the $a$-component of pseudospin-1/2 for Ce$^{3+}$ ion $j$ in its local $\{a$, $b$, $c\}$ coordinate frame, and ${S_j}^{\pm} = {S_j}^{b} \pm \mathrm{i} {S_j}^{c}$. Further detail is given in the SM~\cite{SM}.

Figure~\ref{Figure2}(b) shows the goodness-of-fit measure for our sixth-order NLC fitting to the measured $C_\mathrm{mag}$ of Ce$_2$Hf$_2$O$_7$, denoted as $\langle \delta^2/\epsilon^2 \rangle_{C}$. The plot of $\langle \delta^2/\epsilon^2 \rangle_{C}$ in Fig.~\ref{Figure2}(b) shows two regions of parameter space where $\langle \delta^2/\epsilon^2 \rangle_{C}$ has local minima. The best-fitting parameters from these regions are $(J_a,J_b,J_c) = (0.050, 0.021, 0.004)$~meV (labeled as A) and $(J_a,J_b,J_c) = (0.051, 0.008, -0.018)$~meV (labeled as B).

Figure~\ref{Figure2}(b) also shows the ground state phase diagram predicted for dipole-octupole pyrochlores at the nearest-neighbor level~\cite{Benton2020}, over the non-trivial ($J_a > 0$) portion of phase space, with regions attributed to $0$-flux [U(1)$_0$] and $\pi$-flux [U(1)$_\pi$] QSIs as well as a large region corresponding to all-in all-out order. The ground state phase is either dipolar or octupolar in nature depending on the permutation relating $(J_a,J_b,J_c)$ to $(J_{\tilde{x}}, J_{\tilde{y}}, J_{\tilde{z}})$~\cite{Smith2022, Smith2025}. From Fig.~\ref{Figure2}(b), it is clear that the A parameters (B parameters), and the surrounding region of reasonably-fitting parameters, fall within the region predicted to contain a U(1)$_\pi$ QSI (ordered) ground state. It can also be seen that both parameter sets produce good agreement with the $C_\mathrm{mag}$ data above $\sim$ 0.25~K [Fig.~\ref{Figure2}(a)], so this analysis cannot distinguish between these possibilities.

The NLC calculations are accurate above the low-temperature cutoff in each case (further details in SM~\cite{SM}). Accordingly, the disagreement between the NLC calculations and the measured data below $T \sim 0.25$~K, but still above the low-temperature cutoff of $T\sim0.05$~K ($T \sim 0.15$~K) for the A parameters (B parameters), suggests that interactions beyond the nearest-neighbor XYZ Hamiltonian are significant at these low temperatures.


\begin{figure}[!t]
\linespread{1}
\par
\includegraphics[width=3.4in]{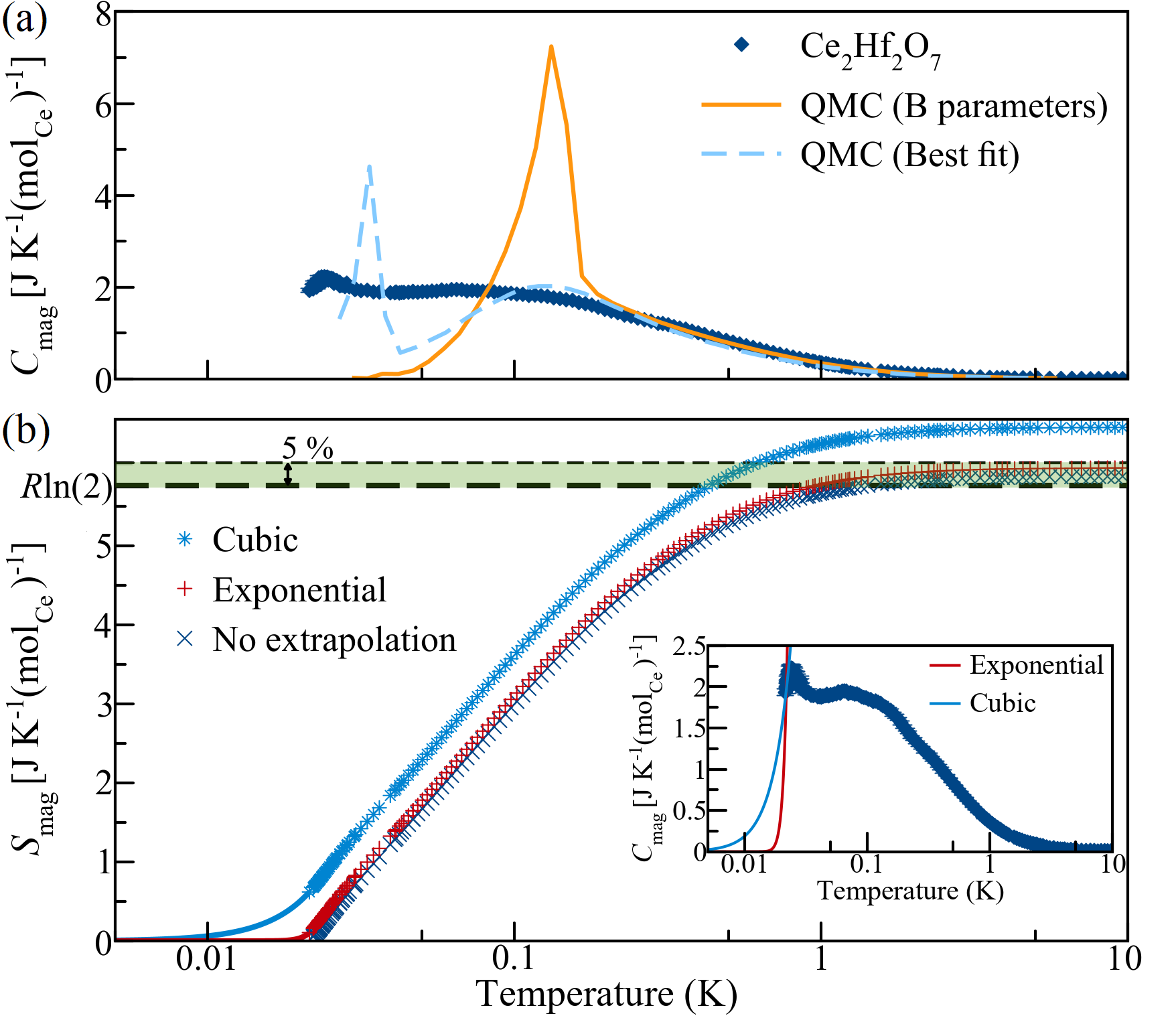}
\par
\caption{(a) The $C_\mathrm{mag}$ measured from Ce$_2$Hf$_2$O$_7$ in this work compared to QMC calculations using the B parameters in the ordered regime from our NLC fitting, $(J_a,J_b,J_c) = (0.051, 0.008, -0.018)$~meV and using the best-fit parameters obtained from our QMC fitting of $C_\mathrm{mag}$, $(J_a,J_b,J_c) = (0.046, -0.003, -0.010)$~meV. (b)~The entropy recovered from the measured $C_\mathrm{mag}$ of Ce$_2$Hf$_2$O$_7$ via $S_\mathrm{mag} = \int_{0}^T\frac{C_\mathrm{mag}}{T}dT$, using the best-fit cubic and exponential low-temperature extrapolations of $C_\mathrm{mag}$, and without extrapolation. The inset to (b) shows these best-fit cubic and exponential extrapolations of $C_\mathrm{mag}$ to zero at $T = 0$~K.} 
\label{Figure3}
\end{figure}


Unbiased quantum Monte Carlo (QMC) calculations are not possible throughout the majority of the disordered regime of the XYZ pyrochlore phase diagram, due to the sign problem~\cite{Banerjee2008, Huang2014, Benton2020, Smith2025}. However, QMC calculations can still be performed for the majority of the ordered regime and specifically for the B parameters [Fig.~\ref{Figure3}(a)]. The QMC calculations of $C_\mathrm{mag}$ using the B parameters show a well-defined peak at $\sim$ 0.15~K, indicative of a phase transition to an ordered state. However, this $C_\mathrm{mag}$ anomaly is larger by more than a factor of 10 in the QMC calculation compared to experiment. It also occurs a factor of $\sim$ 5 too high in temperature. Figure~\ref{Figure3}(a) also shows QMC calculations of $C_{\mathrm{mag}}$ using the best-fit parameters from our QMC fitting (see SM~\cite{SM}), $(J_a,J_b,J_c) = (0.046, -0.003, -0.010)$~meV. Notably, even for these best-fit parameters, there is a large discrepancy between the calculations and measurements below $\sim0.1$~K. 


\begin{figure*}[t]
\linespread{1}
\par
\includegraphics[width=7.2in]{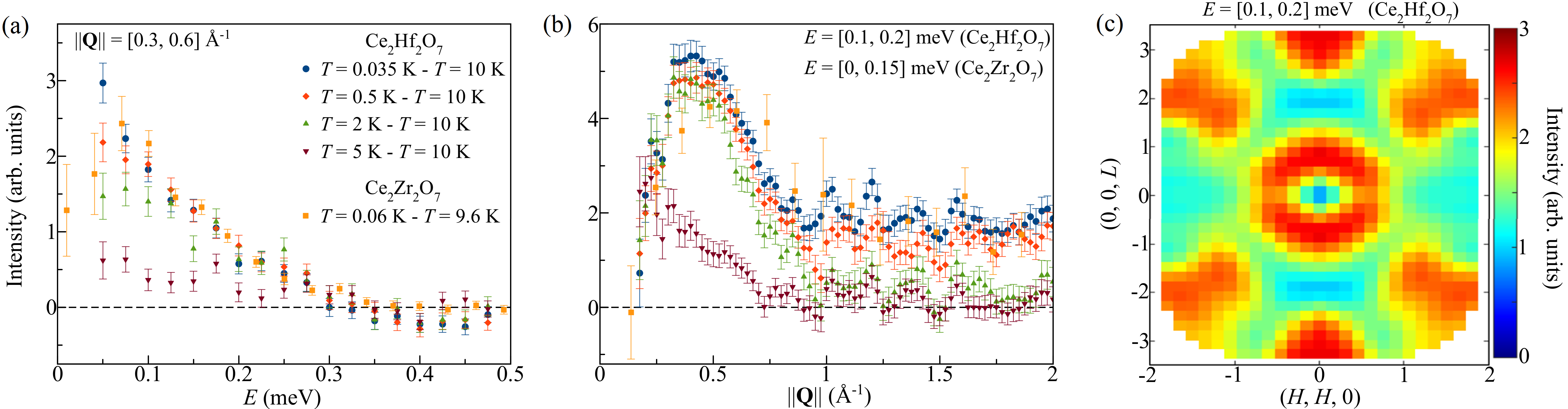}
\par
\caption{(a) and (b) show the powder-averaged neutron scattering signal measured from single crystal Ce$_2$Hf$_2$O$_7$ for (a) $||\mathbf{Q}||$ integration over the range $[0.3,0.6]$~$\angstrom^{-1}$ and (b) energy integration over the range $E=[0.1,0.2]$~meV, at temperatures between $T=0.035$~K and $T=5$~K with a $T=10$~K dataset subtracted. We compare this with the $T=0.06$~K - $T=9.6$~K temperature-difference neutron scattering signal from powder Ce$_2$Zr$_2$O$_7$ for (a) $||\mathbf{Q}||$ integration over $[0.3,0.6]$~$\angstrom^{-1}$ and (b) energy integration over $[0,0.15]$~meV. (c) The symmetrized $T=0.035$~K - $T=5$~K temperature-difference neutron scattering signal measured in the $(H,H,L)$ plane from single crystal Ce$_2$Hf$_2$O$_7$, with an energy integration over $[0.1, 0.2]$~meV and a $(K,\bar{K}, 0)$ integration from $K = -0.1$ to 0.1.}
\label{Figure4}
\end{figure*}


Our measurements of Ce$_2$Hf$_2$O$_7$'s $C_\mathrm{mag}$ account for the full $R\ln(2)$ entropy expected for pseudospin-1/2 degrees of freedom. This markedly differs from earlier $C_\mathrm{mag}$ measurements on Ce-based pyrochlores~\cite{Gao2019, Sibille2020, Smith2022, Poree2023, Poree2023b, Yahne2024}, where at best (in the case of Ce$_2$Zr$_2$O$_7$~\cite{Gao2019, Smith2022}) only $\sim$ 79$\%$ of $R\ln(2)$ is accounted for by the measurements, and the rest must be accounted for by extrapolating the measured $C_\mathrm{mag}$ to zero temperature.

This is illustrated in the inset to Fig.~\ref{Figure3}(b) where the measured $C_\mathrm{mag}$ from Ce$_2$Hf$_2$O$_7$ is shown along with two possible simple extrapolation schemes: one (cubic) corresponding to gapless excitations and one (exponential) corresponding to gapped excitations. The corresponding entropy accounted for by this data is shown in Fig.~\ref{Figure3}(b). It is clear that the data itself, without extrapolation, accounts for $\sim$ $R\ln(2)$ in entropy within 2$\%$. The exponential extrapolation in Fig.~\ref{Figure3}(b) uses a gap energy of 0.025~meV, the value determined from the high energy-resolution inelastic neutron scattering measurements on Ce$_2$Hf$_2$O$_7$ in Ref.~\cite{Poree2023b}. Any low-temperature extrapolation of the $C_\mathrm{mag}$ data consistent with $R\ln(2)$ entropy must be very abrupt in order to not overshoot $R\ln(2)$.

We also performed both low and high energy neutron spectroscopy on Ce$_2$Hf$_2$O$_7$. The high energy neutron spectroscopy, shown in the SM~\cite{SM}, informs on the CEF states and is largely consistent with previously published work~\cite{Poree2022} and with a dipole-octupole CEF ground state doublet. The low-energy spectroscopy on our Ce$_2$Hf$_2$O$_7$ single crystal was performed with the LET spectrometer at the ISIS Neutron Source down to $T = 0.035$~K, and can be compared to earlier measurements from Ce$_2$Zr$_2$O$_7$.

Fig.~\ref{Figure4}(a) and~\ref{Figure4}(b) show the powder-averaged inelastic neutron scattering from our single crystal of Ce$_2$Hf$_2$O$_7$ for a $||\mathbf{Q}||$ integration over $[0.3,0.6]$~$\angstrom^{-1}$ and for energy integration over $[0.1,0.2]$~meV, respectively, between $T=0.035$~K and $T=5$~K with a $T = 10$~K dataset subtracted. These integration ranges were chosen as to cover the dominant magnetic spectral range. The $||\mathbf{Q}||$-integration shows a quasielastic signal that grows from $E\sim 0.25$~meV down to the elastic resolution near $E = 0.05$~meV, and the energy-integrated signal shows a peak in $||\mathbf{Q}||$ near 0.5~$\angstrom^{-1}$. We compare this with low-energy inelastic scattering data for the established QSI Ce$_2$Zr$_2$O$_7$ at $T = 0.06$~K with a $T = 9.6$~K dataset subtracted~\cite{Gaudet2019, Smith2022} and indeed the agreement between the two is very good.  

Fig.~\ref{Figure4}(c) shows the inelastic neutron scattering in the $(H,H,L)$ plane from single crystal Ce$_2$Hf$_2$O$_7$ at $T = 0.035$~K with a $T = 5$~K dataset subtracted, for energy integration over $E = [0.1, 0.2]$~meV. The {\bf Q}-dependence of this low-energy spectral weight reveals a snowflakelike pattern similar to that associated with various pyrochlore spin liquids, including spin ices and $\mathbb{Z}_2$ spin liquids~\cite{Fennell2009, Clancy2009, Benton2012, Gao2019, Gaudet2019, Smith2022, Kim2022, Desrochers2022, Smith2024, Desrochers2024a, Desrochers2024b}. We conclude that both Ce$_2$Hf$_2$O$_7$ at $T \sim 0.035$~K and Ce$_2$Zr$_2$O$_7$ at $T\sim 0.06$~K display similar dynamic spin liquid correlations and no obvious magnetic Bragg peaks (see SM for the latter~\cite{SM}). For Ce$_2$Hf$_2$O$_7$, the existence of these spin liquid correlations at temperatures above the low-temperature peak in the heat capacity suggests the presence of a classical spin liquid regime at intermediate temperatures above $T \sim 0.025$~K, which is distinct from the zero-entropy quantum ground state below  $T \sim 0.025$~K~\cite{Kato2015, Huang2018a, Huang2020}.

While this tends to suggest that the A parameters in the $\pi$-flux QSI ground state regime are appropriate to Ce$_2$Hf$_2$O$_7$, NLC calculations using both the A and B parameters can reasonably account for the diffuse inelastic scattering shown in Fig.~\ref{Figure4}(c). Indeed, similar diffuse scattering is predicted by our NLC calculations for all permutations relating $(J_{\tilde{x}}, J_{\tilde{y}}, J_{\tilde{z}})$ to $(J_a, J_b, J_c)$ for the A parameters. However, our NLC calculations above the ordering transition for the B parameters also provide a reasonable description of the measured diffuse scattering for some permutations of the B parameters (see SM~\cite{SM}). We also investigate which permutations of the A and B parameters are reasonable using NLC calculations to fit the magnetic susceptibility of Ce$_2$Hf$_2$O$_7$ in the SM~\cite{SM}. 

A scenario consistent with the A parameters is that the small peak at $T_2 \sim 0.025$~K does not indicate a phase transition, but rather a cross-over between two distinct disordered phases~\cite{Kato2015, Huang2018a, Huang2020}. In this scenario, the phase below $T_2$ is a QSL ground state with entropy that rapidly approaches zero below $T_2$. 

A cubic heat capacity below $T_2$ would be appropriate for emergent photon excitations of a QSI-type QSL ground state~\cite{Li2017, Kato2015, Huang2020}. However, depending on their effective speed of light, their $T^3$ contribution may only enter at very low temperatures~\cite{Benton2012}. Furthermore, interactions between visons and photons can also cause the photons to develop an effective temperature-dependent gap~\cite{Kwasigroch2020}. Additionally, recent work~\cite{Desrochers2022} has investigated the XYZ Hamiltonian at the mean-field level and has shown that the QSI ground states are in close competition with gapped $\mathbb{Z}_2$ QSLs over a large region of parameter space containing the A parameters from our NLC fitting. These authors reason that the gapped $\mathbb{Z}_2$ QSL phases should be considered on equal footing with the QSI ground states predicted for the XYZ Hamiltonian using mean-field methods [see Fig.~\ref{Figure2}(b)], as fluctuations beyond the mean-field level could easily alter the relative energies of these phases. Interestingly, some of these $\mathbb{Z}_2$ QSLs show a snowflakelike pattern of neutron scattering in the $(H,H,L)$ plane similar to spin ices and to our measurements in Fig.~\ref{Figure4}(c)~\cite{Desrochers2022}. 

To conclude, our measurements and analysis on high-quality single crystal Ce$_2$Hf$_2$O$_7$ show that this dipole-octupole pyrochlore enters its quantum magnetic ground state below $T_2 \sim 0.025$~K, signified by a sharp peak in $C_\mathrm{mag}$. The ground state appears to have gapped excitations, consistent with certain forms of a QSL phase or an ordered phase. Comparison with both QMC and NLC calculations suggests that terms beyond the nearest-neighbor XYZ Hamiltonian have some effect on the magnetic behavior below $\sim 0.25$~K. In the QSL ground state scenario, both $\mathbb{Z}_2$ QSL and $\pi$-flux QSI ground states are viable. At intermediate temperatures above the sharp peak in $C_\mathrm{mag}$, Ce$_2$Hf$_2$O$_7$ strongly resembles a classical spin liquid phase, implying that $T_2$ is a crossover between a classical and a quantum spin liquid in the QSL ground state scenario.

\begin{acknowledgments}
This work was supported by the Natural Sciences and Engineering Research Council of Canada. We greatly appreciate the technical support from Marek~Kiela and Jim~Garrett at McMaster University. We thank Pascal Manuel, Dmitry Khalyavin, and Fabio Orlandi at the ISIS Neutron and Muon Source for technical support and for feedback on the manuscript. Work in Los Alamos was supported by the U.S. Department of Energy, Office of Science, National Quantum Information Science Research Centers, and Quantum Science Center (A.W., S.L., and R.M.). A portion of this research used resources at the Spallation Neutron Source, a DOE Office of Science User Facility operated by the Oak Ridge National Laboratory. Beamtime at the Spallation Neutron Source was allocated to SEQUOIA spectrometer on proposal number IPTS-28896. We gratefully acknowledge the Science and Technology Facilities Council (STFC) for access to neutron beamtime at ISIS allocated under proposal numbers RB2220644~\cite{LET_experiment_DOI} and RB2220630~\cite{WISH_experiment_DOI}. We gratefully acknowledge the Institut Laue-Langevin for neutron beamtime allocated under proposal number 4-05-852~\cite{ILL_experiment_DOI}. This work was supported in part by the Deutsche Forschungsgemeinschaft under grants SFB 1143 (project-id 247310070) and the cluster of excellence ct.qmat (EXC 2147, project-id 390858490). This work was also supported in part by NSF Grant No. DMR-1752759 and AFOSR Grant No. FA9550-20-1-02. We thank the Max Planck Institute for the Physics of Complex Systems for its computing resources.
\end{acknowledgments}

%

\renewcommand{\figurename}{FIG. S\!\!}
\setcounter{figure}{0}

\clearpage

\section{SUPPLEMENTAL MATERIAL:}

\section{Details of Powder Synthesis, Single Crystal Growth, Phase Characterization, and Crystallinity Characterization}

Single crystals of Ce$_2$Hf$_2$O$_7$ were obtained through optical floating zone growth from polycrystalline feed stock. CeO$_2$ (99.995$\%$) powder was heated in air at 500$^\circ$C for 5~hours to ensure proper oxygen stoichiometry and stoichiometric mixtures of CeO$_2$ and HfN (99.5$\%$) were then mixed in a ball mill before being subsequently heated in air to 900$^\circ$C in a covered alumina crucible for 5~hours. The solidified rods were then re-ground in a ball mill and repressed into rods to be used as feed and seed stock. The rods were then heated to 1550$^\circ$C for 3~hours in an atmosphere containing a ratio of 90/10 argon to hydrogen. This argon-hydrogen annealing process was repeated, with regrinding and reformation of the rods between each 3~hour cycle, until achieving phase purity (typically three cycles). During the floating zone growth we used a growth rate of $\sim$7.5 mm/hour while counter-rotating feed and seed stock at 10 rpm in an argon atmosphere with a pressure of 0.3~MPa. 

\begin{figure}[!t]
\linespread{1}
\par
\includegraphics[width=2.85in]{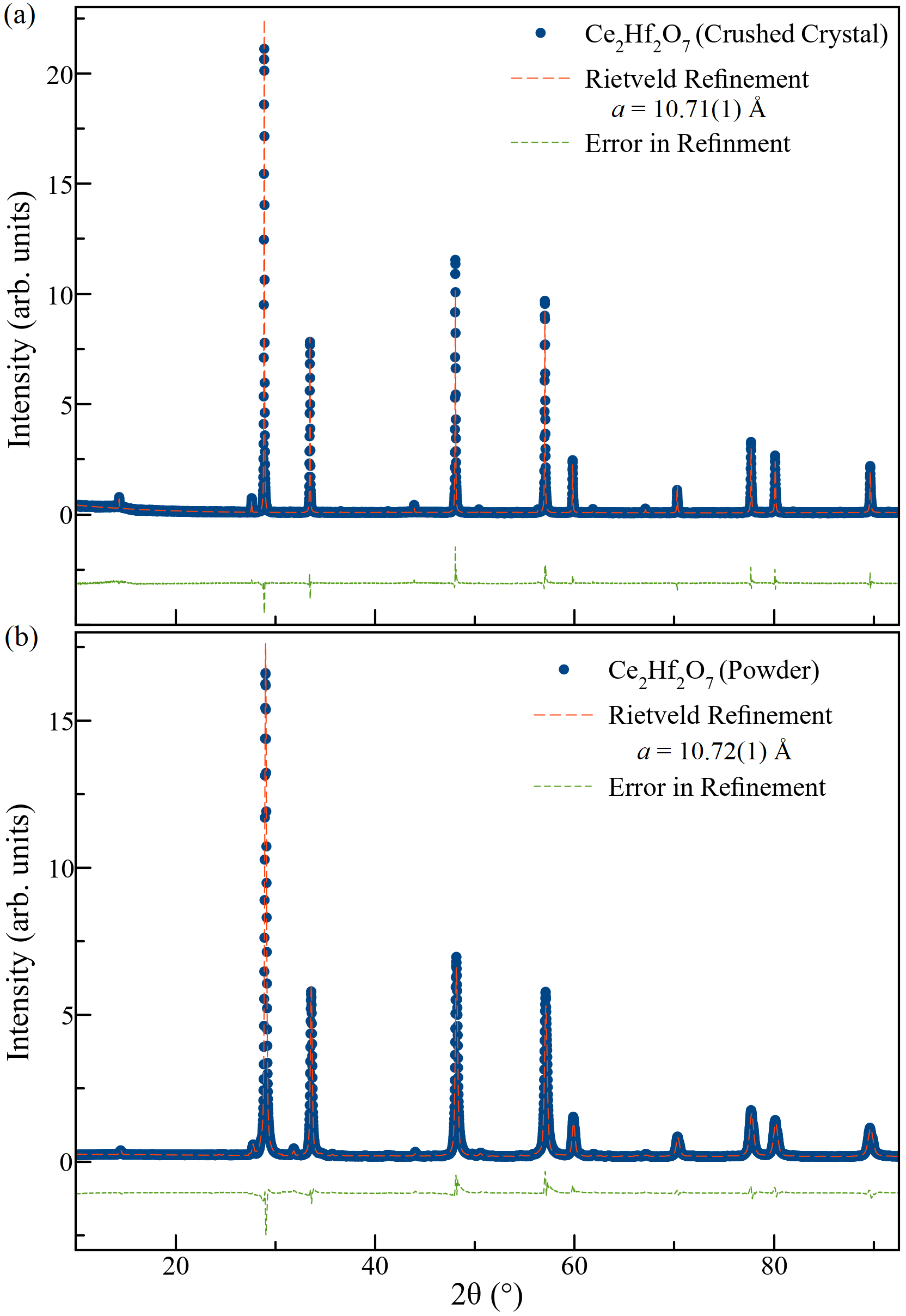}
\par
\caption{The x-ray diffraction pattern measured from (a)~a powdered piece of our single crystal sample and from (b)~one of our powder samples formed using standard solid state synthesis techniques, using incident x-rays with a wavelength of $\lambda = 1.5406~\angstrom$. The lines in each fit show the results of Rietveld refinement to the pyrochlore structure (red) and the difference between the measured and refined diffraction patterns (green), where the latter has been shifted downwards for visibility.} 
\label{FigureS1}
\end{figure}

\begin{figure*}[!t]
\linespread{1}
\par
\includegraphics[width=6in]{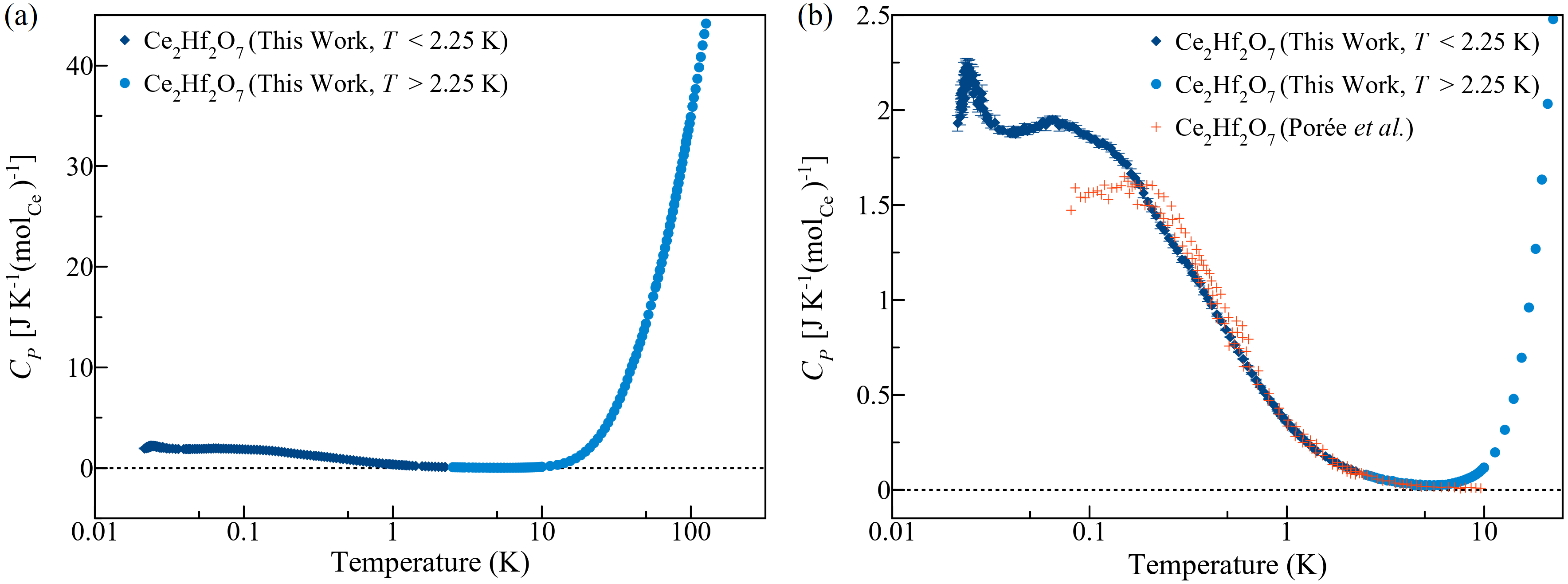}
\par
\caption{(a) The heat capacity ($C_P$) measured from single crystal Ce$_2$Hf$_2$O$_7$ in this work, with the $T < 2.25$~K data shown in dark blue and the $T > 2.25$~K data shown in light blue, on a logarithmic temperature scale. (b) The $T < 25$~K regime of the $C_P$ data measured in this work (blue) along with the $C_{\mathrm{mag}}$ data reported for single crystal Ce$_2$Hf$_2$O$_7$ by Por\'ee \emph{et al.}~\cite{Poree2023b} (red). We connect the $T < 2.25$~K (dark blue) portion of our measured data to the $T > 2.25$~K portion of the $C_{\mathrm{mag}}$ data reported for single crystal Ce$_2$Hf$_2$O$_7$ by Por\'ee \emph{et al.}~\cite{Poree2023b} (red), and we use this connection as the $C_\mathrm{mag}$ measured from our single crystal of Ce$_2$Hf$_2$O$_7$ (shown in Figs.~2 and 3 of the main text).} 
\label{FigureS2}
\end{figure*}

A single crystal of Ce$_2$Hf$_2$O$_7$ obtained with this protocol is shown in Fig.~1(a) of the main text. The single crystal is a yellow-green color and is transparent. Refinements of the pyrochlore crystal structure to the powder x-ray diffraction data measured from a crushed piece of single crystal Ce$_2$Hf$_2$O$_7$ and powder Ce$_2$Hf$_2$O$_7$ (before floating zone growth) are shown in Figs.~S1(a) and S1(b), respectively. The cubic lattice constants obtained from these refinements are $a = 10.71(1)~\angstrom$ and $10.72(1)~\angstrom$ for the crushed single crystal sample and the solid-state-synthesized powder sample, respectively. 

Our neutron Laue diffraction measurements are shown in the inset to Fig.~1(b) in main text and confirm the high-quality crystallinity of our Ce$_2$Hf$_2$O$_7$ sample. Specifically, these measurements show the three-fold-symmetric neutron Laue pattern expected for neutrons incident along $(1,1,1)$, with no signs of scattering from additional crystal grains. These measurements used the OrientExpress instrument at the Institut Laue-Langevin with a sample-to-detector distance of 70.65~mm~\cite{ILL_experiment_DOI}. 

\section{Heat Capacity Measurements}

We used quasi-adiabatic techniques to measure the specific heat of single crystal Ce$_2$Hf$_2$O$_7$. The sample was mounted on a sapphire platform with GE varnish. The heater, a metal film chip resistor, was glued to the opposite side of the platform.  A calibrated ruthenium oxide chip resistance thermometer was mounted directly on the sample. The lowest achieved sample temperature is determined by the combination of the background heat load to the sample stage (due to vibration and electrical noise) and the heat conductance of the link. Large sample mass (57.7~mg) and high specific heat allowed the use of a strong heat link to the bath: a 1-inch long 0.003-inch diameter Au-7\%Cu wire, glued directly to the sample, resulting in the lowest achieved temperature of roughly 20 mK. Large heat capacity of the sample also resulted in a very long temperature-relaxation time constant of several hours. As a result, it was possible to improve the precision of the measurement by increasing the averaging time of a Lakeshore 370 resistance bridge, used to measure thermometer’s resistance, up to 60 seconds. It took over six hours to acquire a full temperature decay curve of 400 points (1 min per point), used to determine specific heat at each temperature. We also took advantage of a very slow thermal relaxation by increasing the time of a heat pulse to eight minutes. This resulted in reduced thermal non-equilibrium within the sample during and immediately after the heat pulse. As a result, we were able to reduce the uncertainty of specific heat to less than 3\%, enabling us to resolve the specific heat anomaly (about 10\% increase above the background value) at 24.5~mK.

The full heat capacity ($C_P$) measurements on our single crystal sample of Ce$_2$Hf$_2$O$_7$ are shown in Fig.~S2 and extend between $T = 0.02$~K and $T = 125$~K. Figure~S2(a) shows the full temperature-range of the measurements and Figure~S2(b) shows the low-temperature regime below $T = 25$~K, both on logarithmic temperature scales. Above $T = 5$~K, the measured heat capacity from Ce$_2$Hf$_2$O$_7$ begins to increase due to the contribution from thermally-excited phonons. This phonon contribution is negligible for temperatures below $\sim$ 5 K. For temperatures above $T=2.25$~K we use earlier measurements on a different sample of Ce$_2$Hf$_2$O$_7$~\cite{Poree2023b} [red in Fig.~S2(a)], which utilized a subtraction of the phonon contribution to isolate the magnetic contribution to the heat capacity ($C_\mathrm{mag}$). Our new data connects smoothly to the data of Ref.~\cite{Poree2023b} for a decade in temperature from $T\sim0.5$~K to $T\sim5$~K, and Fig.~2(a) of the main text shows this composite $C_\mathrm{mag}$ dataset from 0.02~K to 8~K. 

The inset to Fig.~3(b) of the main text shows the best-fitting cubic and exponential low-temperature extrapolations to the $C_{\mathrm{mag}}$ reported for Ce$_2$Hf$_2$O$_7$ in this work, along with the entropy recovered via $S_\mathrm{mag} = \int_{0}^T\frac{C_\mathrm{mag}}{T}dT$ using each of these extrapolations to give $C_{\mathrm{mag}}$ below the lowest-temperature data point. An appropriate extrapolation must give a $C_{\mathrm{mag}}/T$ that approaches zero as temperature approaches absolute zero. The use of low-temperature extrapolations of $C_{\mathrm{mag}}$ can be important in recovering the full $R\ln(2)$ entropy associated with a CEF ground state doublet~\cite{Smith2022}, especially considering the weighting by $1/T$ in $C_{\mathrm{mag}}/T$.


\begin{figure}[t]
\linespread{1}
\par
\includegraphics[width=3.4in]{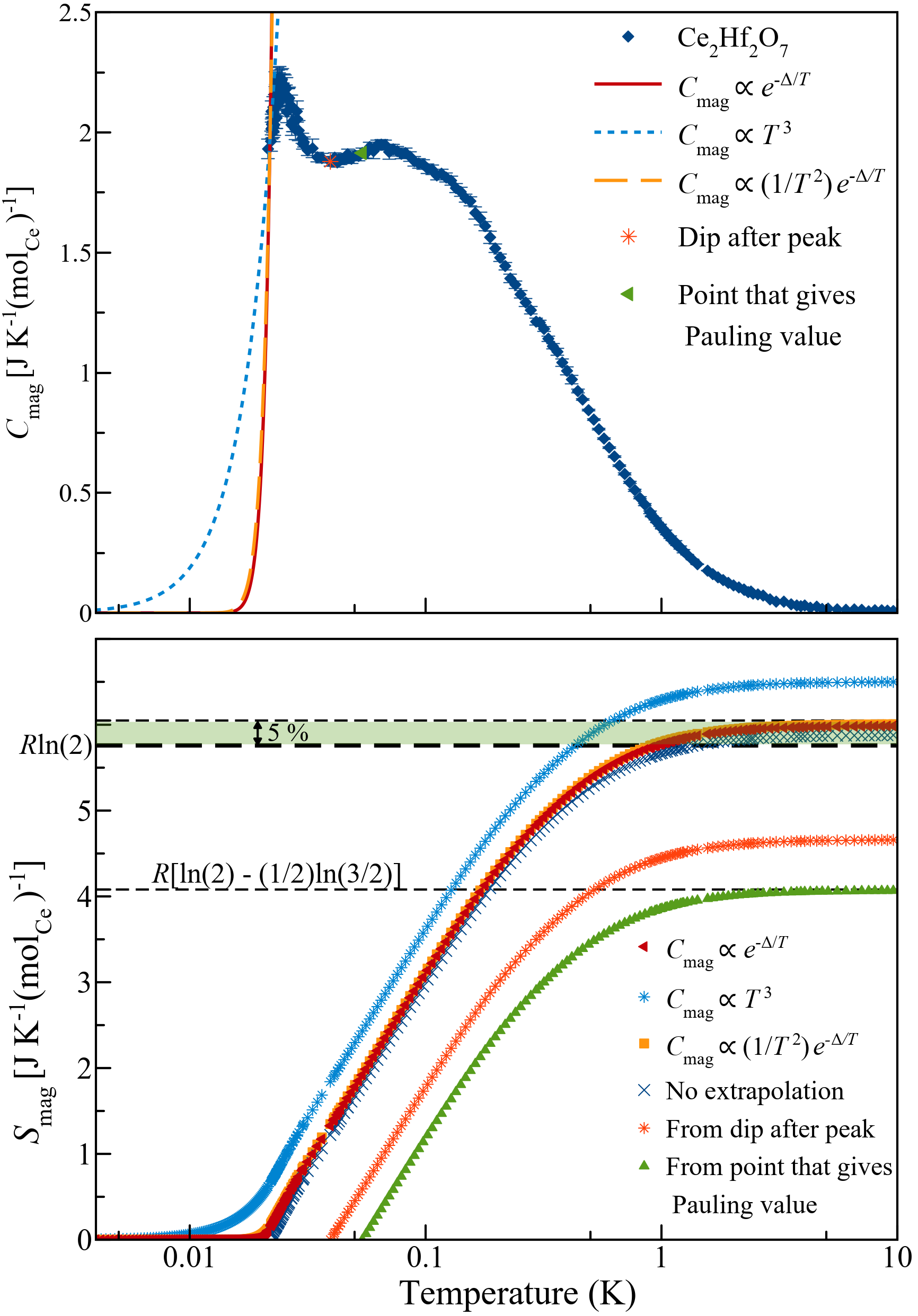}
\par
\caption{(a)~The $C_\mathrm{mag}$ measured from Ce$_2$Hf$_2$O$_7$ in this work and various extrapolations of $C_\mathrm{mag}$ to zero at $T = 0$~K. Specifically, we show the best-fit cubic ($C_{\mathrm{mag}} \propto T^3$) and exponential ($C_{\mathrm{mag}} \propto e^{-\Delta/T}$) low-temperature extrapolations of $C_{\mathrm{mag}}$ as well as the best-fit extrapolation of the form $C_{\mathrm{mag}} \propto (1/T^2)e^{-\Delta/T}$. The gap energy was set to $\Delta = 0.025$~meV for the exponential extrapolation and the extrapolation of the of the form $C_{\mathrm{mag}} \propto (1/T^2)e^{-\Delta/T}$. (b) The entropy recovered for Ce$_2$Hf$_2$O$_7$ in this work via $S_\mathrm{mag} = \int_{0}^T\frac{C_\mathrm{mag}}{T}dT$, using each of the extrapolations in (a) to give $C_{\mathrm{mag}}$ below the lowest-temperature data point. (a) also shows the minimum of the dip in $C_{\mathrm{mag}}$ that occurs between the sharp low-temperature peak and the broad hump at higher temperature (red asterisk), as well as the point in $C_{\mathrm{mag}}$ (green triangle) for which the entropy recovered above this point to $T = 10$~K is the Pauling entropy $R[\ln(2) - (1/2)\ln(3/2)]$. (b) shows the entropy recovered above these two aforementioned points in $C_{\mathrm{mag}}$.} 
\label{FigureS3}
\end{figure}


Notably, even without extrapolation of $C_{\mathrm{mag}}$ below the lowest-temperature data point, the recovered entropy from the lowest-temperature data point to $T = 10$~K exceeds the value of $R\ln(2)$ expected for a CEF ground state doublet, by about 2\% of $R\ln(2)$, despite the fact that the conclusion of a CEF ground state doublet in Ce$_2$Hf$_2$O$_7$ is a robust conclusion (see Ref.~\cite{Poree2022} and below in this SM). This apparent inconsistency may be due to the fact that La$_2$Hf$_2$O$_7$ measurements were used to estimate the phonon contribution in the measured heat capacity from Ce$_2$Hf$_2$O$_7$. In fact, a slight overestimation of $R\ln(2)$ is generally consistent with expectations based on the fact that La$^{3+}$ is lighter than Ce$^{3+}$: Because of this mass difference, one would generally expect some phonons to be at slightly higher energies for La$_2$Hf$_2$O$_7$ compared to the analogous phonons for Ce$_2$Hf$_2$O$_7$, and this would be consistent with an undersubtraction of the phonon contribution for Ce$_2$Hf$_2$O$_7$ when using this method. However, there are also other factors that may cause or contribute to this. For example, any overestimation of $C_{\mathrm{mag}}$ (but still within the error bars) would lead to a corresponding overestimation of the entropy integral of $C_{\mathrm{mag}}/T$. Due to experimental factors like these, it is common for measured entropies to vary from their expected values within about 5\%. Because of this, we have labeled both $R\ln(2)$ and $1.05R\ln(2)$ in Figures 3 and S3. 

We expand on our extrapolation and entropy analysis in Figure~S3, where we now include an extrapolation of the form $C_{\mathrm{mag}} \propto (1/T^2)e^{-\Delta/T}$ along with the cubic ($C_{\mathrm{mag}}~\propto~T^3$) and exponential ($C_{\mathrm{mag}}~\propto~e^{-\Delta/T}$) low-temperature extrapolations of $C_{\mathrm{mag}}$. Figure~S3(a) shows the best-fitting low-temperature extrapolation for each of these forms and Figure~S3(b) shows the entropy recovered via $S_\mathrm{mag} = \int_{0}^T\frac{C_\mathrm{mag}}{T}dT$ using each of these extrapolations to give $C_{\mathrm{mag}}$ below the lowest-temperature data point. Both the exponential extrapolation and the $C_{\mathrm{mag}} \propto (1/T^2)e^{-\Delta/T}$ extrapolation give $R\ln(2)$ entropy within 5\% at $T=10$~K, while the cubic extrapolation results in a significant overestimation of the expected $R\ln(2)$ entropy at $T=10$~K. 

For both the exponential extrapolation and the extrapolation of the form $C_{\mathrm{mag}} \propto (1/T^2)e^{-\Delta/T}$, the gap value was set to $\Delta = 0.025$~meV, consistent with the gap value measured via high energy-resolution neutron scattering measurements on Ce$_2$Hf$_2$O$_7$ in Ref.~\cite{Poree2023b}. Attempts to fit the gap value $\Delta$ using these extrapolations yield best-fit $\Delta$ values that are significantly smaller than $\Delta = 0.025$~meV, and these fits correspondingly result in significant overestimation of the expected $R\ln(2)$ entropy at $T=10$~K.

We end our entropy-analysis with two points of interest in the $C_{\mathrm{mag}}$ reported for Ce$_2$Hf$_2$O$_7$ in this work. First, we examine the entropy recovered after the dip in $C_{\mathrm{mag}}$ that occurs between the sharp low-temperature peak and the broad hump at higher temperature. In further detail, the point at the minimum of this dip is shown in red in Fig.~S3(a), and the entropy recovered at temperatures above this point is shown in Fig.~S3(b). We also highlight the point in $C_{\mathrm{mag}}$ for which the entropy recovered above that point to $T = 10$~K is the Pauling entropy $R[\ln(2) - (1/2)\ln(3/2)]$ associated with the classical spin ice degeneracy. This $C_{\mathrm{mag}}$ point is shown in green in Fig.~S3(a) and the corresponding recovery of the Pauling entropy from that point to $T = 10$~K is shown in Fig.~S3(b).

\begin{figure*}[t]
\linespread{1}
\par
\includegraphics[width=7.2in]{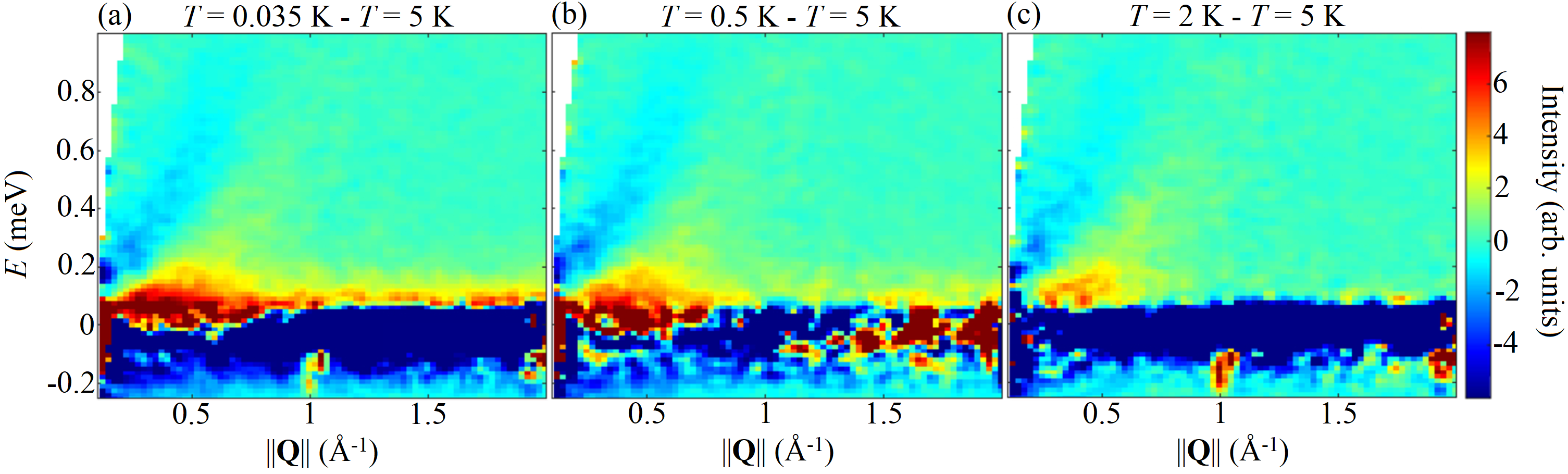}
\par
\caption{The temperature evolution of the low-energy inelastic neutron scattering in the powder-averaged spectra from our single crystal sample of Ce$_2$Hf$_2$O$_7$ aligned in the $(H,H,L)$ scattering plane with an incident energy of $E_\mathrm{i} = 3.7$meV. Specifically, this shows datasets measured at $T$ =  0.035~K~(a), 0.5~K~(b), and 2~K~(c) with a $T = 5$~K dataset subtracted in each case.} 
\label{FigureS4}
\end{figure*}

\section{Low-Energy Inelastic Neutron Scattering}

In this work we present low-energy inelastic neutron scattering data from a time-of-flight neutron scattering experiment on Ce$_2$Hf$_2$O$_7$ using the LET instrument at the ISIS Neutron and Muon Source. This time-of-flight experiment used incident neutron energies of $E_{\mathrm{i}} = 1.77$~meV and $E_{\mathrm{i}} = 3.7$~meV with 240~Hz chopper frequency, yielding energy resolutions of $\sim$0.04~meV ($E_{\mathrm{i}} = 1.77$~meV) and $\sim$0.08~meV ($E_{\mathrm{i}} = 3.7$~meV) at the elastic line. For this experiment, our $\sim$3.5~gram single crystal sample of Ce$_2$Hf$_2$O$_7$ was mounted in a copper sample holder and aligned in the $(H,H,L)$ scattering plane. For each temperature chosen for the experiment, the sample was rotated in the $(H,H,L)$ plane in 2$^\circ$ steps through a total of 360$^\circ$ and the data was subsequently symmetrized. This symmetrization process helps wash out scattering from the sample holder and sample environment equipment, in favor of scattering from Ce$_2$Hf$_2$O$_7$ (which obeys the applied symmetries). This symmetrization process is further discussed in the supplemental material of Ref.~\cite{Gaudet2019}. The Horace software package was used in analyzing the time-of-flight neutron scattering data presented in this work~\cite{Ewings2016}. 

The energy cuts through the powder-averaged data in Fig.~4(a) of the main text use the $E_{\mathrm{i}} = 1.77$~meV dataset, which has the lower energy-resolution of the two datasets. The $||\mathbf{Q}||$-cuts through the powder-averaged data in Fig.~4(b) of the main text use the $E_{\mathrm{i}} = 3.7$~meV dataset, which has the higher neutron flux of the two incident energies. The powder-averaged datasets were achieved through a directional average of the single crystal data, over the direction of $\mathbf{Q}$ for each $||\mathbf{Q}||$.

Fig.~4(c) of the main text shows the symmetrized $T=0.035$~K - $T=5$~K temperature-difference inelastic neutron scattering signal in the $(H,H,L)$ plane measured from our single crystal sample of Ce$_2$Hf$_2$O$_7$ with an incident energy of $E_{\mathrm{i}} = 3.7$~meV and for an integration in neutron energy transfer over the range $E = [0.1, 0.2]$~meV. This dataset in the $(H,H,L)$ plane uses an integration in the out-of-plane direction, $(K,\bar{K}, 0)$, over the range $K = [-0.1, 0.1]$. 

The onset of the magnetic inelastic neutron scattering signal from Ce$_2$Hf$_2$O$_7$ with decreasing temperature is shown by the energy cuts and $||\mathbf{Q}||$-cuts of the powder-averaged temperature-difference data in Fig.~4(a,b) of the main text. We also show this onset of inelastic signal with decreasing temperature in Fig.~S4, which shows the full powder-averaged neutron scattering signal measured from Ce$_2$Hf$_2$O$_7$ at $T=0.035$~K (a), $T=0.5$~K (b), and $T=2$~K (c) with a $T=5$~K dataset subtracted in each case. Specifically, Fig.~S4 shows positive net scattering with energy-center near $E=0.1$~meV, which onsets by $T=2$~K and grows in intensity with decreasing temperature. Importantly, Fig.~S4 shows no signs for magnetic Bragg scattering at any $||\mathbf{Q}||$; Magnetic Bragg scattering would appear as peaks in the net scattering, centered on $E = 0$~meV and specific $||\mathbf{Q}||$ values and onsetting with decreasing temperature.

The energy integration used in Fig.~4 of the main text, over the range $E = [0.1, 0.2]$~meV, was chosen to cover the dominant portion of the positive net scattering in the temperature-difference inelastic neutron scattering signal, while avoiding negative net scattering centered on $E = 0$~meV that likely results from subtraction of paramagnetic elastic scattering at high temperature. This negative net scattering is shown as dark blue in the powder-averaged temperature-difference data of Fig.~S4. The energy integration over the range $E = [0.1, 0.2]$~meV also helps avoid noise resulting from the imperfect subtraction of elastic coherent scattering and nuclear Bragg scattering, which are far more intense than the weak magnetic signal from Ce$^{3+}$'s small ($\lesssim 1.29 \mu_B$) magnetic moment in Ce$_2$Hf$_2$O$_7$ at low-temperature.

Ref.~\cite{Kim2022} uses exact diagonalization and semiclassical molecular dynamics calculations to compute the diffuse neutron scattering signals in the $(H,H,L)$ plane for the four QSI phases present in the ground state phase diagram predicted for dipolar-octupolar pyrochlores at the nearest-neighbor level: the U(1)$_0$ and U(1)$_\pi$ QSIs, each of which can be dipolar or octupolar in nature. The U(1)$_0$ and U(1)$_\pi$ QSIs are distinguished based on whether their U(1) flux is equal to 0 or $\pi$ when a spinon traverses a hexagonal plaquette in the pyrochlore lattice, and the dipolar and octupolar QSIs are distinguished by whether their emergent electric field transforms under time-reversal and site symmetry as a dipole or octupole~\cite{Benton2020, Patri2020, Huang2020}. The diffuse neutron scattering signals predicted in Ref.~\cite{Kim2022} for both U(1)$_\pi$ QSIs and the dipolar U(1)$_0$ QSI are consistent with the snowflake pattern of scattering in the $(H,H,L)$ plane measured from Ce$_2$Hf$_2$O$_7$. However, Ref.~\cite{Kim2022} predicts a pattern of scattering for the octupolar U(1)$_0$ QSI that is inverted compared to the other QSIs, with a snowflake pattern of scattering that is \textit{less} intense than the scattering nearby in reciprocal space. Accordingly, the predicted result for an octupolar U(1)$_0$ QSI is inconsistent with our measured result [Fig.~4(c) of the main text]. 

\section{Numerical Linked Cluster Calculations of \texorpdfstring{$C_{\mathrm{mag}}$}~}

We begin this section by discussing how we arrive at the pseudospin interaction Hamiltonian relevant for Ce$_2$Hf$_2$O$_7$ at the nearest-neighbor level in zero-field with directional ambiguity removed, $\mathcal{H}_{\mathrm{ABC}}$ (Eq.~1 of the main text). In general, when a description in terms of pseudospin is permitted by a CEF ground state that is well-separated in energy from the excited CEF states, the symmetry of the crystal electric field ground state dictates the general form of the pseudospin interaction Hamiltonian~\cite{RauReview2019}. The symmetry-allowed pseudospin-1/2 interaction Hamiltonian at the nearest-neighbor level for pyrochlores with a dipole-octupole CEF ground state doublet is given by~\cite{Huang2014, RauReview2019}:

\begin{equation}\label{eq:1}
\begin{split}
    \mathcal{H}_\mathrm{DO} & = \sum_{\langle ij \rangle}[     J_{x}{S_i}^{x}{S_j}^{x} + J_{y}{S_i}^{y}{S_j}^{y} + J_{z}{S_i}^{z}{S_j}^{z} \\ 
    & + J_{xz}({S_i}^{x}{S_j}^{z} + {S_i}^{z}{S_j}^{x})] - g_z \mu_\mathrm{B} \sum_{i} \mathbf{h} \cdot\hat{{\bf z}}_i \; {S_i}^{z}  \;,
\end{split}
\end{equation}

\noindent where ${S_{i}}^{\alpha}$ ($\alpha = x$, $y$, $z$) are the pseudospin components of the rare-earth atom $i$ in the local $\{x$, $y$, $z\}$ coordinate frame. This coordinate frame is defined locally for each ion $i$ with the $\mathbf{z}_i$ anisotropy axis along the threefold rotation axis through rare-earth site $i$ and with $\mathbf{y}_i$ along one of the symmetrically equivalent twofold rotation axes through rare-earth site $i$, where $\mathbf{x}_i = \mathbf{y}_i \cross \mathbf{z}_i$. The second sum represents the Zeeman interaction between the rare-earth ion and the magnetic field $\mathbf{h}$. The anisotropic $g$-factor $g_z$ is determined by the CEF ground state doublet, which gives $g_z = 2.57$ for the pure $|m_J = \pm 3/2 \rangle$ ground state doublet estimated for Ce$^{3+}$ in Ce$_2$Hf$_2$O$_7$ (see Ref.~\cite{Poree2023b} and below in this SM). 

This nearest-neighbor exchange Hamiltonian can then be simplified via rotation of each local $\{x, y, z\}$ coordinate frame by $\theta$ about the respective local $y$-axis, where $\theta$ is given by~\cite{Huang2014, Benton2016}:

\begin{equation} \label{eq:2}
    \theta = \frac{1}{2}\tan^{-1}\bigg(\frac{2J_{xz}}{J_{x}-J_{z}}\bigg)\;.
\end{equation}

These rotations yield new local coordinate frames which are commonly denoted as the local $\{\tilde{x},\tilde{y},\tilde{z}\}$ coordinate frames, and the new Hamiltonian in the $\{\tilde{x},\tilde{y},\tilde{z}\}$ coordinate frames is the ``XYZ'' Hamiltonian~\cite{Huang2014}:

\begin{equation} \label{eq:3}
\begin{split}
    \mathcal{H}_\mathrm{XYZ} & = \sum_{\langle ij \rangle}[     J_{\tilde{x}}{S_i}^{\tilde{x}}{S_j}^{\tilde{x}} + J_{\tilde{y}}{S_i}^{\tilde{y}}{S_j}^{\tilde{y}} + J_{\tilde{z}}{S_i}^{\tilde{z}}{S_j}^{\tilde{z}}] \\ 
    & - g_z \mu_\mathrm{B} \sum_{i} \mathbf{h} \cdot\hat{{\bf z}}_i({S_i}^{\tilde{z}}\cos\theta + {S_i}^{\tilde{x}}\sin\theta) \;.
\end{split}
\end{equation}

For zero field, defining $\{a$, $b$, $c\}$ to be the permutation of $\{\tilde{x},\tilde{y},\tilde{z}\}$ that satisfies $|J_a|\geq|J_b|,|J_c|$ and $J_b \geq J_c$ then gives the Hamiltonian $\mathcal{H}_\mathrm{ABC}$:

\begin{equation} \label{eq:4}
\begin{split}
    \mathcal{H}_\mathrm{ABC} & = \sum_{<ij>}[J_{a}{S_i}^{a}{S_j}^{a} + J_{b}{S_i}^{b}{S_j}^{b} + J_{c}{S_i}^{c}{S_j}^{c}] \\
    & = \sum_{\langle ij \rangle}[J_{a}{S_i}^{a}{S_j}^{a} - J_{\pm}({S_i}^{+}{S_j}^{-} + {S_i}^{-}{S_j}^{+}) \\
    & + J_{\pm\pm}({S_i}^{+}{S_j}^{+} + {S_i}^{-}{S_j}^{-})] \;,
\end{split}
\end{equation}\\

\noindent where $J_\pm = -\frac{1}{4}(J_b + J_c)$ and $J_{\pm\pm} = \frac{1}{4}(J_b - J_c)$. This is the Hamiltonian we use for our numerical linked cluster (NLC) calculations and it is also shown in Eq.~1 of the main text.

We use the NLC method with the Hamiltonian $\mathcal{H}_\mathrm{ABC}$ to calculate $C_{\mathrm{mag}}$ over the available parameter space of $\mathcal{H}_\mathrm{ABC}$. We compare these calculations with the $C_{\mathrm{mag}}$ reported for Ce$_2$Hf$_2$O$_7$ in this work and determine values of $(J_a, J_b, J_c)$ that give the best agreement between the calculation and measurement. The NLC method calculates $C_{\mathrm{mag}}$ (or other physical quantities) by first calculating the contributions from different sized clusters of tetrahedra in the pyrochlore lattice, and ignoring the contributions from larger clusters which only become relevant at temperatures below a low-temperature cutoff. The order of these quantum NLC calculations refers to the maximum number of tetrahedra considered in a cluster, NLC calculations up to seventh order were preformed to model $C_{\mathrm{mag}}$. Further details of the NLC method are provided in Ref.~\cite{Applegate2012, Tang2013, Tang2015, Schafer2020, schaefer_magnetic_2022} for example. The methodology specific to the seventh-order calculations is described in Ref.~\cite{Schafer2020}. 

Throughout this paper, we compare the magnetic heat capacity calculated using sixth-order NLC calculations, $C_\mathrm{mag}^{\mathrm{NLC},6}$, to the magnetic heat capacity measured from single crystal Ce$_2$Hf$_2$O$_7$, $C_{\mathrm{mag}}^\mathrm{exp}$, using the goodness-of-fit measure,
\begin{equation}\label{eq:5}
\left\langle \frac{\delta^2}{\epsilon^2} \right\rangle_{\hspace*{-3pt} C} = \sum_{T_\mathrm{exp}} \frac{[C_\mathrm{mag}^{\mathrm{NLC},6}(T_\mathrm{exp})-C_{\mathrm{mag}}^\mathrm{exp}(T_\mathrm{exp})]^2}{\epsilon_{C, \mathrm{NLC},6}(T_\mathrm{exp})^2 + \epsilon_{C, \mathrm{exp}}(T_\mathrm{exp})^2}  ~, 
\end{equation}

where $\epsilon_{C, \mathrm{exp}}(T_\mathrm{exp})$ is the experimental uncertainty on the measured heat capacity at temperature $T_\mathrm{exp}$, and $\epsilon_{C, \mathrm{NLC},6}(T_\mathrm{exp})$ is the uncertainty associated with the sixth-order NLC calculations at temperature $T_\mathrm{exp}$,

\vspace*{-5pt}

\begin{equation}\label{eq:6}
\epsilon_{C, \mathrm{NLC},6}(T_\mathrm{exp}) 
= \mathrm{max}_{T \geq T_\mathrm{exp}}~|C_\mathrm{mag}^{\mathrm{NLC},6}(T) - C_\mathrm{mag}^{\mathrm{NLC},5}(T)| , 
\end{equation}
\vspace*{1pt}

where $C_\mathrm{mag}^{\mathrm{NLC},5}$ is the magnetic heat capacity calculated using fifth-order NLC calculations.

We first used sixth-order NLC calculations, with Euler transformations to improve convergence (see Ref.~\cite{Smith2022} for example), in order to fit the zero-field heat capacity measured from Ce$_2$Hf$_2$O$_7$ and determine the best-fitting exchange parameters $J_a$, $J_b$, and $J_c$. Specifically, $C_{\mathrm{mag}}$ curves were calculated for values of $(J_a, J_b, J_c)$ over the entire available parameter space, and we compare the NLC-calculated heat capacity for each parameter set to the heat capacity measured from Ce$_2$Hf$_2$O$_7$ using the goodness-of-fit measure $\langle \delta^2/\epsilon^2 \rangle_{C}$ in Eq.~\ref{eq:5}. The overall energy scale of the exchange parameters was fit to the high-temperature tail of the heat capacity so as to minimize $\langle \delta^2/\epsilon^2 \rangle_{C}$ summed over the range from $T_\mathrm{exp}$ = 1.5~K to 3~K (see Ref.~\cite{Yahne2024} for example). The exchange parameters $J_a$, $J_b$, and $J_c$ are then determined according to minimization of $\langle \delta^2/\epsilon^2 \rangle_{C}$ summed over the range from $T_\mathrm{exp}$ = 0.1~K to 1.5~K. For most parameter sets, and specifically those corresponding to a QSI ground state in the nearest-neighbor ground state phase diagram [Fig.~2(b) of the main text], this restricts the fit to the regime where the NLC calculations converge. The value of $\langle \delta^2/\epsilon^2 \rangle_{C}$ over the available parameter space is shown in Fig.~2(b) of the main text. The results of this fitting procedure yields the best fitting parameters $(J_a,J_b,J_c) = (0.050, 0.021, 0.004)$~meV [labeled as A in Fig.~2(b) of the main text] and the local minimum $(J_a,J_b,J_c) = (0.051, 0.008, -0.018)$~meV [labeled as B in Fig.~2(b) of the main text]. 

The seventh-order NLC calculations using the A and B parameters are shown in Fig.~2(a) of the main text. The sixth and seventh order NLC calculations for the A parameters are converged with one another down to $T \sim 0.05$~K, while for the B parameters they diverge from one another below $T \sim 0.15$~K. This difference in convergence is due to the different phases described by these parameters. 

\section{Numerical Linked Cluster Calculations of $S(\mathbf{Q})$}

In this section we discuss our sixth-order NLC calculations of the equal-time structure factor $S(\mathbf{Q})$. The details of these calculations are described in Refs.~\cite{Smith2024, schaefer_NLCE_corr_2024}. Here we compute $S(\mathbf{Q})$ for different permutations of the A and B parameter sets for the XYZ Hamiltonian, obtained from our NLC fitting to the $C_{\mathrm{mag}}$ measured from Ce$_2$Hf$_2$O$_7$, and for different values of the $\theta$ parameter (see Eqs.~\ref{eq:2} and \ref{eq:3}). We allow $\theta$ to vary in the range from 0 to $\pi$/4. This is enough to cover all distinguishable scenarios, since changing the sign of $\theta$ does not affect any quantity considered here, and shifting $\theta$ to $\theta + \pi$/2 is the same as reversing the sign of $\theta$ and swapping the values of $J_{\tilde{x}}$ and $J_{\tilde{z}}$, which is already covered by considering all six permutations of exchange parameters. 

Fig.~S5 shows the NLC-calculated $S(\mathbf{Q})$ for the six different permutations of the A parameters, $(J_a,J_b,J_c) = (0.050, 0.021, 0.004)$~meV, for $\theta = 0$. Specifically, Fig.~S5 shows the $S(\mathbf{Q})$ calculated for the A parameters with $\theta = 0$ and for $(J_{\tilde{x}}, J_{\tilde{y}}, J_{\tilde{z}})$ equal to $(J_a, J_b, J_c)$ [Fig.~S5(a)], $(J_a, J_c, J_b)$ [Fig.~S5(b)], $(J_b, J_a, J_c)$ [Fig.~S5(c)], $(J_b, J_c, J_a)$ [Fig.~S5(d)], $(J_c, J_a, J_b)$ [Fig.~S5(e)], and $(J_c, J_b, J_a)$ [Fig.~S5(f)]. Fig.~S6 and Fig.~S7 show the NLC-calculated $S(\mathbf{Q})$ for these same permutations of the A parameters for $\theta = 0.125\pi$ and $\theta = 0.25\pi$, respectively. The NLC-predicted scattering for the A parameters agrees reasonably-well with the measured data for all permutations of the A parameters and all values of $\theta$.


\begin{figure}[t]
\linespread{1}
\par
\includegraphics[width=3.4in]{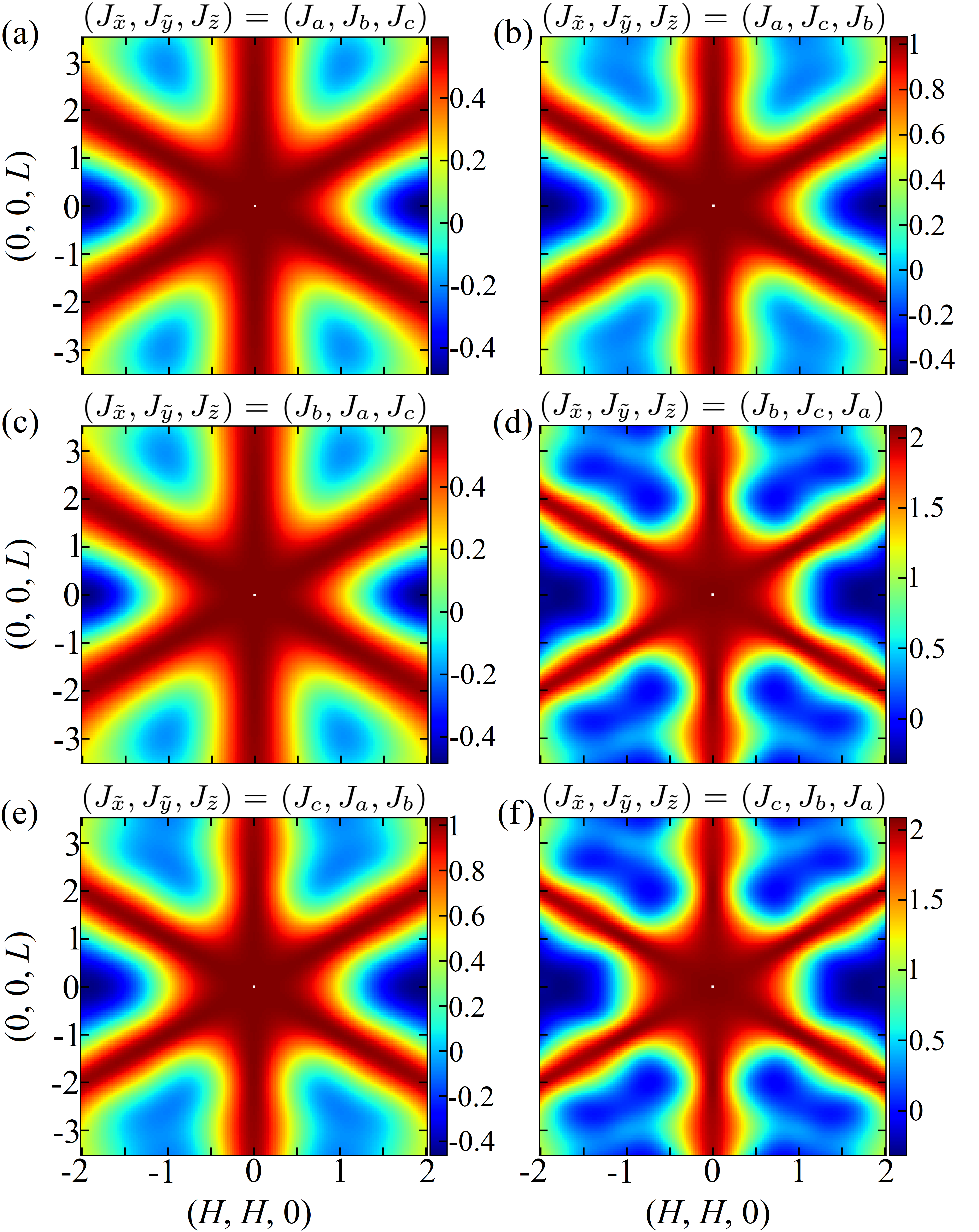}
\par
\caption{The equal-time structure factor in the $(H,H,L)$ plane of reciprocal space at $T = 0.3$~K with the corresponding $T = 5$~K calculation subtracted, predicted according to sixth-order NLC using $\theta = 0$ with the different permutations of the A parameters, $(J_a,J_b,J_c) = (0.050, 0.021, 0.004)$~meV. Specifically, we show this calculation for $(J_{\tilde{x}}, J_{\tilde{y}}, J_{\tilde{z}})$ equal to (a)~$(J_a, J_b, J_c)$, (b)~$(J_a, J_c, J_b)$, (c) $(J_b, J_a, J_c)$, (d) $(J_b, J_c, J_a)$, (e) $(J_c, J_a, J_b)$, and (f) $(J_c, J_b, J_a)$.} 
\label{FigureS5}
\end{figure}


\begin{figure}[t]
\linespread{1}
\par
\includegraphics[width=3.4in]{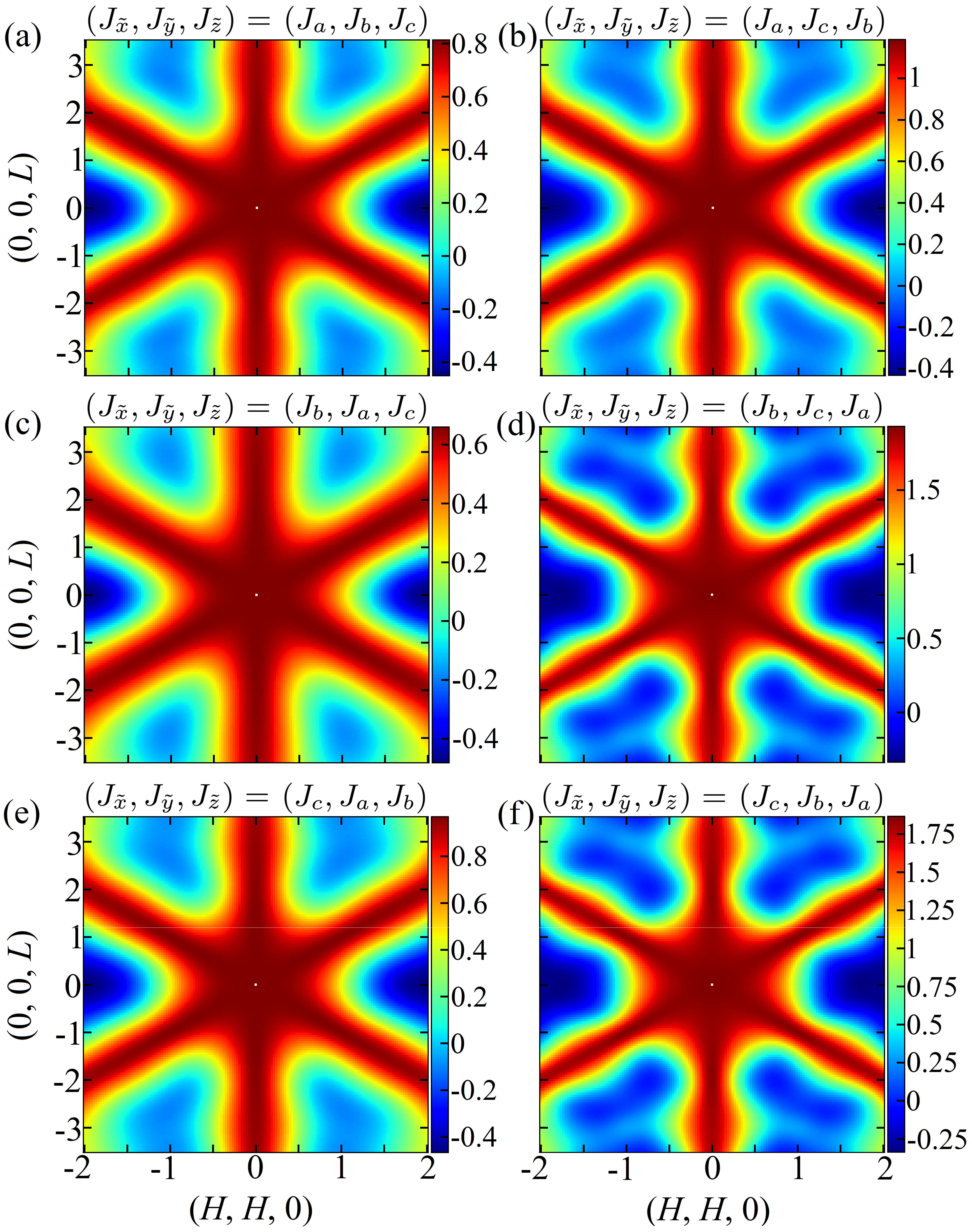}
\par
\caption{The equal-time structure factor in the $(H,H,L)$ plane of reciprocal space at $T = 0.3$~K with the corresponding $T = 5$~K calculation subtracted, predicted according to sixth-order NLC using $\theta = 0.125\pi$ with the different permutations of the A parameters, $(J_a,J_b,J_c) = (0.050, 0.021, 0.004)$~meV. Specifically, we show this calculation for $(J_{\tilde{x}}, J_{\tilde{y}}, J_{\tilde{z}})$ equal to (a)~$(J_a, J_b, J_c)$, (b)~$(J_a, J_c, J_b)$, (c) $(J_b, J_a, J_c)$, (d) $(J_b, J_c, J_a)$, (e) $(J_c, J_a, J_b)$, and (f) $(J_c, J_b, J_a)$.} 
\label{FigureS6}
\end{figure}



\begin{figure}[t]
\linespread{1}
\par
\includegraphics[width=3.4in]{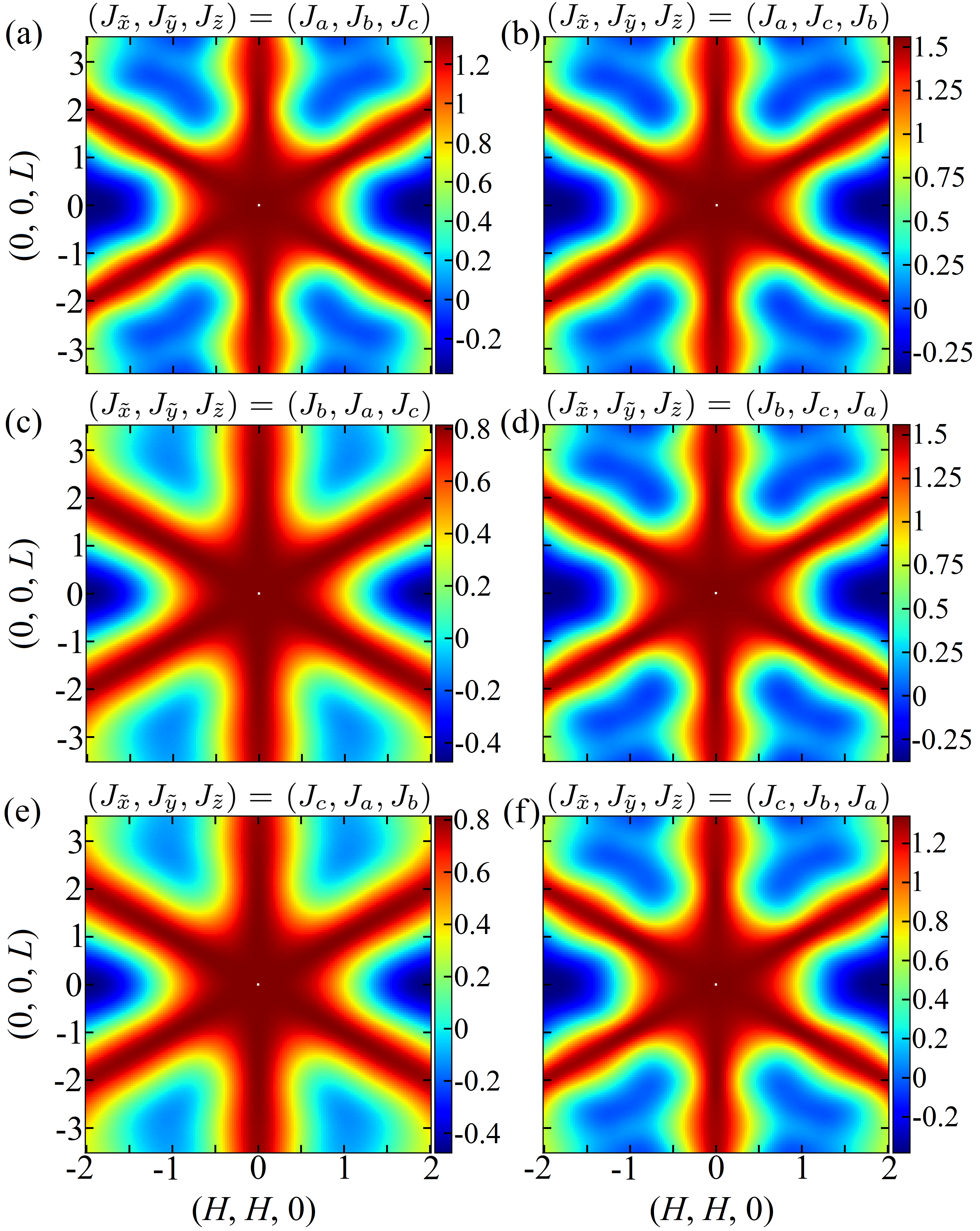}
\par
\caption{The equal-time structure factor in the $(H,H,L)$ plane of reciprocal space at $T = 0.3$~K with the corresponding $T = 5$~K calculation subtracted, predicted according to sixth-order NLC using $\theta = 0.25\pi$ with the different permutations of the A parameters, $(J_a,J_b,J_c) = (0.050, 0.021, 0.004)$~meV. Specifically, we show this calculation for $(J_{\tilde{x}}, J_{\tilde{y}}, J_{\tilde{z}})$ equal to (a)~$(J_a, J_b, J_c)$, (b)~$(J_a, J_c, J_b)$, (c) $(J_b, J_a, J_c)$, (d) $(J_b, J_c, J_a)$, (e) $(J_c, J_a, J_b)$, and (f) $(J_c, J_b, J_a)$. } 
\label{FigureS7}
\end{figure}


\begin{figure}[t]
\linespread{1}
\par
\includegraphics[width=3.4in]{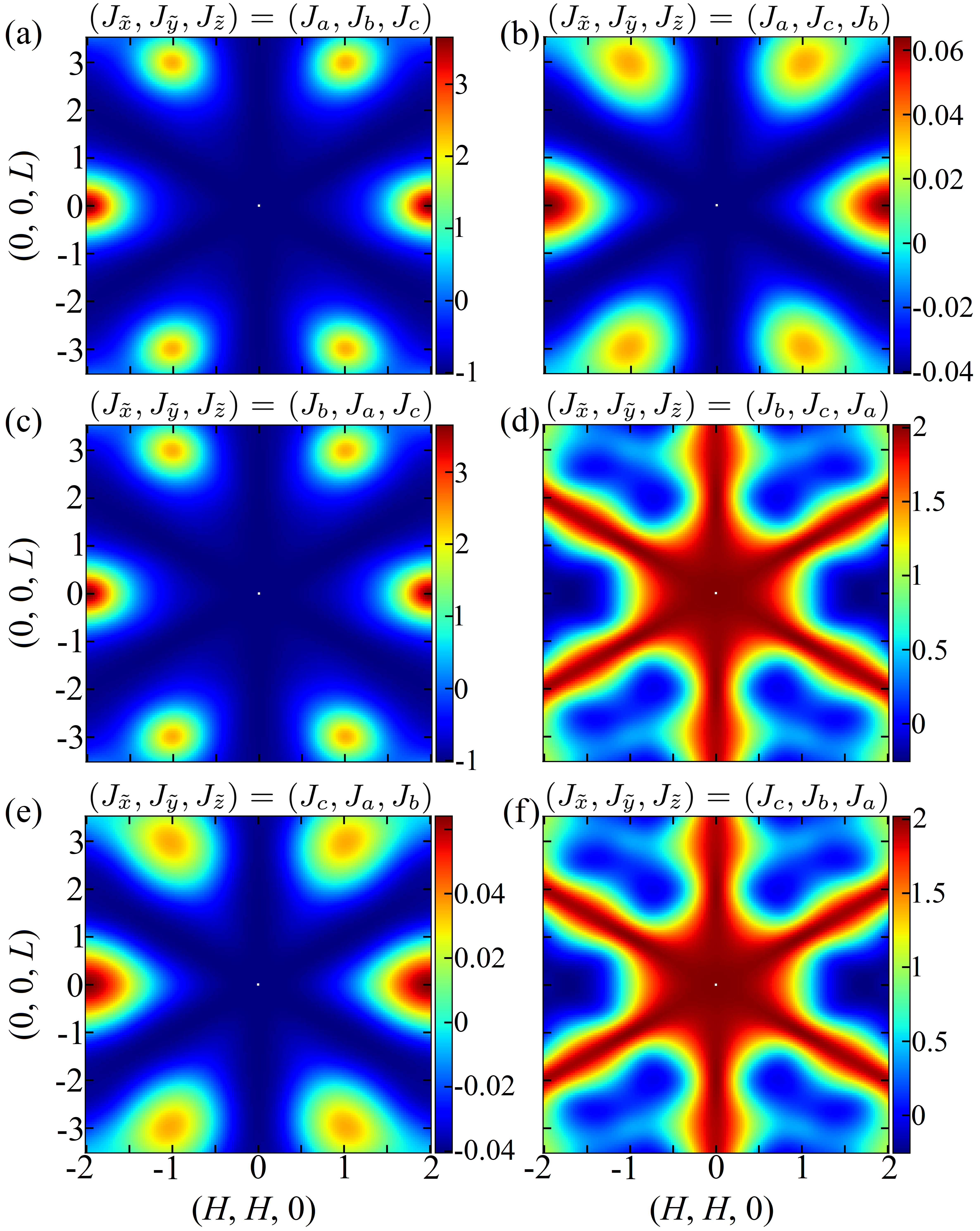}
\par
\caption{The equal-time structure factor in the $(H,H,L)$ plane of reciprocal space at $T = 0.3$~K with the corresponding $T = 5$~K calculation subtracted, predicted according to sixth-order NLC using $\theta = 0$ with the different permutations of the B parameters, $(J_a,J_b,J_c) = (0.051, 0.008, -0.018)$~meV. Specifically, we show this calculation for $(J_{\tilde{x}}, J_{\tilde{y}}, J_{\tilde{z}})$ equal to (a)~$(J_a, J_b, J_c)$, (b)~$(J_a, J_c, J_b)$, (c) $(J_b, J_a, J_c)$, (d) $(J_b, J_c, J_a)$, (e) $(J_c, J_a, J_b)$, and (f) $(J_c, J_b, J_a)$.} 
\label{FigureS8}
\end{figure}


\begin{figure}[t]
\linespread{1}
\par
\includegraphics[width=3.4in]{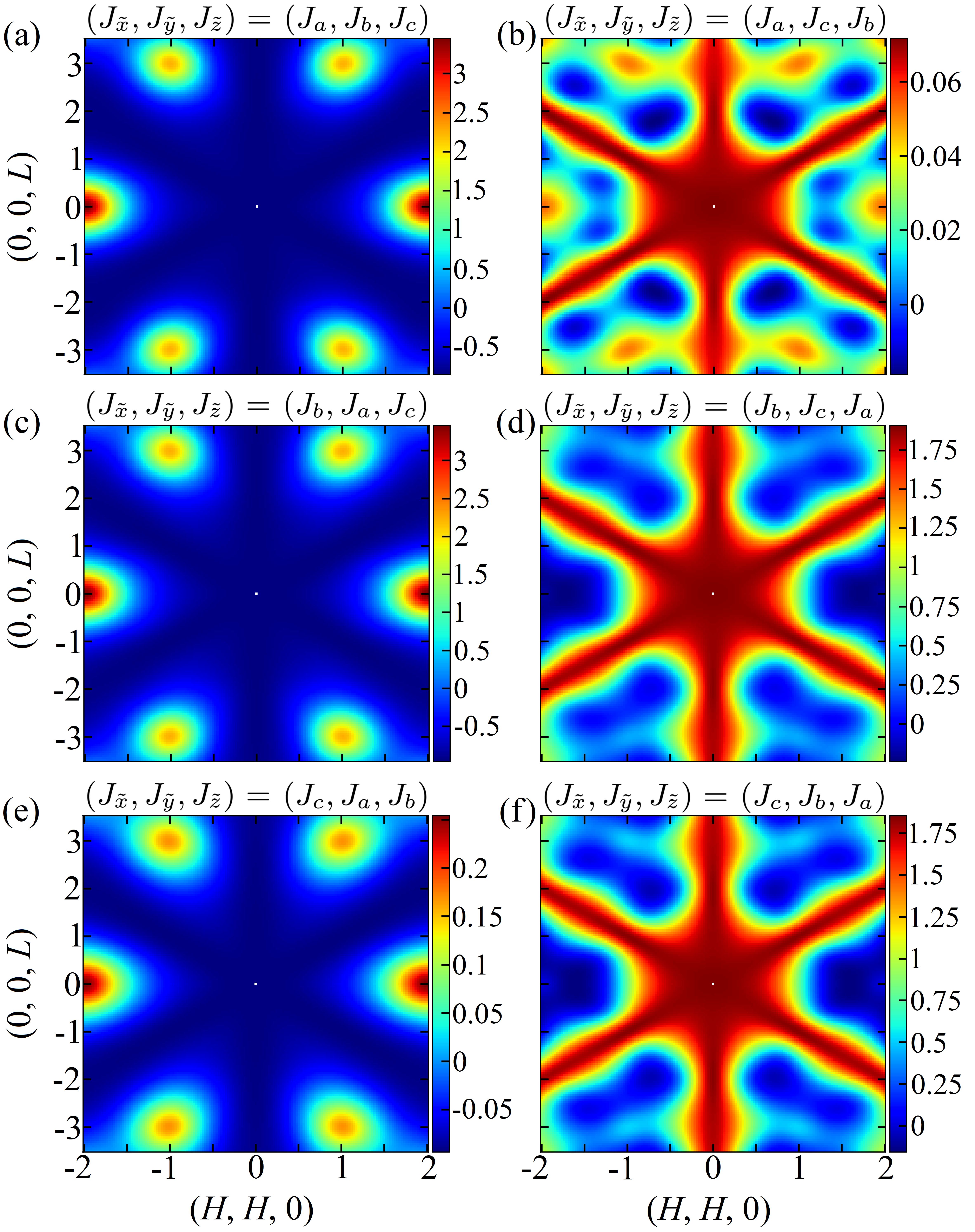}
\par
\caption{The equal-time structure factor in the $(H,H,L)$ plane of reciprocal space at $T = 0.3$~K with the corresponding $T = 5$~K calculation subtracted, predicted according to sixth-order NLC using $\theta = 0.075\pi$ with the different permutations of the B parameters, $(J_a,J_b,J_c) = (0.051, 0.008, -0.018)$~meV. Specifically, we show this calculation for $(J_{\tilde{x}}, J_{\tilde{y}}, J_{\tilde{z}})$ equal to (a)~$(J_a, J_b, J_c)$, (b)~$(J_a, J_c, J_b)$, (c) $(J_b, J_a, J_c)$, (d) $(J_b, J_c, J_a)$, (e) $(J_c, J_a, J_b)$, and (f) $(J_c, J_b, J_a)$.} 
\label{FigureS9}
\end{figure}


\begin{figure}[t]
\linespread{1}
\par
\includegraphics[width=3.4in]{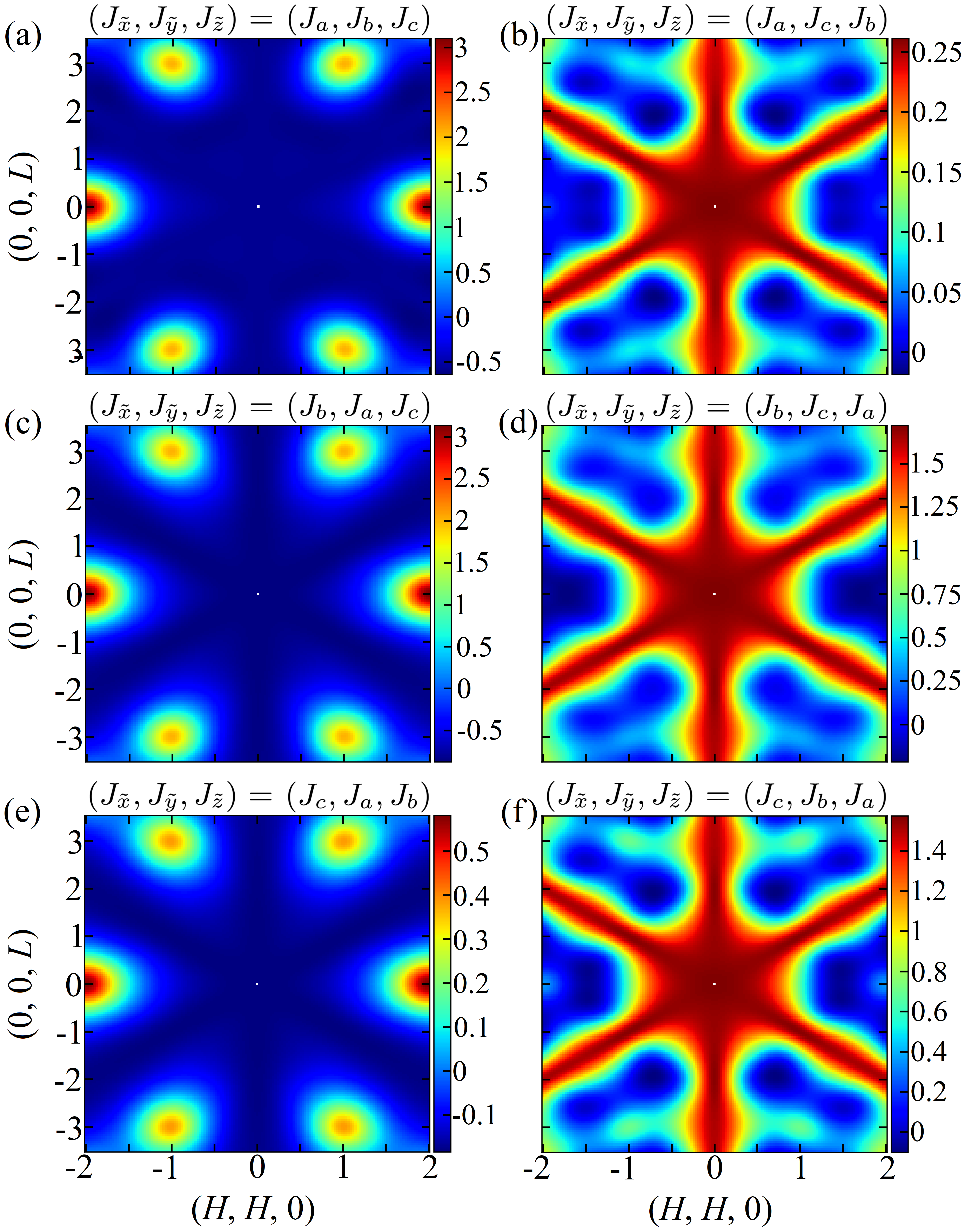}
\par
\caption{The equal-time structure factor in the $(H,H,L)$ plane of reciprocal space at $T = 0.3$~K with the corresponding $T = 5$~K calculation subtracted, predicted according to sixth-order NLC using $\theta = 0.125\pi$ with the different permutations of the B parameters, $(J_a,J_b,J_c) = (0.051, 0.008, -0.018)$~meV. Specifically, we show this calculation for $(J_{\tilde{x}}, J_{\tilde{y}}, J_{\tilde{z}})$ equal to (a)~$(J_a, J_b, J_c)$, (b)~$(J_a, J_c, J_b)$, (c) $(J_b, J_a, J_c)$, (d) $(J_b, J_c, J_a)$, (e) $(J_c, J_a, J_b)$, and (f) $(J_c, J_b, J_a)$.} 
\label{FigureS10}
\end{figure}


\begin{figure}[t]
\linespread{1}
\par
\includegraphics[width=3.4in]{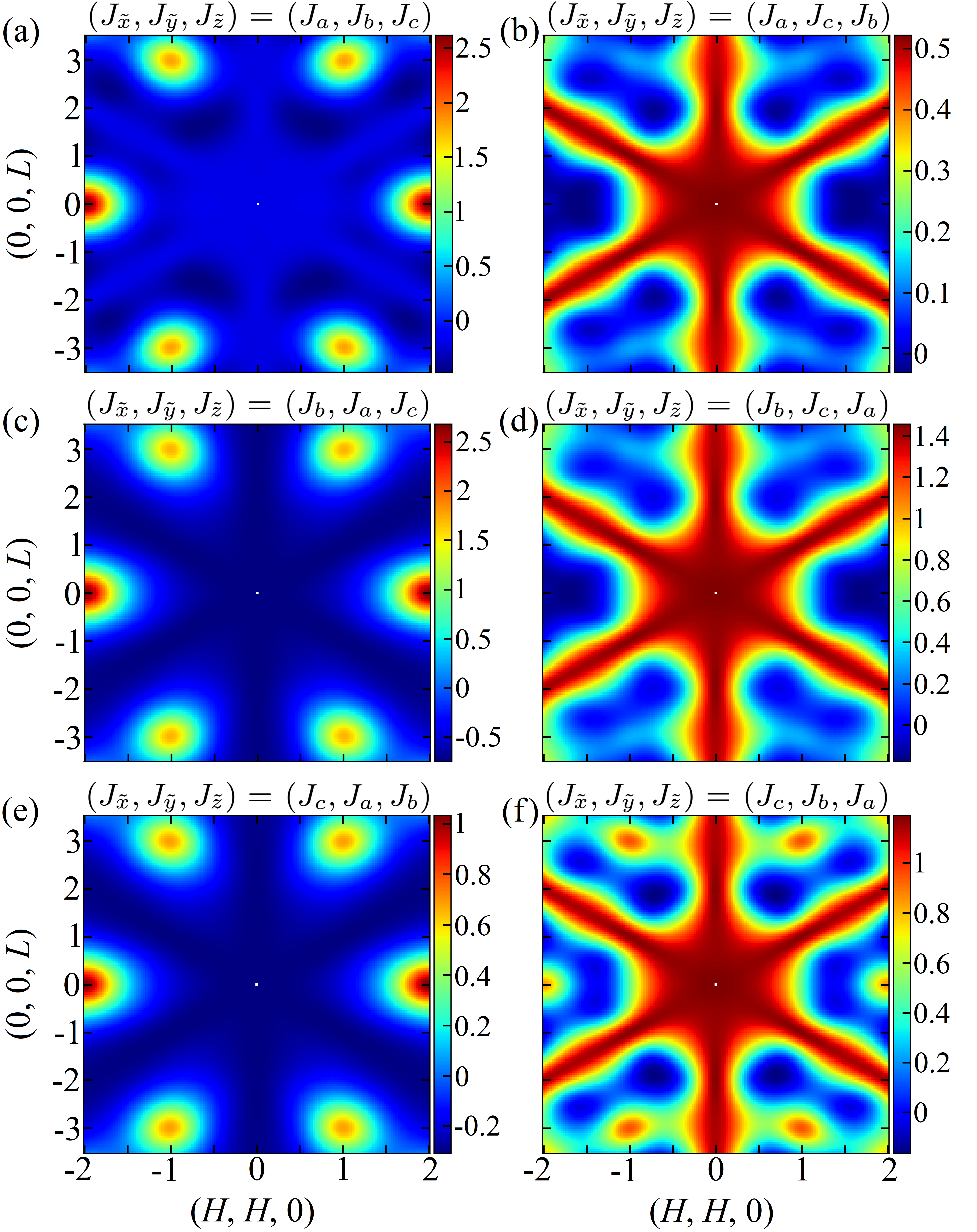}
\par
\caption{The equal-time structure factor in the $(H,H,L)$ plane of reciprocal space at $T = 0.3$~K with the corresponding $T = 5$~K calculation subtracted, predicted according to sixth-order NLC using $\theta = 0.175\pi$ with the different permutations of the B parameters, $(J_a,J_b,J_c) = (0.051, 0.008, -0.018)$~meV. Specifically, we show this calculation for $(J_{\tilde{x}}, J_{\tilde{y}}, J_{\tilde{z}})$ equal to (a)~$(J_a, J_b, J_c)$, (b)~$(J_a, J_c, J_b)$, (c) $(J_b, J_a, J_c)$, (d) $(J_b, J_c, J_a)$, (e) $(J_c, J_a, J_b)$, and (f) $(J_c, J_b, J_a)$.} 
\label{FigureS11}
\end{figure}


\begin{figure}[t]
\linespread{1}
\par
\includegraphics[width=3.4in]{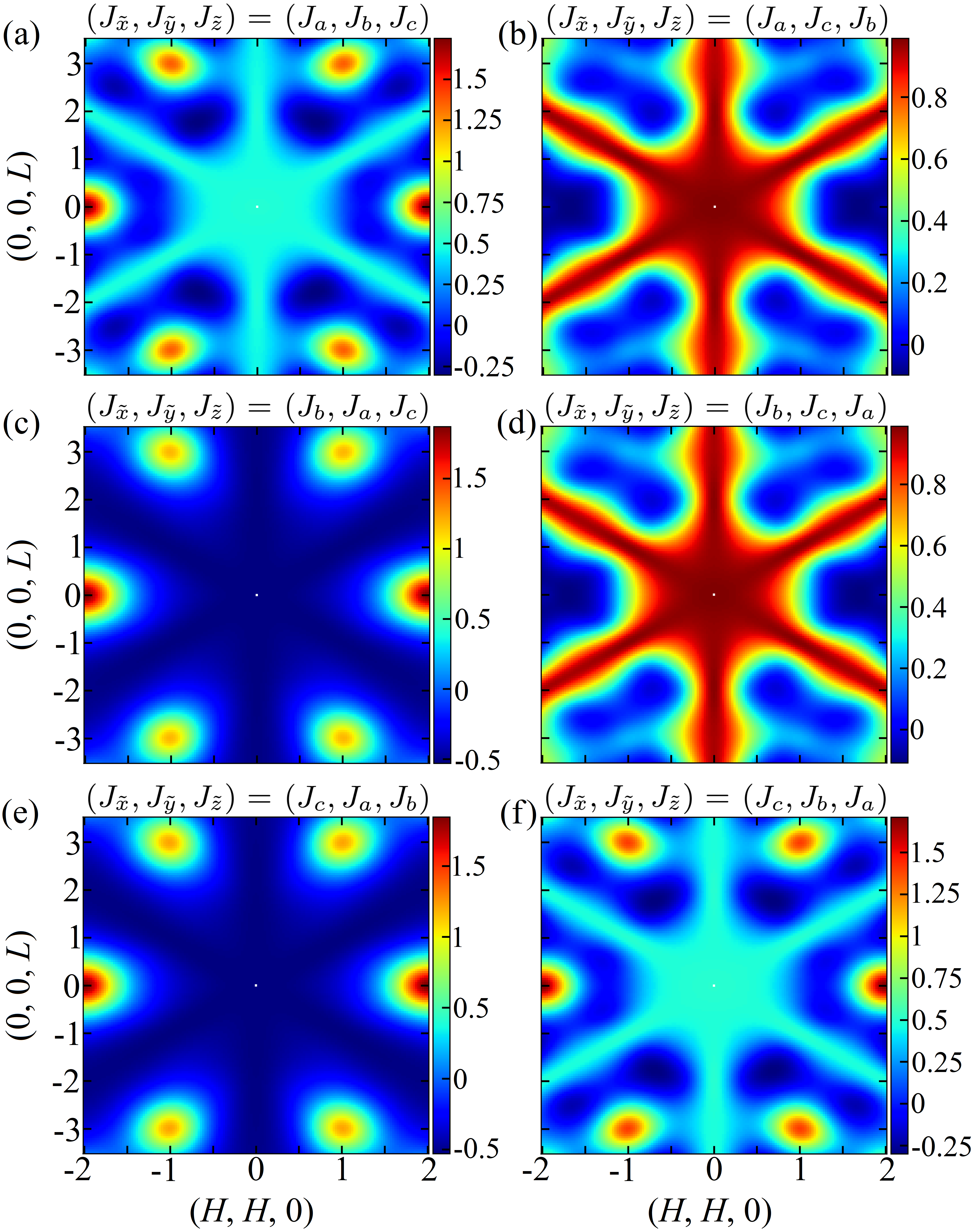}
\par
\caption{The equal-time structure factor in the $(H,H,L)$ plane of reciprocal space at $T = 0.3$~K with the corresponding $T = 5$~K calculation subtracted, predicted according to sixth-order NLC using $\theta = 0.25\pi$ with the different permutations of the B parameters, $(J_a,J_b,J_c) = (0.051, 0.008, -0.018)$~meV. Specifically, we show this calculation for $(J_{\tilde{x}}, J_{\tilde{y}}, J_{\tilde{z}})$ equal to (a)~$(J_a, J_b, J_c)$, (b)~$(J_a, J_c, J_b)$, (c) $(J_b, J_a, J_c)$, (d) $(J_b, J_c, J_a)$, (e) $(J_c, J_a, J_b)$, and (f) $(J_c, J_b, J_a)$.} 
\label{FigureS12}
\end{figure}

Fig.~S8 (Fig.~S9, Fig.~S10, Fig.~S11, Fig.~S12) shows the NLC-calculated $S(\mathbf{Q})$ for the six different permutations of the B parameters, $(J_a,J_b,J_c) = (0.051, 0.008, -0.018)$~meV, for $\theta = 0$ ($\theta = 0.075\pi$, $\theta = 0.125\pi$, $\theta = 0.175\pi$, $\theta = 0.25\pi$). The NLC-predicted scattering for the B parameters agrees reasonably-well with the measured data for $(J_{\tilde{x}}, J_{\tilde{y}}, J_{\tilde{z}})$ equal to $(J_a, J_c, J_b)$ when $0.075\pi \lesssim  \theta \leq \pi/4$, for $(J_{\tilde{x}}, J_{\tilde{y}}, J_{\tilde{z}})$ equal to $(J_b, J_c, J_a)$ for all values of $\theta$, and for $(J_{\tilde{x}}, J_{\tilde{y}}, J_{\tilde{z}})$ equal to $(J_c, J_b, J_a)$ when $0 \leq \theta \lesssim 0.175\pi$.

\section{Numerical Linked Cluster Calculations of Magnetic Susceptibility}

Our magnetic susceptibility measurements were taken on a 58~mg single crystal sample of Ce$_2$Hf$_2$O$_7$ using a Quantum Design magnetic property measurement system magnetometer equipped with a $^3$He insert, with a magnetic field of $h=0.01$~T along the $[1,1,0]$ direction. The measured magnetic susceptibility from Ce$_2$Hf$_2$O$_7$ is shown in Fig.~S13(e,f) and shows no indication of long-ranged magnetic order down to the lowest-temperature data point at $T \sim 0.5$~K. In this section, we discuss our fourth-order NLC fits to this experimental magnetic susceptibility data from Ce$_2$Hf$_2$O$_7$. 

The fits to the measured magnetic susceptibility were performed for each permutation of the A and B parameters of the XYZ Hamilton and for varying values of $\theta$ and $g_z$ (see Eq.~\ref{eq:3}). We compare the magnetic susceptibility calculated using fourth-order NLC calculations, $\chi^{\mathrm{NLC},4}$, to the magnetic susceptibility measured from single crystal Ce$_2$Hf$_2$O$_7$, $\chi^\mathrm{exp}$, using the goodness-of-fit measure,
\begin{equation}\label{eq:7}
\left\langle \frac{\delta^2}{\epsilon^2} \right\rangle_{\hspace*{-3pt} \chi} = \sum_{T_\mathrm{exp}} \frac{[\chi^{\mathrm{NLC},4}(T_\mathrm{exp})-\chi^\mathrm{exp}(T_\mathrm{exp})]^2}{\epsilon_{\chi, \mathrm{NLC},4}(T_\mathrm{exp})^2 + \epsilon_{\chi, \mathrm{exp}}(T_\mathrm{exp})^2}  ~, 
\end{equation}

where $\epsilon_{\chi, \mathrm{exp}}(T_\mathrm{exp})$ is the experimental uncertainty on the measured magnetic susceptibility at temperature $T_\mathrm{exp}$, and $\epsilon_{\chi, \mathrm{NLC},4}(T_\mathrm{exp})$ is the uncertainty associated with the fourth-order NLC calculations at temperature $T_\mathrm{exp}$,

\vspace*{-5pt}

\begin{equation}\label{eq:8}
\epsilon_{\chi, \mathrm{NLC},4}(T_\mathrm{exp}) 
= \mathrm{max}_{T \geq T_\mathrm{exp}}~|\chi^{\mathrm{NLC},4}(T) - \chi^{\mathrm{NLC},3}(T)| , 
\end{equation}
\vspace*{1pt}

where $\chi^{\mathrm{NLC},3}$ is the magnetic susceptibility calculated using third-order NLC calculations.

We compute the magnetic susceptibility for a magnetic field of strength $h=0.01$~T along the $[1,1,0]$ direction and compare these calculations with the magnetic susceptibility measured from Ce$_2$Hf$_2$O$_7$ for the same field strength and direction, with $T_\mathrm{exp}$ ranging from 0.45~K to 30~K.

Figure~S13(a,b) shows the $\theta$-dependence of the goodness-of-fit parameter $\langle  \delta^2/\epsilon^2  \rangle_{\chi}$ for each permutation of the A [Fig.~S13(a)] and B [Fig.~S13(b)] parameters. Figure~S13(c,d) shows the $\theta$-dependence of the best-fit anisotropic g-factor $g_z$ for each permutation of the A [Fig.~S13(c)] and B [Fig.~S13(d)] parameters. 

For the A parameters, given by $(J_a,J_b,J_c) = (0.050, 0.021, 0.004)$~meV, the permutations $(J_{\tilde{x}}, J_{\tilde{y}}, J_{\tilde{z}}) = (J_a, J_b, J_c)$ and $(J_b, J_a, J_c)$ fit the measured magnetic susceptibility the best, and equally well, with the best fits for these permutations corresponding to $\theta = 0$ and $g_z = 2.11$. However, the minimum in the goodness-of-fit parameter is relatively shallow and broad for these permutations such that a wide range of $\theta$ provides reasonable descriptions of the measured magnetic susceptibility in each case. Additionally, the permutation $(J_{\tilde{x}}, J_{\tilde{y}}, J_{\tilde{z}}) = (J_c, J_a, J_b)$ is able to provide a reasonable description of the measured magnetic susceptibility for values of $\theta$ near 0.25$\pi$. 

For the B parameters, given by $(J_a,J_b,J_c) = (0.051, 0.008, -0.018)$~meV, the permutations $(J_{\tilde{x}}, J_{\tilde{y}}, J_{\tilde{z}}) = (J_a, J_b, J_c)$ and $(J_c, J_a, J_b)$ fit the measured magnetic susceptibility the best, and equally well, with the best fits for these permutations corresponding to $\theta = 0.16\pi$ and $\theta = 0.19\pi$, respectively, with a best-fit $g_z$ value of $g_z = 2.15$ in each case. However, the minimum in the goodness-of-fit parameter is relatively shallow and broad for these permutations such that a wide range of $\theta$ provides reasonable descriptions of the measured magnetic susceptibility in each case. Additionally, the permutations $(J_{\tilde{x}}, J_{\tilde{y}}, J_{\tilde{z}}) = (J_b, J_a, J_c)$ and $(J_a, J_c, J_b)$ are able to provide a reasonable description of the measured magnetic susceptibility for values of $\pi/8 \lesssim \theta \leq \pi/4$ and $0 \leq \theta \lesssim \pi/8$, respectively. 

Figure~S13(e,f) shows the experimental magnetic susceptibility data used for this fitting procedure: The measured magnetic susceptibility from Ce$_2$Hf$_2$O$_7$ in a magnetic field of strength $h=0.01$~T along the $(1,1,0)$ direction. Figure~S13(e,f) also shows the magnetic susceptibility predicted according to our fourth order NLC calculations using the best-fit values of $\theta$ and $g_z$ for various permutations of the A [Fig.~S13(e)] and B [Fig.~S13(f)] parameters.

Notably, the B parameters are only able to provide simultaneous reasonable descriptions of the measured magnetic susceptibility and the measured diffuse scattering signal (see previous section of SM) for the permutation $(J_{\tilde{x}}, J_{\tilde{y}}, J_{\tilde{z}}) = (J_a, J_c, J_b)$ and for $0.075\pi \lesssim  \theta \lesssim 0.125\pi$, with the corresponding best-fit $g_z$ values being $g_z \sim 2.07$.


\begin{figure*}[t]
\linespread{1}
\par
\includegraphics[width=6in]{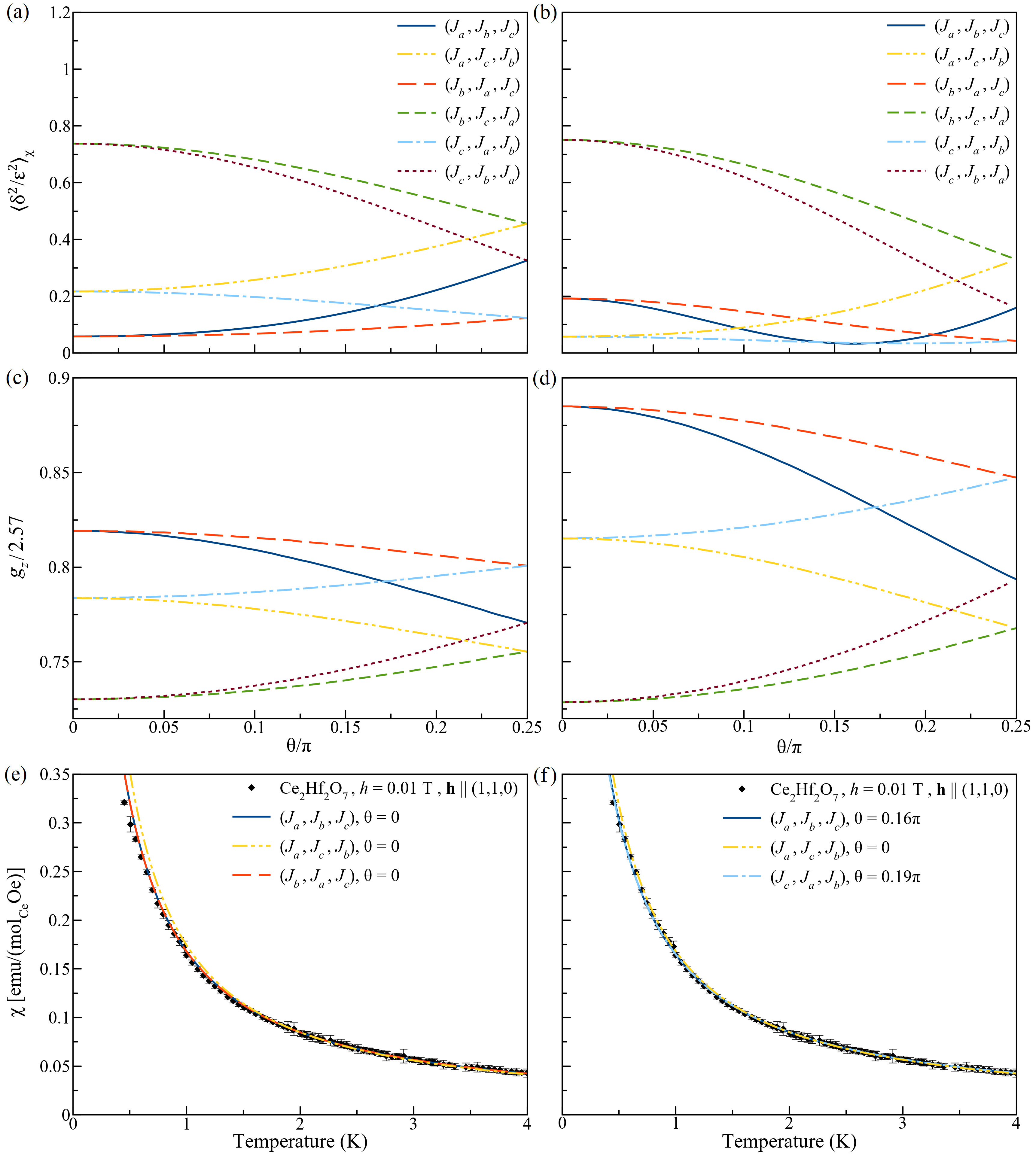}
\par
\caption{(a) and (b) show the $\theta$-dependence of the goodness-of-fit parameter for our fourth order NLC fitting to the measured magnetic susceptibility of Ce$_2$Hf$_2$O$_7$, for each distinct permutation of the (a) A and (b) B exchange parameters. (c) and (d) show the $\theta$-dependence of the best-fit reduced anisotropic g-factor $g_z/2.57$ for our NLC fitting to the measured magnetic susceptibility of Ce$_2$Hf$_2$O$_7$, for each distinct permutation of the (c) A and (d) B exchange parameters. (e) and (f) show the measured magnetic susceptibility from Ce$_2$Hf$_2$O$_7$ in a magnetic field of strength $h=0.01$~T along the $(1,1,0)$ direction, as well as the magnetic susceptibility predicted according to our fourth order NLC calculations using the best-fit value of $\theta$ for various permutations of the (e) A and (f) B parameters.}
\label{FigureS13}
\end{figure*}


\section{Quantum Monte Carlo \\ Simulations of \texorpdfstring{$C_{\mathrm{mag}}$}~}

We have performed quantum Monte Carlo (QMC) simulations using the stochastic series expansion method \cite{Sandvik1999} to calculate $C_\mathrm{mag}$ for various parameter sets in the unfrustrated regime of parameter space for $\mathcal{H}_\mathrm{ABC}$ ($J_{\pm} > 0$ in Eq.~1 of the main text). Specifically, this was done for parameters in the unfrustrated region that provide reasonable agreement for the comparison of the NLC calculations of $C_\mathrm{mag}$ with the $C_\mathrm{mag}$ measured from Ce$_2$Hf$_2$O$_7$.

We compare the magnetic heat capacity calculated using our QMC simulations, $C_\mathrm{mag}^{\mathrm{QMC}}$, to the magnetic heat capacity measured from single crystal Ce$_2$Hf$_2$O$_7$, $C_{\mathrm{mag}}^\mathrm{exp}$, using the goodness-of-fit measure,

\begin{equation}
    \chi_C^2 = \sum_{T_\mathrm{exp}}\frac{[C_{\mathrm{mag}}^\mathrm{QMC}(T_\mathrm{exp}) - C_{\mathrm{mag}}^\mathrm{exp}(T_\mathrm{exp})]^2}{\epsilon_{C, \mathrm{exp}}(T_\mathrm{exp})^2}
\end{equation}

where $\epsilon_{C, \mathrm{exp}}(T_\mathrm{exp})$ is the experimental uncertainty on the measured heat capacity at temperature $T_\mathrm{exp}$.

Figure~S14 shows the goodness-of-fit parameter $\chi_C^2$ for this comparison of the QMC calculations with the measured $C_{\mathrm{mag}}$ of Ce$_2$Hf$_2$O$_7$ for $T_\mathrm{exp} \in [0.037, 5.5]$~K, with the boundary between the ordered regime and U(1)$_\pi$ QSI regime shown as a solid red line. The best-fitting parameter set obtained from this QMC fitting procedure with $T_\mathrm{exp} \in [0.037, 5.5]$~K is also shown as a red cross in Figure~S14 and corresponds to $(J_a,J_b,J_c) = (0.046, -0.003, -0.010)$~meV. The B parameters obtained from our NLC fitting (the best-fit parameters in the unfrustrated regime from our NLC fitting), $(J_a,J_b,J_c) = (0.051, 0.008, -0.018)$~meV, are also shown in Figure~S14 for comparison. Figure~3(a) of the main text shows our QMC simulations of $C_{\mathrm{mag}}$ using the B parameters and using the best-fit parameters from our QMC fitting, compared to the $C_\mathrm{mag}$ measured from Ce$_2$Hf$_2$O$_7$ in this work. It is worth mentioning that both $\chi_C^2$ and the best-fit parameter set obtained from this fitting procedure have a significant dependence on the low-temperature cutoff used for $T_\mathrm{exp}$, but ultimately no reasonable fits are obtained regardless of the choice of low-temperature cutoff. 


\begin{figure}[!h]
\linespread{1}
\par
\includegraphics[width=3.4in]{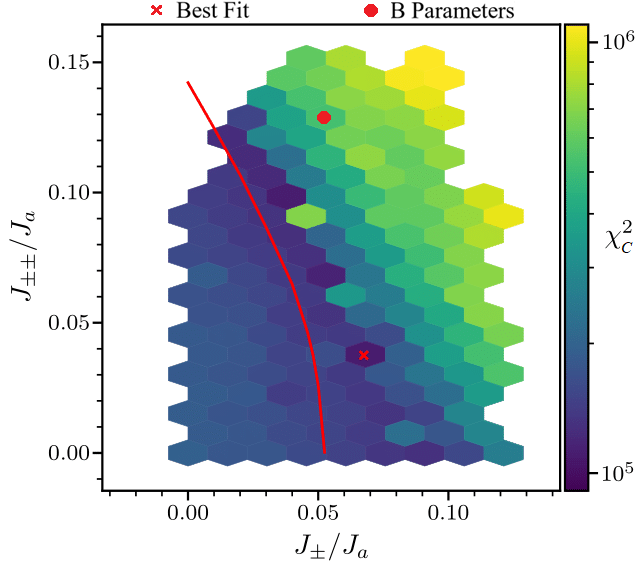}
\par
\caption{(a) The goodness-of-fit parameter $\chi_C^2$ for our QMC calculations of $C_{\mathrm{mag}}$ compared to that measured from Ce$_2$Hf$_2$O$_7$, for $T_\mathrm{exp} \in [0.037, 5.5]$~K (see SM text). Specifically, we show the dependence of $\chi_C^2$ on $J_\pm/J_a$ and $J_{\pm\pm}/J_a$ over the region of phase space in which our QMC calculations were performed. The red cross shows the parameter set that gives the best agreement between the QMC calculations and the measured data from Ce$_2$Hf$_2$O$_7$ for $T_\mathrm{exp} \in [0.037, 5.5]$~K, and the red circle shows the B parameters obtained from our NLC fitting. The red line in the plot shows the boundary between the ordered and disordered regimes of the ground state phase diagram as predicted at the nearest-neighbor level in Ref.~\cite{Benton2020}, also shown in Fig.~2(b) of the main text.} 
\label{FigureS14}
\end{figure}


\section{High-Energy Inelastic Neutron Scattering and CEF Analysis}

We have performed high-energy inelastic neutron scattering measurements on a $\sim$6.5~g powder sample of Ce$_2$Hf$_2$O$_7$ using the SEQUOIA high-resolution inelastic chopper spectrometer (see Ref.~\cite{Granroth2010}) at the Spallation Neutron Source of Oak Ridge National Laboratory, with neutron incident energies $E_\mathrm{i} = 150$ and 750~meV, yielding energy resolutions of $\sim$10~meV ($E_{\mathrm{i}} = 150$~meV) and $\sim$55~meV ($E_{\mathrm{i}} = 750$~meV) at the elastic line. The high-flux configuration of the SEQUOIA instrument was used and the sample was measured in a cylindrical aluminum sample can with 0.25~inch diameter.

We first discuss the results of our high-energy inelastic neutron scattering measurements at $T=5$~K on a powder sample of Ce$_2$Hf$_2$O$_7$ with incident energy of 150~meV. This incident energy was used to probe the transitions from the CEF ground state doublet to the first and second excited state doublets. Fig.~S15(a) shows the $E_\mathrm{i} = 150$~meV neutron scattering powder spectra measured from Ce$_2$Hf$_2$O$_7$ at $T=5$~K, with the subtraction of a dataset measured on an empty sample holder. We observe two clear CEF excitations in the $E_\mathrm{i} = 150$~meV powder spectra, highlighted by black arrows at $E \sim 58$~meV and $E \sim 111$~meV in Fig.~S15(a), identifiable by their lack of dispersion and the fact that their intensity decreases with increasing $||\mathbf{Q}||$ consistent with the Ce$^{3+}$ magnetic form factor.

The spin and angular momentum quantum numbers for 4\textit{f}$^1$ Ce$^{3+}$, $S = 1/2$ and $L = 3$, give a spin-orbit ground state manifold with angular momentum quantum number of $J = |L-S| = 5/2$ according to Hund's rules, and one excited spin-orbit manifold with $J=L+S=7/2$. The spin-orbit gap for Ce$^{3+}$ is on the order of 200~meV (Ref.~\cite{Freeman1962}) and dominates the weaker CEF splitting, allowing for a reasonable approximation using the Russel-Saunders coupling scheme where the CEF splitting does not induce mixing between the $J = 5/2$ and $J = 7/2$ states, which we employ in our upcoming analysis.


\begin{figure*}[t]
\linespread{1}
\par
\includegraphics[width=6in]{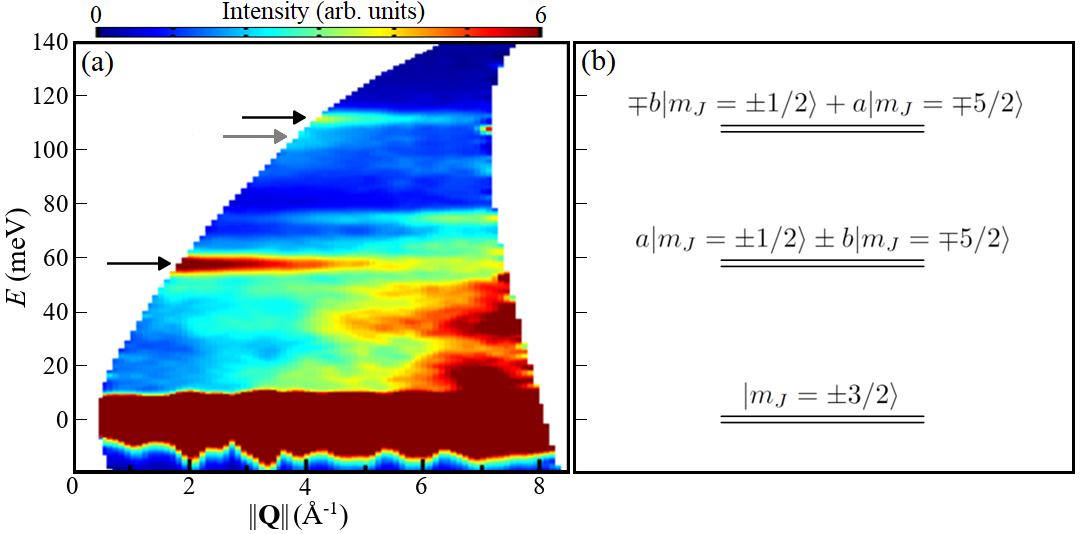}
\par
\caption{(a) Inelastic neutron scattering powder spectra measured from a powder sample of Ce$_2$Hf$_2$O$_7$ at $T=5$~K with an incident energy of $E_\mathrm{i} = 150$~meV. A $T=5$~K dataset measured from the empty sample-holder was subtracted to help isolate the scattering from Ce$_2$Hf$_2$O$_7$. Two strong excitations, at $E \sim 58$~meV and $E \sim 111$~meV, can be identified as CEF excitations due to their dispersionless nature and the fact that their intensity decreases with increasing $||\mathbf{Q}||$, consistent with the Ce$^{3+}$ magnetic form factor. A third, weaker, dispersionless excitation also appears to be consistent with the Ce$^{3+}$ magnetic form factor, and is visible at $E \sim 104$~meV. We attribute this excitation at $E \sim 104$~meV to a potential vibronic bound state between CEF excitation and phonon (see SM text). (b) The best-fitting CEF scheme from our refinement of the neutron scattering powder spectra in (a), showing the CEF energy levels and eigenstates within the $J=5/2$ spin-orbit ground state manifold.} 
\label{FigureS15}
\end{figure*}


The temperature used, $T=5$~K, is sufficiently low enough to avoid significant thermal population of excited CEF states, such that the only CEF transitions with observable intensity are transitions originating from the CEF ground state. Furthermore, given that Ce$^{3+}$ is a Kramer's ion, the CEF states cannot be split any further than doublets. Therefore, considering there are $2J+1 = 6$ states in total within the $J = 5/2$ manifold, the two strong CEF transitions detected in our $E_\mathrm{i} = 150$~meV data should constitute all transitions from the CEF ground state to excited states within the $J=5/2$ spin-orbit manifold. 

However, we also identify a third, weaker, dispersionless excitation at $E \sim 104$~meV which also appears to decrease in intensity with increasing $||\mathbf{Q}||$, highlighted by the grey arrow in Fig.~S15(a). We attribute this excitation at $E \sim 104$~meV to a potential vibronic bound state between CEF excitation and phonon. This is consistent with the high-energy inelastic neutron scattering data reported for Ce$_2$Hf$_2$O$_7$ in Ref.~\cite{Poree2022}, which shows a similar dispersionless excitation around $E \sim 100$~meV with intensity that decreases with increasing $||\mathbf{Q}||$. In fact, a similar potential vibronic bound state was also detected at $E \sim 100$~meV in the inelastic neutron scattering data reported for Ce$_2$Zr$_2$O$_7$, albeit, more clearly in Ref.~\cite{Gaudet2019} than in Ref.~\cite{Gao2019}. In contrast to this, the high-energy inelastic neutron scattering data reported for Ce$_2$Sn$_2$O$_7$, in Ref.~\cite{Sibille2020}, shows no signs for a potential vibronic bound state near $E \sim 100$~meV. Significant magnetoelastic coupling leading to the formation of a clear vibronic bound state has been reported for other rare-earth pyrochlore materials such as Ho$_2$Ti$_2$O$_7$~\cite{Gaudet2018, Ruminy2017}, Tb$_2$Ti$_2$O$_7$~\cite{Fennell2014}, and Pr$_2$Zr$_2$O$_7$~\cite{Xu2021}, for example, as well as other cerium-based magnets~\cite{Thalmeier1982, Thalmeier1984, Schedler2003, Loewenhaupt2003, Chapon2006, Adroja2012, Anand2021}. 

To account for the possibility that the $E \sim 104$~meV excitation measured from our Ce$_2$Hf$_2$O$_7$ sample is indeed a bound state between the second excited CEF doublet and a phonon, we do our CEF analysis both including this potential vibronic bound state among the CEF excitations and excluding it. Specifically, we fit the $E_{\mathrm{i}} = 150$~meV spectra in Fig.~S15(a) to determining the energy values and intensity ratio for the two lowest lying CEF excitations, $E_1$, $E_2$, and $I_1/I_2$, with and without the vibronic bound state included in determining $E_2$ and $I_2$. For the analysis with the vibronic bound state included, $I_2$ is the sum of intensities of the potential vibronic bound state at $E \sim 104$~meV and the strong CEF excitation at $E \sim 111$~meV, and $E_2$ is determined by an intensity-weighted average of the energies for each of these excitations. We use the Stevens operator formalism within the $J=5/2$ spin-orbit ground state manifold, and for each analysis, the parameters of the CEF Hamiltonian are refined in order to determine the CEF Hamiltonian that best-reproduces the measured values of $E_1$, $E_2$, and $I_1/I_2$. 

The CEF Hamiltonian for Ce$^{3+}$, within the $J=5/2$ spin-orbit ground state manifold, is given in the Stevens operator formalism as~{\cite{Prather1961, Hutchings1964}:

\begin{equation} \label{eq:9}
    \mathcal{H}_\mathrm{CEF} = B_2^0 \hat{O}_2^0 + B_4^0 \hat{O}_4^0 + B_4^3 \hat{O}_4^3,\; 
\end{equation}

\noindent  where $\hat{O}_n^m$ are Stevens operators, which are polynomials of order $n$ in the total angular momentum operators~\cite{Stevens1952, Hutchings1964}. It is worth mentioning that the CEF Hamiltonian in the Stevens operator formalism usually contains six terms for the $D_{3d}$ symmetry corresponding to the magnetic site in the pyrochlore lattice~\cite{Prather1961}. However, the restriction $n \leq 2J$ renders $B_6^0, B_6^3, B_6^6 = 0$ within the $J=5/2$ spin-orbit ground state manifold~\cite{Hutchings1964}. 

We use the SPECTRE program (see Ref.~\cite{SPECTRE}) to refine the values of $B_2^0$, $B_4^0$, and $B_4^3$ to the measured values of $E_1$, $E_2$, and $I_1/I_2$. Fig.~S16 shows energy cuts through our $E_\mathrm{i} = 150$~meV data [Fig.~S15(a)] with $||\mathbf{Q}||$-integration over $||\mathbf{Q}|| = [4.5, 5.5]$~$\angstrom^{-1}$, where Lorentzian fits to the intensity from the two strong CEF excitations at $E \sim 58$~meV and $E \sim 111$~meV are shown in green, and the Lorentzian fit to intensity from the potential vibronic bound state at $E \sim 104$~meV is shown in yellow. The curves in blue and purple show the Lorentzian and Gaussian lineshapes used to fit the intensity from phonons and the elastic intensity, respectively. Table I shows the energy values and intensity ratio for the two lowest lying CEF excitations ($E_1$, $E_2$, and $I_1/I_2$) as obtained from the fit in Fig.~S16, for our CEF analyses with and without the potential vibronic bound state included in the determining $E_2$ and $I_2$. Table I also shows the best-fit energy values and intensity ratio for the two lowest lying CEF excitations for our refinement of the CEF Hamiltonian with and without the potential vibronic bound state included.


\begin{figure}[t]
\linespread{1}
\par
\includegraphics[width=3.4 in]{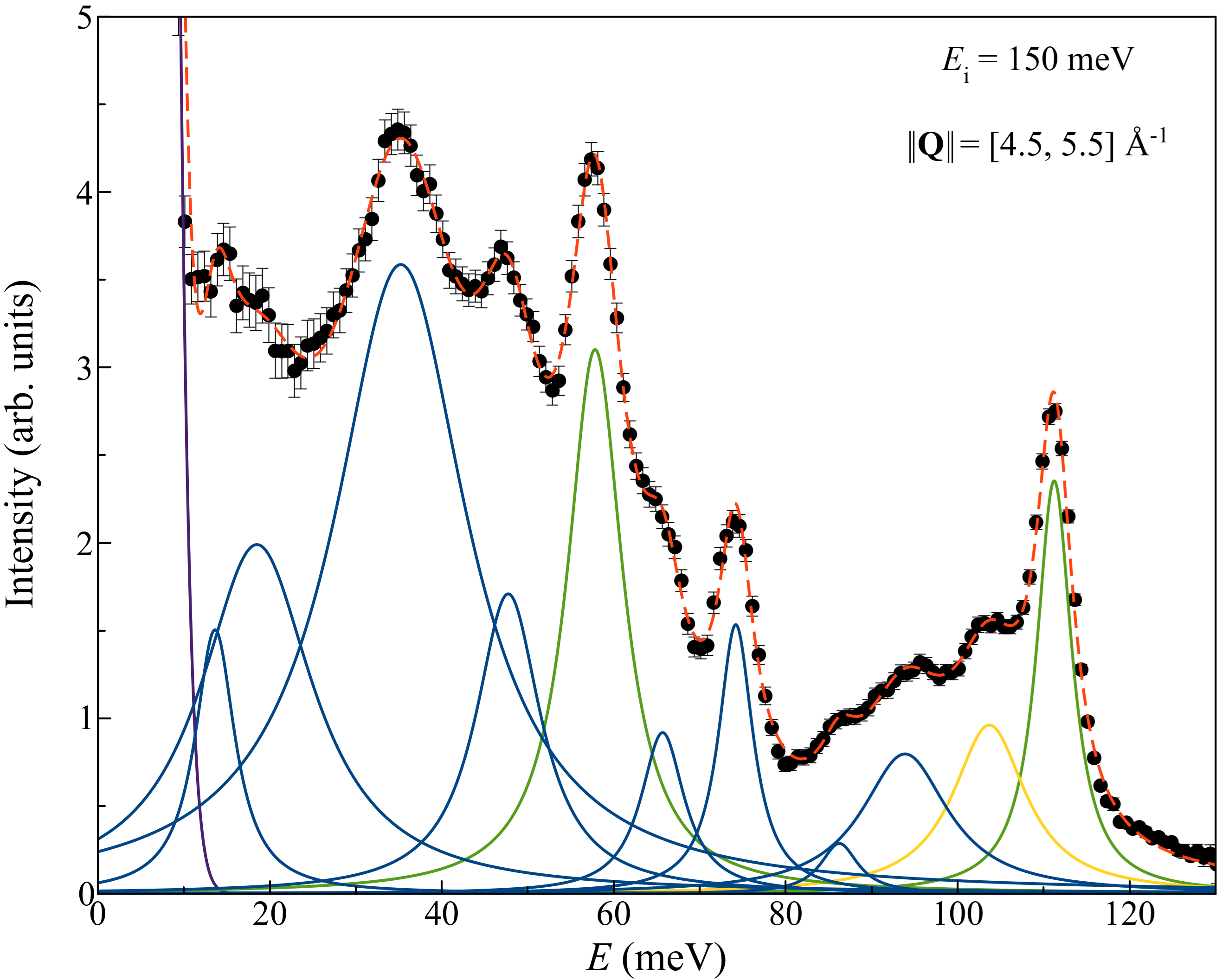}
\par
\caption{Energy cuts through the $E_\mathrm{i}$ = 150~meV neutron scattering spectra measured from a powder sample of Ce$_2$Hf$_2$O$_7$ [shown in Fig.~S15(a)], with integration in $||\mathbf{Q}||$ from 4.5 to 5.5~$\angstrom^{-1}$. The dashed line shows the fit that was used to determine the values of $E_1$, $E_2$, and $I_1/I_2$ for our refinement of the CEF parameters. The solid lines show the individual Lorentzian functions included in the fit and attributed to CEF excitations (green), a potential vibronic bound state (yellow), and phonons (blue), as well as the Gaussian form included in the fit and attributed to the elastic intensity (purple).} 
\label{FigureS16}
\end{figure}


\begin{figure}[!h]
\linespread{1}
\par
\includegraphics[width=3.4in]{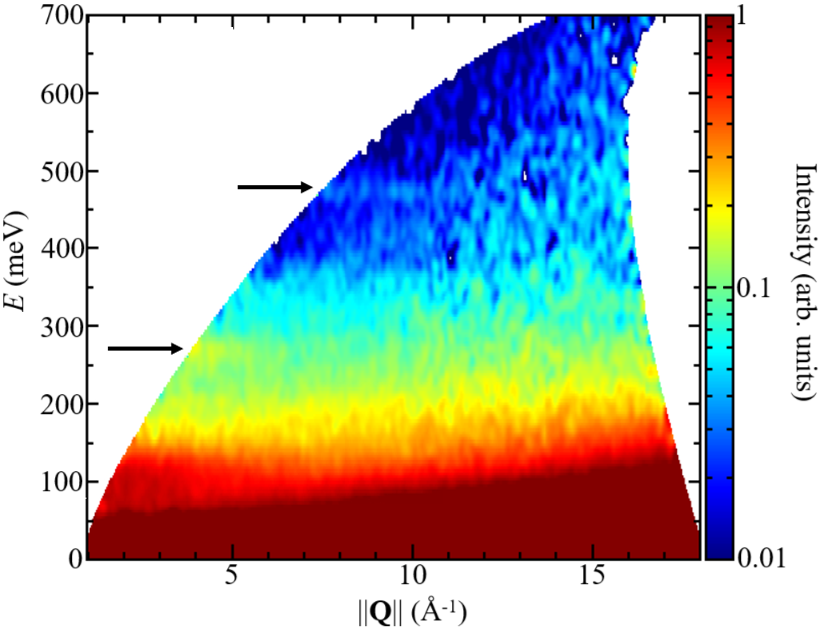}
\par
\caption{The inelastic neutron scattering powder spectra measured from a powder sample of Ce$_2$Hf$_2$O$_7$ at $T=5$~K with an incident energy of $E_\mathrm{i} = 750$~meV. A $T=5$~K dataset measured from the empty sample-holder was subtracted to help isolate the scattering from Ce$_2$Hf$_2$O$_7$. Two CEF excitations, at $E \sim 270$~meV and $E \sim 475$~meV, can be identified due to their dispersionless nature and the $||\mathbf{Q}||$-dependence of their intensity.} 
\label{FigureS17}
\end{figure}


Importantly, both of our analyses yield pure $|m_J = \pm 3/2 \rangle$ CEF ground state doublets, which have dipole-octupole symmetry and Ising single ion anisotropy with a corresponding anisotropic g-factor of $g_z = 2.57$. This is consistent with the CEF ground state estimated for Ce$_2$Hf$_2$O$_7$ in Ref.~\cite{Poree2022}, where Por\'ee \emph{et al.} include the possibility of mixing with the $J = 7/2$ states in the CEF ground state and conclude that the CEF ground state doublet contains only $|m_J = \pm 3/2 \rangle$ terms.

\begin{table}[!b]
\label{TableI}
\begin{tabular}{|c|c|c|c|c|c|}
\hline
 & Observed & Best-Fit & Observed & Best-Fit \\
& (no VBS) & (no VBS) & (with VBS) & (with VBS)\\\hline

\begin{tabular}[c]{@{}c@{}}  \end{tabular}          
$E_1$ (meV) & 57.8(1) & 57.83 & 57.8(1) & 57.83 \\ \hline

\begin{tabular}[c]{@{}c@{}}  \end{tabular}          
$E_2$ (meV) & 111.2(1) & 111.23 & 107.8(4) & 107.77 \\ \hline

\begin{tabular}[c]{@{}c@{}}  \end{tabular}          
$I_1/I_2$ & 2.1(4) & 1.60 & 1.1(2) & 1.12 \\ \hline

\end{tabular}
\caption{The observed and best-fit values of $E_1$, $E_2$, and $I_1/I_2$ with the potential vibronic bound state excluded (no~VBS) and included (with VBS) in the analysis of the neutron scattering powder spectra from Ce$_2$Hf$_2$O$_7$.}
\end{table}

\begin{table}[b!]
\label{TableII}
\begin{tabular}{|c|c|c|c|c|}
\hline
 & $B_2^0$ (meV) & $B_4^0$ (meV) & $B_4^3$ (meV)  \\ \hline

\begin{tabular}[c]{@{}c@{}} \end{tabular}          
No VBS & 3.866 & 0.270 & 0.000 \\ \hline

\begin{tabular}[c]{@{}c@{}}  \end{tabular}          
With VBS & 1.642 & 0.288 & 2.552 \\ \hline

\end{tabular}
\caption{The best-fitting CEF parameters from our refinement to the neutron scattering data with the potential vibronic bound state excluded (no VBS) and included (with VBS) in determining $E_1$, $E_2$, and $I_1/I_2$.}
\end{table}

Fig.~S15(b) shows the resulting CEF scheme from our analyses. For each of our CEF analyses, the CEF ground state is a $|m_J = \pm 3/2 \rangle$ doublet, which is a dipole-octupole doublet~\cite{Huang2014}. Also for each analysis, the first excited CEF doublet has the form $a|m_J = \pm 1/2 \rangle \pm b|m_J=\mp5/2\rangle$, and the second excited CEF doublet has the similar form $a|m_J=\pm 1/2 \rangle \pm b|m_J=\mp 5/2\rangle $, where $a=1$ ($a = 0.789$) and $b=0$ ($b = 0.614$) for our analysis with the potential vibronic bound state excluded (included). The corresponding best-fit CEF parameters, $B_2^0$, $B_4^0$, and $B_4^3$, are shown in Table~II for each of our analyses. Interestingly, the fitting process results in a much better fit to the measured data, specifically the intensity ratio $I_1/I_2$, when the signal at $E \sim 104$~meV is treated as a vibronic bound state.  

Fig.~S17 shows the $E_\mathrm{i} = 750$~meV neutron scattering powder spectra measured from Ce$_2$Hf$_2$O$_7$ at $T=5$~K, with the subtraction of a dataset measured on an empty sample holder, where we use arrows to highlight two CEF excitations (at $E \sim 265$~meV and $E \sim 475$~meV) from the CEF ground state to the $J = 7/2$ spin-orbit manifold. The CEF excitation at $E \sim 265$~meV is also reported in similar inelastic neutron scattering measurements on Ce$_2$Hf$_2$O$_7$ in Ref.~\cite{Poree2022}, where a highest incident energy of $E_\mathrm{i} = 450$~meV was used, which is insufficient to view the $E \sim 475$~meV transition that we detect in this work. Similarly, Ref.~\cite{Gaudet2019} reports a CEF excitation for Ce$_2$Zr$_2$O$_7$ at $E \sim 270$~meV but uses a highest incident energy of $E_\mathrm{i} = 500$~meV, making a weak CEF transition near $E$~$\sim$~475~meV undetectable due to noise near the edge of the measurement range. The high energy inelastic neutron scattering data reported on Ce$_2$Sn$_2$O$_7$ in Ref.~\cite{Sibille2020} finds transitions at $E \sim 262$~meV and $E \sim 430$~meV, nearby to the transitions from the CEF ground state to the $J = 7/2$ spin-orbit manifold that we measure here for Ce$_2$Hf$_2$O$_7$. The measured gap of $E \sim 265$~meV to the $J = 7/2$ spin-orbit manifold that we measure here for Ce$_2$Hf$_2$O$_7$ is consistent with magnitude of the spin orbit gap expected for Ce$^{3+}$~\cite{Freeman1962}, and justifies our approximation of including only the $J=5/2$ manifold in our CEF analysis of the lower-lying states.

We were unable to resolve any other transitions to the $J=7/2$ manifold. Furthermore, the weak intensity of the excitations at $E \sim 265$~meV and $E \sim 475$~meV competing with phonon intensity, and the kinematic restriction limiting the measurement of low $||\mathbf{Q}||$ at high $E$, does not allow for an accurate determination of the intensity ratio for the transitions at $E \sim 265$~meV and $E \sim 475$~meV or the intensity ratio between one of these excitations and a lower lying CEF excitation. Unfortunately, with only the two energy values of two CEF transitions to the $J=7/2$ manifold, any attempt to fit the six CEF parameters would be under-constrained. Ref.~\cite{Poree2022} reports high-energy inelastic neutron scattering measurements on powder Ce$_2$Hf$_2$O$_7$, including an estimation of the intensity ratios between the $E \sim 265$~meV excitation and the two lower-lying excitations at $E \sim 58$ and 111~meV. The analysis of Ref.~\cite{Poree2022} includes the $J = 7/2$ manifold and yields a ground state doublet that contains only $m_J = \pm 3/2$ states, and is predominately made up of $J=5/2$ states with only a small amount of mixing with $J=7/2$ states.  


\begin{thebibliography}{71}%
\makeatletter
\providecommand \@ifxundefined [1]{%
 \@ifx{#1\undefined}
}%
\providecommand \@ifnum [1]{%
 \ifnum #1\expandafter \@firstoftwo
 \else \expandafter \@secondoftwo
 \fi
}%
\providecommand \@ifx [1]{%
 \ifx #1\expandafter \@firstoftwo
 \else \expandafter \@secondoftwo
 \fi
}%
\providecommand \natexlab [1]{#1}%
\providecommand \enquote  [1]{``#1''}%
\providecommand \bibnamefont  [1]{#1}%
\providecommand \bibfnamefont [1]{#1}%
\providecommand \citenamefont [1]{#1}%
\providecommand \href@noop [0]{\@secondoftwo}%
\providecommand \href [0]{\begingroup \@sanitize@url \@href}%
\providecommand \@href[1]{\@@startlink{#1}\@@href}%
\providecommand \@@href[1]{\endgroup#1\@@endlink}%
\providecommand \@sanitize@url [0]{\catcode `\\12\catcode `\$12\catcode `\&12\catcode `\#12\catcode `\^12\catcode `\_12\catcode `\%12\relax}%
\providecommand \@@startlink[1]{}%
\providecommand \@@endlink[0]{}%
\providecommand \url  [0]{\begingroup\@sanitize@url \@url }%
\providecommand \@url [1]{\endgroup\@href {#1}{\urlprefix }}%
\providecommand \urlprefix  [0]{URL }%
\providecommand \Eprint [0]{\href }%
\providecommand \doibase [0]{https://doi.org/}%
\providecommand \selectlanguage [0]{\@gobble}%
\providecommand \bibinfo  [0]{\@secondoftwo}%
\providecommand \bibfield  [0]{\@secondoftwo}%
\providecommand \translation [1]{[#1]}%
\providecommand \BibitemOpen [0]{}%
\providecommand \bibitemStop [0]{}%
\providecommand \bibitemNoStop [0]{.\EOS\space}%
\providecommand \EOS [0]{\spacefactor3000\relax}%
\providecommand \BibitemShut  [1]{\csname bibitem#1\endcsname}%
\let\auto@bib@innerbib\@empty
\bibitem [{\citenamefont {Hermele}\ \emph {et~al.}(2004)\citenamefont {Hermele}, \citenamefont {Fisher},\ and\ \citenamefont {Balents}}]{Hermele2004}%
  \BibitemOpen
  \bibfield  {author} {\bibinfo {author} {\bibfnamefont {M.}~\bibnamefont {Hermele}}, \bibinfo {author} {\bibfnamefont {M.~P.~A.}\ \bibnamefont {Fisher}},\ and\ \bibinfo {author} {\bibfnamefont {L.}~\bibnamefont {Balents}},\ }\bibfield  {title} {\bibinfo {title} {\textit{Pyrochlore Photons: The $U(1)$ Spin Liquid in a $S=\frac{1}{2}$ Three-Dimensional Frustrated Magnet}},\ }\href {https://doi.org/10.1103/PhysRevB.69.064404} {\bibfield  {journal} {\bibinfo  {journal} {Phys. Rev. B}\ }\textbf {\bibinfo {volume} {69}},\ \bibinfo {pages} {064404} (\bibinfo {year} {2004})}\BibitemShut {NoStop}%
\bibitem [{\citenamefont {Banerjee}\ \emph {et~al.}(2008)\citenamefont {Banerjee}, \citenamefont {Isakov}, \citenamefont {Damle},\ and\ \citenamefont {Kim}}]{Banerjee2008}%
  \BibitemOpen
  \bibfield  {author} {\bibinfo {author} {\bibfnamefont {A.}~\bibnamefont {Banerjee}}, \bibinfo {author} {\bibfnamefont {S.~V.}\ \bibnamefont {Isakov}}, \bibinfo {author} {\bibfnamefont {K.}~\bibnamefont {Damle}},\ and\ \bibinfo {author} {\bibfnamefont {Y.~B.}\ \bibnamefont {Kim}},\ }\bibfield  {title} {\bibinfo {title} {\textit{Unusual Liquid State of Hard-Core Bosons on the Pyrochlore Lattice}},\ }\href {https://doi.org/10.1103/PhysRevLett.100.047208} {\bibfield  {journal} {\bibinfo  {journal} {Phys. Rev. Lett.}\ }\textbf {\bibinfo {volume} {100}},\ \bibinfo {pages} {047208} (\bibinfo {year} {2008})}\BibitemShut {NoStop}%
\bibitem [{\citenamefont {Lee}\ \emph {et~al.}(2012)\citenamefont {Lee}, \citenamefont {Onoda},\ and\ \citenamefont {Balents}}]{Lee2012}%
  \BibitemOpen
  \bibfield  {author} {\bibinfo {author} {\bibfnamefont {S.}~\bibnamefont {Lee}}, \bibinfo {author} {\bibfnamefont {S.}~\bibnamefont {Onoda}},\ and\ \bibinfo {author} {\bibfnamefont {L.}~\bibnamefont {Balents}},\ }\bibfield  {title} {\bibinfo {title} {\textit{Generic Quantum Spin Ice}},\ }\href {https://doi.org/10.1103/PhysRevB.86.104412} {\bibfield  {journal} {\bibinfo  {journal} {Phys. Rev. B}\ }\textbf {\bibinfo {volume} {86}},\ \bibinfo {pages} {104412} (\bibinfo {year} {2012})}\BibitemShut {NoStop}%
\bibitem [{\citenamefont {Benton}\ \emph {et~al.}(2012)\citenamefont {Benton}, \citenamefont {Sikora},\ and\ \citenamefont {Shannon}}]{Benton2012}%
  \BibitemOpen
  \bibfield  {author} {\bibinfo {author} {\bibfnamefont {O.}~\bibnamefont {Benton}}, \bibinfo {author} {\bibfnamefont {O.}~\bibnamefont {Sikora}},\ and\ \bibinfo {author} {\bibfnamefont {N.}~\bibnamefont {Shannon}},\ }\bibfield  {title} {\bibinfo {title} {\textit{Seeing the Light: Experimental Signatures of Emergent Electromagnetism in a Quantum Spin Ice}},\ }\href {https://doi.org/10.1103/PhysRevB.86.075154} {\bibfield  {journal} {\bibinfo  {journal} {Phys. Rev. B}\ }\textbf {\bibinfo {volume} {86}},\ \bibinfo {pages} {075154} (\bibinfo {year} {2012})}\BibitemShut {NoStop}%
\bibitem [{\citenamefont {Savary}\ and\ \citenamefont {Balents}(2012)}]{Savary2012b}%
  \BibitemOpen
  \bibfield  {author} {\bibinfo {author} {\bibfnamefont {L.}~\bibnamefont {Savary}}\ and\ \bibinfo {author} {\bibfnamefont {L.}~\bibnamefont {Balents}},\ }\bibfield  {title} {\bibinfo {title} {\textit{Coulombic Quantum Liquids in Spin-$1/2$ Pyrochlores}},\ }\href {https://doi.org/10.1103/PhysRevLett.108.037202} {\bibfield  {journal} {\bibinfo  {journal} {Phys. Rev. Lett.}\ }\textbf {\bibinfo {volume} {108}},\ \bibinfo {pages} {037202} (\bibinfo {year} {2012})}\BibitemShut {NoStop}%
\bibitem [{\citenamefont {Gingras}\ and\ \citenamefont {McClarty}(2014)}]{GingrasReview2014}%
  \BibitemOpen
  \bibfield  {author} {\bibinfo {author} {\bibfnamefont {M.~J.~P.}\ \bibnamefont {Gingras}}\ and\ \bibinfo {author} {\bibfnamefont {P.~A.}\ \bibnamefont {McClarty}},\ }\bibfield  {title} {\bibinfo {title} {\textit{Quantum Spin Ice: A Search for Gapless Quantum Spin Liquids in Pyrochlore Magnets}},\ }\href {https://doi.org/10.1088/0034-4885/77/5/056501} {\bibfield  {journal} {\bibinfo  {journal} {Rep. Prog. Phys}\ }\textbf {\bibinfo {volume} {77}},\ \bibinfo {pages} {056501} (\bibinfo {year} {2014})}\BibitemShut {NoStop}%
\bibitem [{\citenamefont {Sibille}\ \emph {et~al.}(2015)\citenamefont {Sibille}, \citenamefont {Lhotel}, \citenamefont {Pomjakushin}, \citenamefont {Baines}, \citenamefont {Fennell},\ and\ \citenamefont {Kenzelmann}}]{Sibille2015}%
  \BibitemOpen
  \bibfield  {author} {\bibinfo {author} {\bibfnamefont {R.}~\bibnamefont {Sibille}}, \bibinfo {author} {\bibfnamefont {E.}~\bibnamefont {Lhotel}}, \bibinfo {author} {\bibfnamefont {V.}~\bibnamefont {Pomjakushin}}, \bibinfo {author} {\bibfnamefont {C.}~\bibnamefont {Baines}}, \bibinfo {author} {\bibfnamefont {T.}~\bibnamefont {Fennell}},\ and\ \bibinfo {author} {\bibfnamefont {M.}~\bibnamefont {Kenzelmann}},\ }\bibfield  {title} {\bibinfo {title} {\textit{Candidate Quantum Spin Liquid in the ${\mathrm{Ce}}^{3+}$ Pyrochlore Stannate ${\mathrm{Ce}}_{2}{\mathrm{Sn}}_{2}{\mathrm{O}}_{7}$}},\ }\href {https://doi.org/10.1103/PhysRevLett.115.097202} {\bibfield  {journal} {\bibinfo  {journal} {Phys. Rev. Lett.}\ }\textbf {\bibinfo {volume} {115}},\ \bibinfo {pages} {097202} (\bibinfo {year} {2015})}\BibitemShut {NoStop}%
\bibitem [{\citenamefont {Gaudet}\ \emph {et~al.}(2019)\citenamefont {Gaudet}, \citenamefont {Smith}, \citenamefont {Dudemaine}, \citenamefont {Beare}, \citenamefont {Buhariwalla}, \citenamefont {Butch}, \citenamefont {Stone}, \citenamefont {Kolesnikov}, \citenamefont {Xu}, \citenamefont {Yahne}, \citenamefont {Ross}, \citenamefont {Marjerrison}, \citenamefont {Garrett}, \citenamefont {Luke}, \citenamefont {Bianchi},\ and\ \citenamefont {Gaulin}}]{Gaudet2019}%
  \BibitemOpen
  \bibfield  {author} {\bibinfo {author} {\bibfnamefont {J.}~\bibnamefont {Gaudet}}, \bibinfo {author} {\bibfnamefont {E.~M.}\ \bibnamefont {Smith}}, \bibinfo {author} {\bibfnamefont {J.}~\bibnamefont {Dudemaine}}, \bibinfo {author} {\bibfnamefont {J.}~\bibnamefont {Beare}}, \bibinfo {author} {\bibfnamefont {C.~R.~C.}\ \bibnamefont {Buhariwalla}}, \bibinfo {author} {\bibfnamefont {N.~P.}\ \bibnamefont {Butch}}, \bibinfo {author} {\bibfnamefont {M.~B.}\ \bibnamefont {Stone}}, \bibinfo {author} {\bibfnamefont {A.~I.}\ \bibnamefont {Kolesnikov}}, \bibinfo {author} {\bibfnamefont {G.}~\bibnamefont {Xu}}, \bibinfo {author} {\bibfnamefont {D.~R.}\ \bibnamefont {Yahne}}, \bibinfo {author} {\bibfnamefont {K.~A.}\ \bibnamefont {Ross}}, \bibinfo {author} {\bibfnamefont {C.~A.}\ \bibnamefont {Marjerrison}}, \bibinfo {author} {\bibfnamefont {J.~D.}\ \bibnamefont {Garrett}}, \bibinfo {author} {\bibfnamefont {G.~M.}\ \bibnamefont {Luke}}, \bibinfo {author} {\bibfnamefont {A.~D.}\ \bibnamefont {Bianchi}},\ and\ \bibinfo
  {author} {\bibfnamefont {B.~D.}\ \bibnamefont {Gaulin}},\ }\bibfield  {title} {\bibinfo {title} {\textit{Quantum Spin Ice Dynamics in the Dipole-Octupole Pyrochlore Magnet $\mathrm{Ce}_2\mathrm{Zr}_2\mathrm{O}_7$}},\ }\href {https://doi.org/10.1103/PhysRevLett.122.187201} {\bibfield  {journal} {\bibinfo  {journal} {Phys. Rev. Lett.}\ }\textbf {\bibinfo {volume} {122}},\ \bibinfo {pages} {187201} (\bibinfo {year} {2019})}\BibitemShut {NoStop}%
\bibitem [{\citenamefont {Gao}\ \emph {et~al.}(2019)\citenamefont {Gao}, \citenamefont {Chen}, \citenamefont {Tam}, \citenamefont {Huang}, \citenamefont {Sasmal}, \citenamefont {Adroja}, \citenamefont {Ye}, \citenamefont {Cao}, \citenamefont {Sala}, \citenamefont {Stone}, \citenamefont {Baines}, \citenamefont {Barker}, \citenamefont {Hu}, \citenamefont {Chung}, \citenamefont {Xu}, \citenamefont {Cheong}, \citenamefont {Nallaiyan}, \citenamefont {Spagna}, \citenamefont {Maple},\ and\ \citenamefont {Dai}}]{Gao2019}%
  \BibitemOpen
  \bibfield  {author} {\bibinfo {author} {\bibfnamefont {B.}~\bibnamefont {Gao}}, \bibinfo {author} {\bibfnamefont {T.}~\bibnamefont {Chen}}, \bibinfo {author} {\bibfnamefont {D.}~\bibnamefont {Tam}}, \bibinfo {author} {\bibfnamefont {C.-L.}\ \bibnamefont {Huang}}, \bibinfo {author} {\bibfnamefont {K.}~\bibnamefont {Sasmal}}, \bibinfo {author} {\bibfnamefont {D.}~\bibnamefont {Adroja}}, \bibinfo {author} {\bibfnamefont {F.}~\bibnamefont {Ye}}, \bibinfo {author} {\bibfnamefont {H.}~\bibnamefont {Cao}}, \bibinfo {author} {\bibfnamefont {G.}~\bibnamefont {Sala}}, \bibinfo {author} {\bibfnamefont {M.}~\bibnamefont {Stone}}, \bibinfo {author} {\bibfnamefont {C.}~\bibnamefont {Baines}}, \bibinfo {author} {\bibfnamefont {J.}~\bibnamefont {Barker}}, \bibinfo {author} {\bibfnamefont {H.}~\bibnamefont {Hu}}, \bibinfo {author} {\bibfnamefont {J.-H.}\ \bibnamefont {Chung}}, \bibinfo {author} {\bibfnamefont {X.}~\bibnamefont {Xu}}, \bibinfo {author} {\bibfnamefont {S.-W.}\ \bibnamefont {Cheong}}, \bibinfo {author}
  {\bibfnamefont {M.}~\bibnamefont {Nallaiyan}}, \bibinfo {author} {\bibfnamefont {S.}~\bibnamefont {Spagna}}, \bibinfo {author} {\bibfnamefont {M.}~\bibnamefont {Maple}},\ and\ \bibinfo {author} {\bibfnamefont {P.}~\bibnamefont {Dai}},\ }\bibfield  {title} {\bibinfo {title} {\textit{Experimental Signatures of a Three-dimensional Quantum Spin Liquid in Effective Spin-$1/2$ $\mathrm{Ce}_2\mathrm{Zr}_2\mathrm{O}_7$ Pyrochlore}},\ }\href {https://doi.org/10.1038/s41567-019-0577-6} {\bibfield  {journal} {\bibinfo  {journal} {Nat. Phys.}\ }\textbf {\bibinfo {volume} {15}},\ \bibinfo {pages} {1052–1057} (\bibinfo {year} {2019})}\BibitemShut {NoStop}%
\bibitem [{\citenamefont {Sibille}\ \emph {et~al.}(2020)\citenamefont {Sibille}, \citenamefont {Gauthier}, \citenamefont {Lhotel}, \citenamefont {Por{\'e}e}, \citenamefont {Pomjakushin}, \citenamefont {Ewings}, \citenamefont {Perring}, \citenamefont {Ollivier}, \citenamefont {Wildes}, \citenamefont {Ritter}, \citenamefont {Hansen}, \citenamefont {Keen}, \citenamefont {Nilsen}, \citenamefont {Keller}, \citenamefont {Petit},\ and\ \citenamefont {Fennell}}]{Sibille2020}%
  \BibitemOpen
  \bibfield  {author} {\bibinfo {author} {\bibfnamefont {R.}~\bibnamefont {Sibille}}, \bibinfo {author} {\bibfnamefont {N.}~\bibnamefont {Gauthier}}, \bibinfo {author} {\bibfnamefont {E.}~\bibnamefont {Lhotel}}, \bibinfo {author} {\bibfnamefont {V.}~\bibnamefont {Por{\'e}e}}, \bibinfo {author} {\bibfnamefont {V.}~\bibnamefont {Pomjakushin}}, \bibinfo {author} {\bibfnamefont {R.~A.}\ \bibnamefont {Ewings}}, \bibinfo {author} {\bibfnamefont {T.~G.}\ \bibnamefont {Perring}}, \bibinfo {author} {\bibfnamefont {J.}~\bibnamefont {Ollivier}}, \bibinfo {author} {\bibfnamefont {A.}~\bibnamefont {Wildes}}, \bibinfo {author} {\bibfnamefont {C.}~\bibnamefont {Ritter}}, \bibinfo {author} {\bibfnamefont {T.~C.}\ \bibnamefont {Hansen}}, \bibinfo {author} {\bibfnamefont {D.~A.}\ \bibnamefont {Keen}}, \bibinfo {author} {\bibfnamefont {G.~J.}\ \bibnamefont {Nilsen}}, \bibinfo {author} {\bibfnamefont {L.}~\bibnamefont {Keller}}, \bibinfo {author} {\bibfnamefont {S.}~\bibnamefont {Petit}},\ and\ \bibinfo {author} {\bibfnamefont
  {T.}~\bibnamefont {Fennell}},\ }\bibfield  {title} {\bibinfo {title} {\textit{A Quantum Liquid of Magnetic Octupoles on the Pyrochlore Lattice}},\ }\href {https://doi.org/10.1038/s41567-020-0827-7} {\bibfield  {journal} {\bibinfo  {journal} {Nat. Phys.}\ }\textbf {\bibinfo {volume} {16}},\ \bibinfo {pages} {546} (\bibinfo {year} {2020})}\BibitemShut {NoStop}%
\bibitem [{\citenamefont {Por\'ee}\ \emph {et~al.}(2022)\citenamefont {Por\'ee}, \citenamefont {Lhotel}, \citenamefont {Petit}, \citenamefont {Krajewska}, \citenamefont {Puphal}, \citenamefont {Clark}, \citenamefont {Pomjakushin}, \citenamefont {Walker}, \citenamefont {Gauthier}, \citenamefont {Gawryluk},\ and\ \citenamefont {Sibille}}]{Poree2022}%
  \BibitemOpen
  \bibfield  {author} {\bibinfo {author} {\bibfnamefont {V.}~\bibnamefont {Por\'ee}}, \bibinfo {author} {\bibfnamefont {E.}~\bibnamefont {Lhotel}}, \bibinfo {author} {\bibfnamefont {S.}~\bibnamefont {Petit}}, \bibinfo {author} {\bibfnamefont {A.}~\bibnamefont {Krajewska}}, \bibinfo {author} {\bibfnamefont {P.}~\bibnamefont {Puphal}}, \bibinfo {author} {\bibfnamefont {A.~H.}\ \bibnamefont {Clark}}, \bibinfo {author} {\bibfnamefont {V.}~\bibnamefont {Pomjakushin}}, \bibinfo {author} {\bibfnamefont {H.~C.}\ \bibnamefont {Walker}}, \bibinfo {author} {\bibfnamefont {N.}~\bibnamefont {Gauthier}}, \bibinfo {author} {\bibfnamefont {D.~J.}\ \bibnamefont {Gawryluk}},\ and\ \bibinfo {author} {\bibfnamefont {R.}~\bibnamefont {Sibille}},\ }\bibfield  {title} {\bibinfo {title} {\textit{Crystal-Field States and Defect Levels in Candidate Quantum Spin Ice ${\mathrm{Ce}}_{2}{\mathrm{Hf}}_{2}{\mathrm{O}}_{7}$}},\ }\href {https://doi.org/10.1103/PhysRevMaterials.6.044406} {\bibfield  {journal} {\bibinfo  {journal} {Phys. Rev.
  Materials}\ }\textbf {\bibinfo {volume} {6}},\ \bibinfo {pages} {044406} (\bibinfo {year} {2022})}\BibitemShut {NoStop}%
\bibitem [{\citenamefont {Curnoe}(2007)}]{Curnoe2007}%
  \BibitemOpen
  \bibfield  {author} {\bibinfo {author} {\bibfnamefont {S.~H.}\ \bibnamefont {Curnoe}},\ }\bibfield  {title} {\bibinfo {title} {\textit{Quantum Spin Configurations in ${\mathrm{Tb}}_{2}{\mathrm{Ti}}_{2}{\mathrm{O}}_{7}$}},\ }\href {https://doi.org/10.1103/PhysRevB.75.212404} {\bibfield  {journal} {\bibinfo  {journal} {Phys. Rev. B}\ }\textbf {\bibinfo {volume} {75}},\ \bibinfo {pages} {212404} (\bibinfo {year} {2007})}\BibitemShut {NoStop}%
\bibitem [{\citenamefont {Onoda}\ and\ \citenamefont {Tanaka}(2011)}]{Onada2011}%
  \BibitemOpen
  \bibfield  {author} {\bibinfo {author} {\bibfnamefont {S.}~\bibnamefont {Onoda}}\ and\ \bibinfo {author} {\bibfnamefont {Y.}~\bibnamefont {Tanaka}},\ }\bibfield  {title} {\bibinfo {title} {\textit{Quantum Fluctuations in the Effective Pseudospin-$\frac{1}{2}$ Model for Magnetic Pyrochlore Oxides}},\ }\href {https://doi.org/10.1103/PhysRevB.83.094411} {\bibfield  {journal} {\bibinfo  {journal} {Phys. Rev. B}\ }\textbf {\bibinfo {volume} {83}},\ \bibinfo {pages} {094411} (\bibinfo {year} {2011})}\BibitemShut {NoStop}%
\bibitem [{\citenamefont {Huang}\ \emph {et~al.}(2014)\citenamefont {Huang}, \citenamefont {Chen},\ and\ \citenamefont {Hermele}}]{Huang2014}%
  \BibitemOpen
  \bibfield  {author} {\bibinfo {author} {\bibfnamefont {Y.-P.}\ \bibnamefont {Huang}}, \bibinfo {author} {\bibfnamefont {G.}~\bibnamefont {Chen}},\ and\ \bibinfo {author} {\bibfnamefont {M.}~\bibnamefont {Hermele}},\ }\bibfield  {title} {\bibinfo {title} {\textit{Quantum Spin Ices and Topological Phases from Dipolar-Octupolar Doublets on the Pyrochlore Lattice}},\ }\href {https://doi.org/10.1103/PhysRevLett.112.167203} {\bibfield  {journal} {\bibinfo  {journal} {Phys. Rev. Lett.}\ }\textbf {\bibinfo {volume} {112}},\ \bibinfo {pages} {167203} (\bibinfo {year} {2014})}\BibitemShut {NoStop}%
\bibitem [{\citenamefont {Rau}\ and\ \citenamefont {Gingras}(2019)}]{RauReview2019}%
  \BibitemOpen
  \bibfield  {author} {\bibinfo {author} {\bibfnamefont {J.~G.}\ \bibnamefont {Rau}}\ and\ \bibinfo {author} {\bibfnamefont {M.~J.}\ \bibnamefont {Gingras}},\ }\bibfield  {title} {\bibinfo {title} {\textit{Frustrated Quantum Rare-Earth Pyrochlores}},\ }\href {https://doi.org/10.1146/annurev-conmatphys-022317-110520} {\bibfield  {journal} {\bibinfo  {journal} {Annu. Rev. Condens. Matter Phys}\ }\textbf {\bibinfo {volume} {10}},\ \bibinfo {pages} {357} (\bibinfo {year} {2019})}\BibitemShut {NoStop}%
\bibitem [{\citenamefont {Benton}(2020)}]{Benton2020}%
  \BibitemOpen
  \bibfield  {author} {\bibinfo {author} {\bibfnamefont {O.}~\bibnamefont {Benton}},\ }\bibfield  {title} {\bibinfo {title} {\textit{Ground-state Phase Diagram of Dipolar-Octupolar Pyrochlores}},\ }\href {https://doi.org/10.1103/PhysRevB.102.104408} {\bibfield  {journal} {\bibinfo  {journal} {Phys. Rev. B}\ }\textbf {\bibinfo {volume} {102}},\ \bibinfo {pages} {104408} (\bibinfo {year} {2020})}\BibitemShut {NoStop}%
\bibitem [{\citenamefont {Patri}\ \emph {et~al.}(2020{\natexlab{a}})\citenamefont {Patri}, \citenamefont {Hosoi},\ and\ \citenamefont {Kim}}]{Patri2020}%
  \BibitemOpen
  \bibfield  {author} {\bibinfo {author} {\bibfnamefont {A.~S.}\ \bibnamefont {Patri}}, \bibinfo {author} {\bibfnamefont {M.}~\bibnamefont {Hosoi}},\ and\ \bibinfo {author} {\bibfnamefont {Y.~B.}\ \bibnamefont {Kim}},\ }\bibfield  {title} {\bibinfo {title} {\textit{Distinguishing Dipolar and Octupolar Quantum Spin Ices using Contrasting Magnetostriction Signatures}},\ }\href {https://doi.org/10.1103/PhysRevResearch.2.023253} {\bibfield  {journal} {\bibinfo  {journal} {Phys. Rev. Research}\ }\textbf {\bibinfo {volume} {2}},\ \bibinfo {pages} {023253} (\bibinfo {year} {2020}{\natexlab{a}})}\BibitemShut {NoStop}%
\bibitem [{\citenamefont {Huang}\ \emph {et~al.}(2020)\citenamefont {Huang}, \citenamefont {Liu}, \citenamefont {Meng}, \citenamefont {Yu}, \citenamefont {Deng},\ and\ \citenamefont {Chen}}]{Huang2020}%
  \BibitemOpen
  \bibfield  {author} {\bibinfo {author} {\bibfnamefont {C.-J.}\ \bibnamefont {Huang}}, \bibinfo {author} {\bibfnamefont {C.}~\bibnamefont {Liu}}, \bibinfo {author} {\bibfnamefont {Z.}~\bibnamefont {Meng}}, \bibinfo {author} {\bibfnamefont {Y.}~\bibnamefont {Yu}}, \bibinfo {author} {\bibfnamefont {Y.}~\bibnamefont {Deng}},\ and\ \bibinfo {author} {\bibfnamefont {G.}~\bibnamefont {Chen}},\ }\bibfield  {title} {\bibinfo {title} {\textit{Extended Coulomb Liquid of Paired Hardcore Boson Model on a Pyrochlore Lattice}},\ }\href {https://doi.org/10.1103/PhysRevResearch.2.042022} {\bibfield  {journal} {\bibinfo  {journal} {Phys. Rev. Research}\ }\textbf {\bibinfo {volume} {2}},\ \bibinfo {pages} {042022(R)} (\bibinfo {year} {2020})}\BibitemShut {NoStop}%
\bibitem [{\citenamefont {Smith}\ \emph {et~al.}(2022)\citenamefont {Smith}, \citenamefont {Benton}, \citenamefont {Yahne}, \citenamefont {Placke}, \citenamefont {Sch\"afer}, \citenamefont {Gaudet}, \citenamefont {Dudemaine}, \citenamefont {Fitterman}, \citenamefont {Beare}, \citenamefont {Wildes}, \citenamefont {Bhattacharya}, \citenamefont {DeLazzer}, \citenamefont {Buhariwalla}, \citenamefont {Butch}, \citenamefont {Movshovich}, \citenamefont {Garrett}, \citenamefont {Marjerrison}, \citenamefont {Clancy}, \citenamefont {Kermarrec}, \citenamefont {Luke}, \citenamefont {Bianchi}, \citenamefont {Ross},\ and\ \citenamefont {Gaulin}}]{Smith2022}%
  \BibitemOpen
  \bibfield  {author} {\bibinfo {author} {\bibfnamefont {E.~M.}\ \bibnamefont {Smith}}, \bibinfo {author} {\bibfnamefont {O.}~\bibnamefont {Benton}}, \bibinfo {author} {\bibfnamefont {D.~R.}\ \bibnamefont {Yahne}}, \bibinfo {author} {\bibfnamefont {B.}~\bibnamefont {Placke}}, \bibinfo {author} {\bibfnamefont {R.}~\bibnamefont {Sch\"afer}}, \bibinfo {author} {\bibfnamefont {J.}~\bibnamefont {Gaudet}}, \bibinfo {author} {\bibfnamefont {J.}~\bibnamefont {Dudemaine}}, \bibinfo {author} {\bibfnamefont {A.}~\bibnamefont {Fitterman}}, \bibinfo {author} {\bibfnamefont {J.}~\bibnamefont {Beare}}, \bibinfo {author} {\bibfnamefont {A.~R.}\ \bibnamefont {Wildes}}, \bibinfo {author} {\bibfnamefont {S.}~\bibnamefont {Bhattacharya}}, \bibinfo {author} {\bibfnamefont {T.}~\bibnamefont {DeLazzer}}, \bibinfo {author} {\bibfnamefont {C.~R.~C.}\ \bibnamefont {Buhariwalla}}, \bibinfo {author} {\bibfnamefont {N.~P.}\ \bibnamefont {Butch}}, \bibinfo {author} {\bibfnamefont {R.}~\bibnamefont {Movshovich}}, \bibinfo {author}
  {\bibfnamefont {J.~D.}\ \bibnamefont {Garrett}}, \bibinfo {author} {\bibfnamefont {C.~A.}\ \bibnamefont {Marjerrison}}, \bibinfo {author} {\bibfnamefont {J.~P.}\ \bibnamefont {Clancy}}, \bibinfo {author} {\bibfnamefont {E.}~\bibnamefont {Kermarrec}}, \bibinfo {author} {\bibfnamefont {G.~M.}\ \bibnamefont {Luke}}, \bibinfo {author} {\bibfnamefont {A.~D.}\ \bibnamefont {Bianchi}}, \bibinfo {author} {\bibfnamefont {K.~A.}\ \bibnamefont {Ross}},\ and\ \bibinfo {author} {\bibfnamefont {B.~D.}\ \bibnamefont {Gaulin}},\ }\bibfield  {title} {\bibinfo {title} {\textit{Case for a ${\mathrm{U}(1)}_{\ensuremath{\pi}}$ Quantum Spin Liquid Ground State in the Dipole-Octupole Pyrochlore ${\mathrm{Ce}}_{2}{\mathrm{Zr}}_{2}{\mathrm{O}}_{7}$}},\ }\href {https://doi.org/10.1103/PhysRevX.12.021015} {\bibfield  {journal} {\bibinfo  {journal} {Phys. Rev. X}\ }\textbf {\bibinfo {volume} {12}},\ \bibinfo {pages} {021015} (\bibinfo {year} {2022})}\BibitemShut {NoStop}%
\bibitem [{\citenamefont {Bhardwaj}\ \emph {et~al.}(2022)\citenamefont {Bhardwaj}, \citenamefont {Zhang}, \citenamefont {Yan}, \citenamefont {Moessner}, \citenamefont {Nevidomskyy},\ and\ \citenamefont {Changlani}}]{Changlani2022}%
  \BibitemOpen
  \bibfield  {author} {\bibinfo {author} {\bibfnamefont {A.}~\bibnamefont {Bhardwaj}}, \bibinfo {author} {\bibfnamefont {S.}~\bibnamefont {Zhang}}, \bibinfo {author} {\bibfnamefont {H.}~\bibnamefont {Yan}}, \bibinfo {author} {\bibfnamefont {R.}~\bibnamefont {Moessner}}, \bibinfo {author} {\bibfnamefont {A.}~\bibnamefont {Nevidomskyy}},\ and\ \bibinfo {author} {\bibfnamefont {H.}~\bibnamefont {Changlani}},\ }\bibfield  {title} {\bibinfo {title} {\textit{Sleuthing out Exotic Quantum Spin Liquidity in the Pyrochlore Magnet ${\mathrm{Ce}}_{2}{\mathrm{Zr}}_{2}{\mathrm{O}}_{7}$}},\ }\href {https://doi.org/10.1038/s41535-022-00458-2} {\bibfield  {journal} {\bibinfo  {journal} {npj Quantum Materials}\ }\textbf {\bibinfo {volume} {7}} (\bibinfo {year} {2022})}\BibitemShut {NoStop}%
\bibitem [{\citenamefont {Gao}\ \emph {et~al.}(2022)\citenamefont {Gao}, \citenamefont {Chen}, \citenamefont {Yan}, \citenamefont {Duan}, \citenamefont {Huang}, \citenamefont {Yao}, \citenamefont {Ye}, \citenamefont {Balz}, \citenamefont {Stewart}, \citenamefont {Nakajima}, \citenamefont {Ohira-Kawamura}, \citenamefont {Xu}, \citenamefont {Xu}, \citenamefont {Cheong}, \citenamefont {Morosan}, \citenamefont {Nevidomskyy}, \citenamefont {Chen},\ and\ \citenamefont {Dai}}]{Gao2022}%
  \BibitemOpen
  \bibfield  {author} {\bibinfo {author} {\bibfnamefont {B.}~\bibnamefont {Gao}}, \bibinfo {author} {\bibfnamefont {T.}~\bibnamefont {Chen}}, \bibinfo {author} {\bibfnamefont {H.}~\bibnamefont {Yan}}, \bibinfo {author} {\bibfnamefont {C.}~\bibnamefont {Duan}}, \bibinfo {author} {\bibfnamefont {C.-L.}\ \bibnamefont {Huang}}, \bibinfo {author} {\bibfnamefont {X.~P.}\ \bibnamefont {Yao}}, \bibinfo {author} {\bibfnamefont {F.}~\bibnamefont {Ye}}, \bibinfo {author} {\bibfnamefont {C.}~\bibnamefont {Balz}}, \bibinfo {author} {\bibfnamefont {J.~R.}\ \bibnamefont {Stewart}}, \bibinfo {author} {\bibfnamefont {K.}~\bibnamefont {Nakajima}}, \bibinfo {author} {\bibfnamefont {S.}~\bibnamefont {Ohira-Kawamura}}, \bibinfo {author} {\bibfnamefont {G.}~\bibnamefont {Xu}}, \bibinfo {author} {\bibfnamefont {X.}~\bibnamefont {Xu}}, \bibinfo {author} {\bibfnamefont {S.-W.}\ \bibnamefont {Cheong}}, \bibinfo {author} {\bibfnamefont {E.}~\bibnamefont {Morosan}}, \bibinfo {author} {\bibfnamefont {A.~H.}\ \bibnamefont {Nevidomskyy}},
  \bibinfo {author} {\bibfnamefont {G.}~\bibnamefont {Chen}},\ and\ \bibinfo {author} {\bibfnamefont {P.}~\bibnamefont {Dai}},\ }\bibfield  {title} {\bibinfo {title} {\textit{Magnetic Field Effects in an Octupolar Quantum Spin Liquid Candidate}},\ }\href {https://doi.org/10.1103/PhysRevB.106.094425} {\bibfield  {journal} {\bibinfo  {journal} {Phys. Rev. B}\ }\textbf {\bibinfo {volume} {106}},\ \bibinfo {pages} {094425} (\bibinfo {year} {2022})}\BibitemShut {NoStop}%
\bibitem [{\citenamefont {Smith}\ \emph {et~al.}(2023)\citenamefont {Smith}, \citenamefont {Dudemaine}, \citenamefont {Placke}, \citenamefont {Sch\"afer}, \citenamefont {Yahne}, \citenamefont {DeLazzer}, \citenamefont {Fitterman}, \citenamefont {Beare}, \citenamefont {Gaudet}, \citenamefont {Buhariwalla}, \citenamefont {Podlesnyak}, \citenamefont {Xu}, \citenamefont {Clancy}, \citenamefont {Movshovich}, \citenamefont {Luke}, \citenamefont {Ross}, \citenamefont {Moessner}, \citenamefont {Benton}, \citenamefont {Bianchi},\ and\ \citenamefont {Gaulin}}]{Smith2023}%
  \BibitemOpen
  \bibfield  {author} {\bibinfo {author} {\bibfnamefont {E.~M.}\ \bibnamefont {Smith}}, \bibinfo {author} {\bibfnamefont {J.}~\bibnamefont {Dudemaine}}, \bibinfo {author} {\bibfnamefont {B.}~\bibnamefont {Placke}}, \bibinfo {author} {\bibfnamefont {R.}~\bibnamefont {Sch\"afer}}, \bibinfo {author} {\bibfnamefont {D.~R.}\ \bibnamefont {Yahne}}, \bibinfo {author} {\bibfnamefont {T.}~\bibnamefont {DeLazzer}}, \bibinfo {author} {\bibfnamefont {A.}~\bibnamefont {Fitterman}}, \bibinfo {author} {\bibfnamefont {J.}~\bibnamefont {Beare}}, \bibinfo {author} {\bibfnamefont {J.}~\bibnamefont {Gaudet}}, \bibinfo {author} {\bibfnamefont {C.~R.~C.}\ \bibnamefont {Buhariwalla}}, \bibinfo {author} {\bibfnamefont {A.}~\bibnamefont {Podlesnyak}}, \bibinfo {author} {\bibfnamefont {G.}~\bibnamefont {Xu}}, \bibinfo {author} {\bibfnamefont {J.~P.}\ \bibnamefont {Clancy}}, \bibinfo {author} {\bibfnamefont {R.}~\bibnamefont {Movshovich}}, \bibinfo {author} {\bibfnamefont {G.~M.}\ \bibnamefont {Luke}}, \bibinfo {author} {\bibfnamefont
  {K.~A.}\ \bibnamefont {Ross}}, \bibinfo {author} {\bibfnamefont {R.}~\bibnamefont {Moessner}}, \bibinfo {author} {\bibfnamefont {O.}~\bibnamefont {Benton}}, \bibinfo {author} {\bibfnamefont {A.~D.}\ \bibnamefont {Bianchi}},\ and\ \bibinfo {author} {\bibfnamefont {B.~D.}\ \bibnamefont {Gaulin}},\ }\bibfield  {title} {\bibinfo {title} {\textit{Quantum Spin Ice Response to a Magnetic Field in the Dipole-Octupole Pyrochlore ${\mathrm{Ce}}_{2}{\mathrm{Zr}}_{2}{\mathrm{O}}_{7}$}},\ }\href {https://doi.org/10.1103/PhysRevB.108.054438} {\bibfield  {journal} {\bibinfo  {journal} {Phys. Rev. B}\ }\textbf {\bibinfo {volume} {108}},\ \bibinfo {pages} {054438} (\bibinfo {year} {2023})}\BibitemShut {NoStop}%
\bibitem [{\citenamefont {Beare}\ \emph {et~al.}(2023)\citenamefont {Beare}, \citenamefont {Smith}, \citenamefont {Dudemaine}, \citenamefont {Sch\"afer}, \citenamefont {Rutherford}, \citenamefont {Sharma}, \citenamefont {Fitterman}, \citenamefont {Marjerrison}, \citenamefont {Williams}, \citenamefont {Aczel}, \citenamefont {Dunsiger}, \citenamefont {Bianchi}, \citenamefont {Gaulin},\ and\ \citenamefont {Luke}}]{Beare2023}%
  \BibitemOpen
  \bibfield  {author} {\bibinfo {author} {\bibfnamefont {J.}~\bibnamefont {Beare}}, \bibinfo {author} {\bibfnamefont {E.~M.}\ \bibnamefont {Smith}}, \bibinfo {author} {\bibfnamefont {J.}~\bibnamefont {Dudemaine}}, \bibinfo {author} {\bibfnamefont {R.}~\bibnamefont {Sch\"afer}}, \bibinfo {author} {\bibfnamefont {M.~R.}\ \bibnamefont {Rutherford}}, \bibinfo {author} {\bibfnamefont {S.}~\bibnamefont {Sharma}}, \bibinfo {author} {\bibfnamefont {A.}~\bibnamefont {Fitterman}}, \bibinfo {author} {\bibfnamefont {C.~A.}\ \bibnamefont {Marjerrison}}, \bibinfo {author} {\bibfnamefont {T.~J.}\ \bibnamefont {Williams}}, \bibinfo {author} {\bibfnamefont {A.~A.}\ \bibnamefont {Aczel}}, \bibinfo {author} {\bibfnamefont {S.~R.}\ \bibnamefont {Dunsiger}}, \bibinfo {author} {\bibfnamefont {A.~D.}\ \bibnamefont {Bianchi}}, \bibinfo {author} {\bibfnamefont {B.~D.}\ \bibnamefont {Gaulin}},\ and\ \bibinfo {author} {\bibfnamefont {G.~M.}\ \bibnamefont {Luke}},\ }\bibfield  {title} {\bibinfo {title}
  {\textit{$\ensuremath{\mu}\mathrm{SR}$ Study of the Dipole-Octupole Quantum Spin Ice Candidate $\mathrm{Ce}_2\mathrm{Zr}_2\mathrm{O}_7$}},\ }\href {https://doi.org/10.1103/PhysRevB.108.174411} {\bibfield  {journal} {\bibinfo  {journal} {Phys. Rev. B}\ }\textbf {\bibinfo {volume} {108}},\ \bibinfo {pages} {174411} (\bibinfo {year} {2023})}\BibitemShut {NoStop}%
\bibitem [{\citenamefont {Smith}\ \emph {et~al.}(2025)\citenamefont {Smith}, \citenamefont {Sch\"afer}, \citenamefont {Dudemaine}, \citenamefont {Placke}, \citenamefont {Yuan}, \citenamefont {Morgan}, \citenamefont {Ye}, \citenamefont {Moessner}, \citenamefont {Benton}, \citenamefont {Bianchi},\ and\ \citenamefont {Gaulin}}]{Smith2024}%
  \BibitemOpen
  \bibfield  {author} {\bibinfo {author} {\bibfnamefont {E.~M.}\ \bibnamefont {Smith}}, \bibinfo {author} {\bibfnamefont {R.}~\bibnamefont {Sch\"afer}}, \bibinfo {author} {\bibfnamefont {J.}~\bibnamefont {Dudemaine}}, \bibinfo {author} {\bibfnamefont {B.}~\bibnamefont {Placke}}, \bibinfo {author} {\bibfnamefont {B.}~\bibnamefont {Yuan}}, \bibinfo {author} {\bibfnamefont {Z.}~\bibnamefont {Morgan}}, \bibinfo {author} {\bibfnamefont {F.}~\bibnamefont {Ye}}, \bibinfo {author} {\bibfnamefont {R.}~\bibnamefont {Moessner}}, \bibinfo {author} {\bibfnamefont {O.}~\bibnamefont {Benton}}, \bibinfo {author} {\bibfnamefont {A.~D.}\ \bibnamefont {Bianchi}},\ and\ \bibinfo {author} {\bibfnamefont {B.~D.}\ \bibnamefont {Gaulin}},\ }\bibfield  {title} {\bibinfo {title} {\textit{Single-Crystal Diffuse Neutron Scattering Study of the Dipole-Octupole Quantum Spin-Ice Candidate ${\mathrm{Ce}}_{2}{\mathrm{Zr}}_{2}{\mathrm{O}}_{7}$: No Apparent Octupolar Correlations Above $T=0.05\text{ }\mathrm{K}$}},\ }\href
  {https://doi.org/10.1103/PhysRevX.15.021033} {\bibfield  {journal} {\bibinfo  {journal} {Phys. Rev. X}\ }\textbf {\bibinfo {volume} {15}},\ \bibinfo {pages} {021033} (\bibinfo {year} {2025})}\BibitemShut {NoStop}%
\bibitem [{\citenamefont {Gao}\ \emph {et~al.}(2025)\citenamefont {Gao}, \citenamefont {Desrochers}, \citenamefont {Tam}, \citenamefont {Kirschbaum}, \citenamefont {Steffens}, \citenamefont {Hiess}, \citenamefont {Nguyen}, \citenamefont {Su}, \citenamefont {Cheong}, \citenamefont {Paschen}, \citenamefont {Kim},\ and\ \citenamefont {Dai}}]{Gao2024}%
  \BibitemOpen
  \bibfield  {author} {\bibinfo {author} {\bibfnamefont {B.}~\bibnamefont {Gao}}, \bibinfo {author} {\bibfnamefont {F.}~\bibnamefont {Desrochers}}, \bibinfo {author} {\bibfnamefont {D.~W.}\ \bibnamefont {Tam}}, \bibinfo {author} {\bibfnamefont {D.~M.}\ \bibnamefont {Kirschbaum}}, \bibinfo {author} {\bibfnamefont {P.}~\bibnamefont {Steffens}}, \bibinfo {author} {\bibfnamefont {A.}~\bibnamefont {Hiess}}, \bibinfo {author} {\bibfnamefont {D.~H.}\ \bibnamefont {Nguyen}}, \bibinfo {author} {\bibfnamefont {Y.}~\bibnamefont {Su}}, \bibinfo {author} {\bibfnamefont {S.-W.}\ \bibnamefont {Cheong}}, \bibinfo {author} {\bibfnamefont {S.}~\bibnamefont {Paschen}}, \bibinfo {author} {\bibfnamefont {Y.~B.}\ \bibnamefont {Kim}},\ and\ \bibinfo {author} {\bibfnamefont {P.}~\bibnamefont {Dai}},\ }\bibfield  {title} {\bibinfo {title} {\textit{Neutron Scattering and Thermodynamic Evidence for Emergent Photons and Fractionalization in a Pyrochlore Spin Ice}},\ }\bibfield  {journal} {\bibinfo  {journal} {Nat. Phys.}\ }\href
  {https://doi.org/10.1038/s41567-025-02922-9} {10.1038/s41567-025-02922-9} (\bibinfo {year} {2025})\BibitemShut {NoStop}%
\bibitem [{\citenamefont {Smith}\ \emph {et~al.}(2024)\citenamefont {Smith}, \citenamefont {Lhotel}, \citenamefont {Petit},\ and\ \citenamefont {Gaulin}}]{Smith2025}%
  \BibitemOpen
  \bibfield  {author} {\bibinfo {author} {\bibfnamefont {E.~M.}\ \bibnamefont {Smith}}, \bibinfo {author} {\bibfnamefont {E.}~\bibnamefont {Lhotel}}, \bibinfo {author} {\bibfnamefont {S.}~\bibnamefont {Petit}},\ and\ \bibinfo {author} {\bibfnamefont {B.~D.}\ \bibnamefont {Gaulin}},\ }\bibfield  {title} {\bibinfo {title} {\textit{Experimental Insights into Quantum Spin Ice Physics in Dipole–Octupole Pyrochlore Magnets}},\ }\href {https://doi.org/10.1146/annurev-conmatphys-041124-015101} {\bibfield  {journal} {\bibinfo  {journal} {Annu. Rev. Condens. Matter Phys.}\ }\textbf {\bibinfo {volume} {16}},\ \bibinfo {pages} {387} (\bibinfo {year} {2024})}\BibitemShut {NoStop}%
\bibitem [{\citenamefont {Desrochers}\ and\ \citenamefont {Kim}(2024{\natexlab{a}})}]{Desrochers2024a}%
  \BibitemOpen
  \bibfield  {author} {\bibinfo {author} {\bibfnamefont {F.}~\bibnamefont {Desrochers}}\ and\ \bibinfo {author} {\bibfnamefont {Y.~B.}\ \bibnamefont {Kim}},\ }\bibfield  {title} {\bibinfo {title} {\textit{Spectroscopic Signatures of Fractionalization in Octupolar Quantum Spin Ice}},\ }\href {https://doi.org/10.1103/PhysRevLett.132.066502} {\bibfield  {journal} {\bibinfo  {journal} {Phys. Rev. Lett.}\ }\textbf {\bibinfo {volume} {132}},\ \bibinfo {pages} {066502} (\bibinfo {year} {2024}{\natexlab{a}})}\BibitemShut {NoStop}%
\bibitem [{\citenamefont {Desrochers}\ and\ \citenamefont {Kim}(2024{\natexlab{b}})}]{Desrochers2024b}%
  \BibitemOpen
  \bibfield  {author} {\bibinfo {author} {\bibfnamefont {F.}~\bibnamefont {Desrochers}}\ and\ \bibinfo {author} {\bibfnamefont {Y.~B.}\ \bibnamefont {Kim}},\ }\bibfield  {title} {\bibinfo {title} {\textit{Finite-Temperature Dynamics in 0-Flux and $\ensuremath{\pi}$-Flux Quantum Spin Ice: Self-Consistent Exclusive Boson Approach}},\ }\href {https://doi.org/10.1103/PhysRevB.109.144410} {\bibfield  {journal} {\bibinfo  {journal} {Phys. Rev. B}\ }\textbf {\bibinfo {volume} {109}},\ \bibinfo {pages} {144410} (\bibinfo {year} {2024}{\natexlab{b}})}\BibitemShut {NoStop}%
\bibitem [{\citenamefont {Por{\'e}e}\ \emph {et~al.}(2025)\citenamefont {Por{\'e}e}, \citenamefont {Yan}, \citenamefont {Desrochers}, \citenamefont {Petit}, \citenamefont {Lhotel}, \citenamefont {Appel}, \citenamefont {Ollivier}, \citenamefont {Kim}, \citenamefont {Nevidomskyy},\ and\ \citenamefont {Sibille}}]{Poree2023}%
  \BibitemOpen
  \bibfield  {author} {\bibinfo {author} {\bibfnamefont {V.}~\bibnamefont {Por{\'e}e}}, \bibinfo {author} {\bibfnamefont {H.}~\bibnamefont {Yan}}, \bibinfo {author} {\bibfnamefont {F.}~\bibnamefont {Desrochers}}, \bibinfo {author} {\bibfnamefont {S.}~\bibnamefont {Petit}}, \bibinfo {author} {\bibfnamefont {E.}~\bibnamefont {Lhotel}}, \bibinfo {author} {\bibfnamefont {M.}~\bibnamefont {Appel}}, \bibinfo {author} {\bibfnamefont {J.}~\bibnamefont {Ollivier}}, \bibinfo {author} {\bibfnamefont {Y.~B.}\ \bibnamefont {Kim}}, \bibinfo {author} {\bibfnamefont {A.~H.}\ \bibnamefont {Nevidomskyy}},\ and\ \bibinfo {author} {\bibfnamefont {R.}~\bibnamefont {Sibille}},\ }\bibfield  {title} {\bibinfo {title} {\textit{Evidence for Fractional Matter Coupled to an Emergent Gauge Field in a Quantum Spin Ice}},\ }\href {https://doi.org/10.1038/s41567-024-02711-w} {\bibfield  {journal} {\bibinfo  {journal} {Nat. Phys.}\ }\textbf {\bibinfo {volume} {21}},\ \bibinfo {pages} {83} (\bibinfo {year} {2025})}\BibitemShut {NoStop}%
\bibitem [{\citenamefont {Yahne}\ \emph {et~al.}(2024)\citenamefont {Yahne}, \citenamefont {Placke}, \citenamefont {Sch\"afer}, \citenamefont {Benton}, \citenamefont {Moessner}, \citenamefont {Powell}, \citenamefont {Kolis}, \citenamefont {Pasco}, \citenamefont {May}, \citenamefont {Frontzek}, \citenamefont {Smith}, \citenamefont {Gaulin}, \citenamefont {Calder},\ and\ \citenamefont {Ross}}]{Yahne2024}%
  \BibitemOpen
  \bibfield  {author} {\bibinfo {author} {\bibfnamefont {D.~R.}\ \bibnamefont {Yahne}}, \bibinfo {author} {\bibfnamefont {B.}~\bibnamefont {Placke}}, \bibinfo {author} {\bibfnamefont {R.}~\bibnamefont {Sch\"afer}}, \bibinfo {author} {\bibfnamefont {O.}~\bibnamefont {Benton}}, \bibinfo {author} {\bibfnamefont {R.}~\bibnamefont {Moessner}}, \bibinfo {author} {\bibfnamefont {M.}~\bibnamefont {Powell}}, \bibinfo {author} {\bibfnamefont {J.~W.}\ \bibnamefont {Kolis}}, \bibinfo {author} {\bibfnamefont {C.~M.}\ \bibnamefont {Pasco}}, \bibinfo {author} {\bibfnamefont {A.~F.}\ \bibnamefont {May}}, \bibinfo {author} {\bibfnamefont {M.~D.}\ \bibnamefont {Frontzek}}, \bibinfo {author} {\bibfnamefont {E.~M.}\ \bibnamefont {Smith}}, \bibinfo {author} {\bibfnamefont {B.~D.}\ \bibnamefont {Gaulin}}, \bibinfo {author} {\bibfnamefont {S.}~\bibnamefont {Calder}},\ and\ \bibinfo {author} {\bibfnamefont {K.~A.}\ \bibnamefont {Ross}},\ }\bibfield  {title} {\bibinfo {title} {\textit{Dipolar Spin Ice Regime Proximate to an
  All-In-All-Out N\'eel Ground State in the Dipolar-Octupolar Pyrochlore ${\mathrm{Ce}}_{2}{\mathrm{Sn}}_{2}{\mathrm{O}}_{7}$}},\ }\href {https://doi.org/10.1103/PhysRevX.14.011005} {\bibfield  {journal} {\bibinfo  {journal} {Phys. Rev. X}\ }\textbf {\bibinfo {volume} {14}},\ \bibinfo {pages} {011005} (\bibinfo {year} {2024})}\BibitemShut {NoStop}%
\bibitem [{\citenamefont {Por\'ee}\ \emph {et~al.}(2023)\citenamefont {Por\'ee}, \citenamefont {Bhardwaj}, \citenamefont {Lhotel}, \citenamefont {Petit}, \citenamefont {Gauthier}, \citenamefont {Yan}, \citenamefont {Pomjakushin}, \citenamefont {Ollivier}, \citenamefont {Quilliam}, \citenamefont {Nevidomskyy}, \citenamefont {Changlani},\ and\ \citenamefont {Sibille}}]{Poree2023b}%
  \BibitemOpen
  \bibfield  {author} {\bibinfo {author} {\bibfnamefont {V.}~\bibnamefont {Por\'ee}}, \bibinfo {author} {\bibfnamefont {A.}~\bibnamefont {Bhardwaj}}, \bibinfo {author} {\bibfnamefont {E.}~\bibnamefont {Lhotel}}, \bibinfo {author} {\bibfnamefont {S.}~\bibnamefont {Petit}}, \bibinfo {author} {\bibfnamefont {N.}~\bibnamefont {Gauthier}}, \bibinfo {author} {\bibfnamefont {H.}~\bibnamefont {Yan}}, \bibinfo {author} {\bibfnamefont {V.}~\bibnamefont {Pomjakushin}}, \bibinfo {author} {\bibfnamefont {J.}~\bibnamefont {Ollivier}}, \bibinfo {author} {\bibfnamefont {J.~A.}\ \bibnamefont {Quilliam}}, \bibinfo {author} {\bibfnamefont {A.~H.}\ \bibnamefont {Nevidomskyy}}, \bibinfo {author} {\bibfnamefont {H.~J.}\ \bibnamefont {Changlani}},\ and\ \bibinfo {author} {\bibfnamefont {R.}~\bibnamefont {Sibille}},\ }\href@noop {} {\bibinfo {title} {\textit{Dipolar-Octupolar Correlations and Hierarchy of Exchange Interactions in $\mathrm{Ce}_2\mathrm{Hf}_2\mathrm{O}_7$}}} (\bibinfo {year} {2023}),\ \Eprint
  {https://arxiv.org/abs/2305.08261} {arXiv:2305.08261 [cond-mat.str-el]} \BibitemShut {NoStop}%
\bibitem [{\citenamefont {Bhardwaj}\ \emph {et~al.}(2025)\citenamefont {Bhardwaj}, \citenamefont {Por\'ee}, \citenamefont {Yan}, \citenamefont {Gauthier}, \citenamefont {Lhotel}, \citenamefont {Petit}, \citenamefont {Quilliam}, \citenamefont {Nevidomskyy}, \citenamefont {Sibille},\ and\ \citenamefont {Changlani}}]{Bhardwaj2024}%
  \BibitemOpen
  \bibfield  {author} {\bibinfo {author} {\bibfnamefont {A.}~\bibnamefont {Bhardwaj}}, \bibinfo {author} {\bibfnamefont {V.}~\bibnamefont {Por\'ee}}, \bibinfo {author} {\bibfnamefont {H.}~\bibnamefont {Yan}}, \bibinfo {author} {\bibfnamefont {N.}~\bibnamefont {Gauthier}}, \bibinfo {author} {\bibfnamefont {E.}~\bibnamefont {Lhotel}}, \bibinfo {author} {\bibfnamefont {S.}~\bibnamefont {Petit}}, \bibinfo {author} {\bibfnamefont {J.~A.}\ \bibnamefont {Quilliam}}, \bibinfo {author} {\bibfnamefont {A.~H.}\ \bibnamefont {Nevidomskyy}}, \bibinfo {author} {\bibfnamefont {R.}~\bibnamefont {Sibille}},\ and\ \bibinfo {author} {\bibfnamefont {H.~J.}\ \bibnamefont {Changlani}},\ }\bibfield  {title} {\bibinfo {title} {\textit{Thermodynamics of the Dipole-Octupole Pyrochlore Magnet ${\mathrm{Ce}}_{2}{\mathrm{Hf}}_{2}{\mathrm{O}}_{7}$ in Applied Magnetic Fields}},\ }\href {https://doi.org/10.1103/PhysRevB.111.155137} {\bibfield  {journal} {\bibinfo  {journal} {Phys. Rev. B}\ }\textbf {\bibinfo {volume} {111}},\ \bibinfo
  {pages} {155137} (\bibinfo {year} {2025})}\BibitemShut {NoStop}%
\bibitem [{SM()}]{SM}%
  \BibitemOpen
  \href@noop {} {}\bibinfo {note} {See Supplemental Material below for further details on experimental methods, theoretical calculations, additional measurements, additional analysis of measurements, and fits of theoretical calculations to measured data.}\BibitemShut {Stop}%
\bibitem [{\citenamefont {Sch\"afer}\ \emph {et~al.}(2020)\citenamefont {Sch\"afer}, \citenamefont {Hagym\'asi}, \citenamefont {Moessner},\ and\ \citenamefont {Luitz}}]{Schafer2020}%
  \BibitemOpen
  \bibfield  {author} {\bibinfo {author} {\bibfnamefont {R.}~\bibnamefont {Sch\"afer}}, \bibinfo {author} {\bibfnamefont {I.}~\bibnamefont {Hagym\'asi}}, \bibinfo {author} {\bibfnamefont {R.}~\bibnamefont {Moessner}},\ and\ \bibinfo {author} {\bibfnamefont {D.~J.}\ \bibnamefont {Luitz}},\ }\bibfield  {title} {\bibinfo {title} {\textit{Pyrochlore $S=\frac{1}{2}$ Heisenberg Antiferromagnet at Finite Temperature}},\ }\href {https://doi.org/10.1103/PhysRevB.102.054408} {\bibfield  {journal} {\bibinfo  {journal} {Phys. Rev. B}\ }\textbf {\bibinfo {volume} {102}},\ \bibinfo {pages} {054408} (\bibinfo {year} {2020})}\BibitemShut {NoStop}%
\bibitem [{\citenamefont {Sch\"afer}(2022)}]{schaefer_magnetic_2022}%
  \BibitemOpen
  \bibfield  {author} {\bibinfo {author} {\bibfnamefont {R.}~\bibnamefont {Sch\"afer}},\ }\emph {\bibinfo {title} {\textit{Magnetic Frustration in Three Dimensions}}},\ \href {https://nbn-resolving.org/urn:nbn:de:bsz:14-qucosa2-829375} {\bibinfo {type} {Dissertation}},\ \bibinfo  {school} {TU Dresden}, \bibinfo {address} {Dresden} (\bibinfo {year} {2022})\BibitemShut {NoStop}%
\bibitem [{\citenamefont {Fennell}\ \emph {et~al.}(2009)\citenamefont {Fennell}, \citenamefont {Deen}, \citenamefont {Wildes}, \citenamefont {Schmalzl}, \citenamefont {Prabhakaran}, \citenamefont {Boothroyd}, \citenamefont {Aldus}, \citenamefont {McMorrow},\ and\ \citenamefont {Bramwell}}]{Fennell2009}%
  \BibitemOpen
  \bibfield  {author} {\bibinfo {author} {\bibfnamefont {T.}~\bibnamefont {Fennell}}, \bibinfo {author} {\bibfnamefont {P.~P.}\ \bibnamefont {Deen}}, \bibinfo {author} {\bibfnamefont {A.~R.}\ \bibnamefont {Wildes}}, \bibinfo {author} {\bibfnamefont {K.}~\bibnamefont {Schmalzl}}, \bibinfo {author} {\bibfnamefont {D.}~\bibnamefont {Prabhakaran}}, \bibinfo {author} {\bibfnamefont {A.~T.}\ \bibnamefont {Boothroyd}}, \bibinfo {author} {\bibfnamefont {R.~J.}\ \bibnamefont {Aldus}}, \bibinfo {author} {\bibfnamefont {D.~F.}\ \bibnamefont {McMorrow}},\ and\ \bibinfo {author} {\bibfnamefont {S.~T.}\ \bibnamefont {Bramwell}},\ }\bibfield  {title} {\bibinfo {title} {\textit{Magnetic Coulomb Phase in the Spin Ice ${\mathrm{Ho}}_{2}{\mathrm{Ti}}_{2}{\mathrm{O}}_{7}$}},\ }\href {https://doi.org/10.1126/science.1177582} {\bibfield  {journal} {\bibinfo  {journal} {Science}\ }\textbf {\bibinfo {volume} {326}},\ \bibinfo {pages} {415} (\bibinfo {year} {2009})}\BibitemShut {NoStop}%
\bibitem [{\citenamefont {Clancy}\ \emph {et~al.}(2009)\citenamefont {Clancy}, \citenamefont {Ruff}, \citenamefont {Dunsiger}, \citenamefont {Zhao}, \citenamefont {Dabkowska}, \citenamefont {Gardner}, \citenamefont {Qiu}, \citenamefont {Copley}, \citenamefont {Jenkins},\ and\ \citenamefont {Gaulin}}]{Clancy2009}%
  \BibitemOpen
  \bibfield  {author} {\bibinfo {author} {\bibfnamefont {J.~P.}\ \bibnamefont {Clancy}}, \bibinfo {author} {\bibfnamefont {J.~P.~C.}\ \bibnamefont {Ruff}}, \bibinfo {author} {\bibfnamefont {S.~R.}\ \bibnamefont {Dunsiger}}, \bibinfo {author} {\bibfnamefont {Y.}~\bibnamefont {Zhao}}, \bibinfo {author} {\bibfnamefont {H.~A.}\ \bibnamefont {Dabkowska}}, \bibinfo {author} {\bibfnamefont {J.~S.}\ \bibnamefont {Gardner}}, \bibinfo {author} {\bibfnamefont {Y.}~\bibnamefont {Qiu}}, \bibinfo {author} {\bibfnamefont {J.~R.~D.}\ \bibnamefont {Copley}}, \bibinfo {author} {\bibfnamefont {T.}~\bibnamefont {Jenkins}},\ and\ \bibinfo {author} {\bibfnamefont {B.~D.}\ \bibnamefont {Gaulin}},\ }\bibfield  {title} {\bibinfo {title} {\textit{Revisiting Static and Dynamic Spin-Ice Correlations in $\mathrm{Ho}_2\mathrm{Ti}_2\mathrm{O}_7$ with Neutron Scattering}},\ }\href {https://doi.org/10.1103/PhysRevB.79.014408} {\bibfield  {journal} {\bibinfo  {journal} {Phys. Rev. B}\ }\textbf {\bibinfo {volume} {79}},\ \bibinfo {pages}
  {014408} (\bibinfo {year} {2009})}\BibitemShut {NoStop}%
\bibitem [{\citenamefont {Hosoi}\ \emph {et~al.}(2022)\citenamefont {Hosoi}, \citenamefont {Zhang}, \citenamefont {Patri},\ and\ \citenamefont {Kim}}]{Kim2022}%
  \BibitemOpen
  \bibfield  {author} {\bibinfo {author} {\bibfnamefont {M.}~\bibnamefont {Hosoi}}, \bibinfo {author} {\bibfnamefont {E.~Z.}\ \bibnamefont {Zhang}}, \bibinfo {author} {\bibfnamefont {A.~S.}\ \bibnamefont {Patri}},\ and\ \bibinfo {author} {\bibfnamefont {Y.~B.}\ \bibnamefont {Kim}},\ }\bibfield  {title} {\bibinfo {title} {\textit{Uncovering Footprints of Dipolar-Octupolar Quantum Spin Ice from Neutron Scattering Signatures}},\ }\href {https://doi.org/10.1103/PhysRevLett.129.097202} {\bibfield  {journal} {\bibinfo  {journal} {Phys. Rev. Lett.}\ }\textbf {\bibinfo {volume} {129}},\ \bibinfo {pages} {097202} (\bibinfo {year} {2022})}\BibitemShut {NoStop}%
\bibitem [{\citenamefont {Desrochers}\ \emph {et~al.}(2022)\citenamefont {Desrochers}, \citenamefont {Chern},\ and\ \citenamefont {Kim}}]{Desrochers2022}%
  \BibitemOpen
  \bibfield  {author} {\bibinfo {author} {\bibfnamefont {F.}~\bibnamefont {Desrochers}}, \bibinfo {author} {\bibfnamefont {L.~E.}\ \bibnamefont {Chern}},\ and\ \bibinfo {author} {\bibfnamefont {Y.~B.}\ \bibnamefont {Kim}},\ }\bibfield  {title} {\bibinfo {title} {\textit{Competing $U$(1) and ${\mathbb{Z}}_{2}$ Dipolar-Octupolar Quantum Spin Liquids on the Pyrochlore Lattice: Application to ${\mathrm{Ce}}_{2}{\mathrm{Zr}}_{2}{\mathrm{O}}_{7}$}},\ }\href {https://doi.org/10.1103/PhysRevB.105.035149} {\bibfield  {journal} {\bibinfo  {journal} {Phys. Rev. B}\ }\textbf {\bibinfo {volume} {105}},\ \bibinfo {pages} {035149} (\bibinfo {year} {2022})}\BibitemShut {NoStop}%
\bibitem [{\citenamefont {Kato}\ and\ \citenamefont {Onoda}(2015)}]{Kato2015}%
  \BibitemOpen
  \bibfield  {author} {\bibinfo {author} {\bibfnamefont {Y.}~\bibnamefont {Kato}}\ and\ \bibinfo {author} {\bibfnamefont {S.}~\bibnamefont {Onoda}},\ }\bibfield  {title} {\bibinfo {title} {\textit{Numerical Evidence of Quantum Melting of Spin Ice: Quantum-to-Classical Crossover}},\ }\href {https://doi.org/10.1103/PhysRevLett.115.077202} {\bibfield  {journal} {\bibinfo  {journal} {Phys. Rev. Lett.}\ }\textbf {\bibinfo {volume} {115}},\ \bibinfo {pages} {077202} (\bibinfo {year} {2015})}\BibitemShut {NoStop}%
\bibitem [{\citenamefont {Huang}\ \emph {et~al.}(2018)\citenamefont {Huang}, \citenamefont {Deng}, \citenamefont {Wan},\ and\ \citenamefont {Meng}}]{Huang2018a}%
  \BibitemOpen
  \bibfield  {author} {\bibinfo {author} {\bibfnamefont {C.-J.}\ \bibnamefont {Huang}}, \bibinfo {author} {\bibfnamefont {Y.}~\bibnamefont {Deng}}, \bibinfo {author} {\bibfnamefont {Y.}~\bibnamefont {Wan}},\ and\ \bibinfo {author} {\bibfnamefont {Z.~Y.}\ \bibnamefont {Meng}},\ }\bibfield  {title} {\bibinfo {title} {\textit{Dynamics of Topological Excitations in a Model Quantum Spin Ice}},\ }\href {https://doi.org/10.1103/PhysRevLett.120.167202} {\bibfield  {journal} {\bibinfo  {journal} {Phys. Rev. Lett.}\ }\textbf {\bibinfo {volume} {120}},\ \bibinfo {pages} {167202} (\bibinfo {year} {2018})}\BibitemShut {NoStop}%
\bibitem [{\citenamefont {Li}\ and\ \citenamefont {Chen}(2017)}]{Li2017}%
  \BibitemOpen
  \bibfield  {author} {\bibinfo {author} {\bibfnamefont {Y.-D.}\ \bibnamefont {Li}}\ and\ \bibinfo {author} {\bibfnamefont {G.}~\bibnamefont {Chen}},\ }\bibfield  {title} {\bibinfo {title} {\textit{Symmetry Enriched U(1) Topological Orders for Dipole-Octupole Doublets on a Pyrochlore Lattice}},\ }\href {https://doi.org/10.1103/PhysRevB.95.041106} {\bibfield  {journal} {\bibinfo  {journal} {Phys. Rev. B}\ }\textbf {\bibinfo {volume} {95}},\ \bibinfo {pages} {041106(R)} (\bibinfo {year} {2017})}\BibitemShut {NoStop}%
\bibitem [{\citenamefont {Kwasigroch}(2020)}]{Kwasigroch2020}%
  \BibitemOpen
  \bibfield  {author} {\bibinfo {author} {\bibfnamefont {M.~P.}\ \bibnamefont {Kwasigroch}},\ }\bibfield  {title} {\bibinfo {title} {\textit{Vison-Generated Photon Mass in Quantum Spin Ice: A Theoretical Framework}},\ }\href {https://doi.org/10.1103/PhysRevB.102.125113} {\bibfield  {journal} {\bibinfo  {journal} {Phys. Rev. B}\ }\textbf {\bibinfo {volume} {102}},\ \bibinfo {pages} {125113} (\bibinfo {year} {2020})}\BibitemShut {NoStop}%
\bibitem [{\citenamefont {Gaulin}\ \emph {et~al.}(2022{\natexlab{a}})\citenamefont {Gaulin}, \citenamefont {Huang}, \citenamefont {Balz},\ and\ \citenamefont {Smith}}]{LET_experiment_DOI}%
  \BibitemOpen
  \bibfield  {author} {\bibinfo {author} {\bibfnamefont {B.~D.}\ \bibnamefont {Gaulin}}, \bibinfo {author} {\bibfnamefont {S.~H.-Y.}\ \bibnamefont {Huang}}, \bibinfo {author} {\bibfnamefont {C.}~\bibnamefont {Balz}},\ and\ \bibinfo {author} {\bibfnamefont {E.~M.}\ \bibnamefont {Smith}},\ }\bibfield  {title} {\bibinfo {title} {\textit{Probing the Magnetic Ground State in the Quantum Spin Liquid Candidate $\mathrm{Ce}_2\mathrm{Hf}_2\mathrm{O}_7$}},\ }\href {https://doi.org/10.5286/ISIS.E.RB2220644} {\bibfield  {journal} {\bibinfo  {journal} {STFC ISIS Neutron and Muon Source, 10.5286/ISIS.E.RB2220644}\ } (\bibinfo {year} {2022}{\natexlab{a}})}\BibitemShut {NoStop}%
\bibitem [{\citenamefont {Gaulin}\ \emph {et~al.}(2022{\natexlab{b}})\citenamefont {Gaulin}, \citenamefont {Manuel}, \citenamefont {Smith},\ and\ \citenamefont {Huang}}]{WISH_experiment_DOI}%
  \BibitemOpen
  \bibfield  {author} {\bibinfo {author} {\bibfnamefont {B.~D.}\ \bibnamefont {Gaulin}}, \bibinfo {author} {\bibfnamefont {P.}~\bibnamefont {Manuel}}, \bibinfo {author} {\bibfnamefont {E.~M.}\ \bibnamefont {Smith}},\ and\ \bibinfo {author} {\bibfnamefont {S.~H.-Y.}\ \bibnamefont {Huang}},\ }\bibfield  {title} {\bibinfo {title} {\textit{Temperature Dependence of Diffuse Scattering in the New Pyrochlore Magnet $\mathrm{Ce}_2\mathrm{Hf}_2\mathrm{O}_7$}},\ }\href {https://doi.org/10.5286/ISIS.E.RB2220630-1} {\bibfield  {journal} {\bibinfo  {journal} {STFC ISIS Neutron and Muon Source, 10.5286/ISIS.E.RB2220630-1}\ } (\bibinfo {year} {2022}{\natexlab{b}})}\BibitemShut {NoStop}%
\bibitem [{\citenamefont {Kermarrec}\ \emph {et~al.}(2023)\citenamefont {Kermarrec}, \citenamefont {Chatterjee}, \citenamefont {Gaulin}, \citenamefont {Huang}, \citenamefont {Schmalz}, \citenamefont {Schmidt}, \citenamefont {Smith}, \citenamefont {Steffens},\ and\ \citenamefont {Wildes}}]{ILL_experiment_DOI}%
  \BibitemOpen
  \bibfield  {author} {\bibinfo {author} {\bibfnamefont {E.}~\bibnamefont {Kermarrec}}, \bibinfo {author} {\bibfnamefont {D.}~\bibnamefont {Chatterjee}}, \bibinfo {author} {\bibfnamefont {B.}~\bibnamefont {Gaulin}}, \bibinfo {author} {\bibfnamefont {S.~H.-Y.}\ \bibnamefont {Huang}}, \bibinfo {author} {\bibfnamefont {K.}~\bibnamefont {Schmalz}}, \bibinfo {author} {\bibfnamefont {W.}~\bibnamefont {Schmidt}}, \bibinfo {author} {\bibfnamefont {E.}~\bibnamefont {Smith}}, \bibinfo {author} {\bibfnamefont {P.}~\bibnamefont {Steffens}},\ and\ \bibinfo {author} {\bibfnamefont {A.}~\bibnamefont {Wildes}},\ }\bibfield  {title} {\bibinfo {title} {\textit{Polarization Analysis of the Diffuse Scattering and Spin Excitations in a Quantum Spin Liquid Candidate Pyrochlore}},\ }\href {https://doi.org/10.5291/ILL-DATA.4-05-852} {\bibfield  {journal} {\bibinfo  {journal} {Institut Laue-Langevin (ILL), 10.5291/ILL-DATA.4-05-852}\ } (\bibinfo {year} {2023})}\BibitemShut {NoStop}%
\bibitem [{\citenamefont {Ewings}\ \emph {et~al.}(2016)\citenamefont {Ewings}, \citenamefont {Buts}, \citenamefont {Le}, \citenamefont {{van Duijn}}, \citenamefont {Bustinduy},\ and\ \citenamefont {Perring}}]{Ewings2016}%
  \BibitemOpen
  \bibfield  {author} {\bibinfo {author} {\bibfnamefont {R.}~\bibnamefont {Ewings}}, \bibinfo {author} {\bibfnamefont {A.}~\bibnamefont {Buts}}, \bibinfo {author} {\bibfnamefont {M.}~\bibnamefont {Le}}, \bibinfo {author} {\bibfnamefont {J.}~\bibnamefont {{van Duijn}}}, \bibinfo {author} {\bibfnamefont {I.}~\bibnamefont {Bustinduy}},\ and\ \bibinfo {author} {\bibfnamefont {T.}~\bibnamefont {Perring}},\ }\bibfield  {title} {\bibinfo {title} {\textit{Horace: Software for the Analysis of Data from Single Crystal Spectroscopy Experiments at Time-of-Flight Neutron Instruments}},\ }\href {https://doi.org/https://doi.org/10.1016/j.nima.2016.07.036} {\bibfield  {journal} {\bibinfo  {journal} {Nuclear Instruments and Methods in Physics Research Section A: Accelerators, Spectrometers, Detectors and Associated Equipment}\ }\textbf {\bibinfo {volume} {834}},\ \bibinfo {pages} {132} (\bibinfo {year} {2016})}\BibitemShut {NoStop}%
\bibitem [{\citenamefont {Patri}\ \emph {et~al.}(2020{\natexlab{b}})\citenamefont {Patri}, \citenamefont {Hosoi},\ and\ \citenamefont {Kim}}]{Kim2020}%
  \BibitemOpen
  \bibfield  {author} {\bibinfo {author} {\bibfnamefont {A.~S.}\ \bibnamefont {Patri}}, \bibinfo {author} {\bibfnamefont {M.}~\bibnamefont {Hosoi}},\ and\ \bibinfo {author} {\bibfnamefont {Y.~B.}\ \bibnamefont {Kim}},\ }\bibfield  {title} {\bibinfo {title} {\textit{Distinguishing Dipolar and Octupolar Quantum Spin Ices using Contrasting Magnetostriction Signatures}},\ }\href {https://doi.org/10.1103/PhysRevResearch.2.023253} {\bibfield  {journal} {\bibinfo  {journal} {Phys. Rev. Research}\ }\textbf {\bibinfo {volume} {2}},\ \bibinfo {pages} {023253} (\bibinfo {year} {2020}{\natexlab{b}})}\BibitemShut {NoStop}%
\bibitem [{\citenamefont {Benton}(2016)}]{Benton2016}%
  \BibitemOpen
  \bibfield  {author} {\bibinfo {author} {\bibfnamefont {O.}~\bibnamefont {Benton}},\ }\bibfield  {title} {\bibinfo {title} {\textit{Quantum Origins of Moment Fragmentation in ${\mathrm{Nd}}_{2}{\mathrm{Zr}}_{2}{\mathrm{O}}_{7}$}},\ }\href {https://doi.org/10.1103/PhysRevB.94.104430} {\bibfield  {journal} {\bibinfo  {journal} {Phys. Rev. B}\ }\textbf {\bibinfo {volume} {94}},\ \bibinfo {pages} {104430} (\bibinfo {year} {2016})}\BibitemShut {NoStop}%
\bibitem [{\citenamefont {Applegate}\ \emph {et~al.}(2012)\citenamefont {Applegate}, \citenamefont {Hayre}, \citenamefont {Singh}, \citenamefont {Lin}, \citenamefont {Day},\ and\ \citenamefont {Gingras}}]{Applegate2012}%
  \BibitemOpen
  \bibfield  {author} {\bibinfo {author} {\bibfnamefont {R.}~\bibnamefont {Applegate}}, \bibinfo {author} {\bibfnamefont {N.~R.}\ \bibnamefont {Hayre}}, \bibinfo {author} {\bibfnamefont {R.~R.~P.}\ \bibnamefont {Singh}}, \bibinfo {author} {\bibfnamefont {T.}~\bibnamefont {Lin}}, \bibinfo {author} {\bibfnamefont {A.~G.~R.}\ \bibnamefont {Day}},\ and\ \bibinfo {author} {\bibfnamefont {M.~J.~P.}\ \bibnamefont {Gingras}},\ }\bibfield  {title} {\bibinfo {title} {\textit{Vindication of $\mathrm{Yb}_2\mathrm{Ti}_2\mathrm{O}_7$ as a Model Exchange Quantum Spin Ice}},\ }\href {https://doi.org/10.1103/PhysRevLett.109.097205} {\bibfield  {journal} {\bibinfo  {journal} {Phys. Rev. Lett.}\ }\textbf {\bibinfo {volume} {109}},\ \bibinfo {pages} {097205} (\bibinfo {year} {2012})}\BibitemShut {NoStop}%
\bibitem [{\citenamefont {Tang}\ \emph {et~al.}(2013)\citenamefont {Tang}, \citenamefont {Khatami},\ and\ \citenamefont {Rigol}}]{Tang2013}%
  \BibitemOpen
  \bibfield  {author} {\bibinfo {author} {\bibfnamefont {B.}~\bibnamefont {Tang}}, \bibinfo {author} {\bibfnamefont {E.}~\bibnamefont {Khatami}},\ and\ \bibinfo {author} {\bibfnamefont {M.}~\bibnamefont {Rigol}},\ }\bibfield  {title} {\bibinfo {title} {\textit{A Short Introduction to Numerical Linked-Cluster Expansions}},\ }\href {https://doi.org/https://doi.org/10.1016/j.cpc.2012.10.008} {\bibfield  {journal} {\bibinfo  {journal} {Comput. Phys. Commun}\ }\textbf {\bibinfo {volume} {184}},\ \bibinfo {pages} {557} (\bibinfo {year} {2013})}\BibitemShut {NoStop}%
\bibitem [{\citenamefont {Tang}\ \emph {et~al.}(2015)\citenamefont {Tang}, \citenamefont {Iyer},\ and\ \citenamefont {Rigol}}]{Tang2015}%
  \BibitemOpen
  \bibfield  {author} {\bibinfo {author} {\bibfnamefont {B.}~\bibnamefont {Tang}}, \bibinfo {author} {\bibfnamefont {D.}~\bibnamefont {Iyer}},\ and\ \bibinfo {author} {\bibfnamefont {M.}~\bibnamefont {Rigol}},\ }\bibfield  {title} {\bibinfo {title} {\textit{Thermodynamics of Two-Dimensional Spin Models with Bimodal Random-Bond Disorder}},\ }\href {https://doi.org/10.1103/PhysRevB.91.174413} {\bibfield  {journal} {\bibinfo  {journal} {Phys. Rev. B}\ }\textbf {\bibinfo {volume} {91}},\ \bibinfo {pages} {174413} (\bibinfo {year} {2015})}\BibitemShut {NoStop}%
\bibitem [{\citenamefont {Sch\"afer}\ and\ \citenamefont {Placke}()}]{schaefer_NLCE_corr_2024}%
  \BibitemOpen
  \bibfield  {author} {\bibinfo {author} {\bibfnamefont {R.}~\bibnamefont {Sch\"afer}}\ and\ \bibinfo {author} {\bibfnamefont {B.}~\bibnamefont {Placke}},\ }\bibfield  {title} {\bibinfo {title} {\textit{Neutron Scattering Signatures of Dipolar-Octupolar Spin Liquids}},\ }\bibinfo {note} {to be published}\BibitemShut {NoStop}%
\bibitem [{\citenamefont {Sandvik}(1999)}]{Sandvik1999}%
  \BibitemOpen
  \bibfield  {author} {\bibinfo {author} {\bibfnamefont {A.~W.}\ \bibnamefont {Sandvik}},\ }\bibfield  {title} {\bibinfo {title} {\textit{Stochastic Series Expansion Method with Operator-Loop Update}},\ }\href {https://doi.org/10.1103/PhysRevB.59.R14157} {\bibfield  {journal} {\bibinfo  {journal} {Phys. Rev. B}\ }\textbf {\bibinfo {volume} {59}},\ \bibinfo {pages} {R14157} (\bibinfo {year} {1999})}\BibitemShut {NoStop}%
\bibitem [{\citenamefont {Granroth}\ \emph {et~al.}(2010)\citenamefont {Granroth}, \citenamefont {Kolesnikov}, \citenamefont {Sherline}, \citenamefont {Clancy}, \citenamefont {Ross}, \citenamefont {Ruff}, \citenamefont {Gaulin},\ and\ \citenamefont {Nagler}}]{Granroth2010}%
  \BibitemOpen
  \bibfield  {author} {\bibinfo {author} {\bibfnamefont {G.~E.}\ \bibnamefont {Granroth}}, \bibinfo {author} {\bibfnamefont {A.~I.}\ \bibnamefont {Kolesnikov}}, \bibinfo {author} {\bibfnamefont {T.~E.}\ \bibnamefont {Sherline}}, \bibinfo {author} {\bibfnamefont {J.~P.}\ \bibnamefont {Clancy}}, \bibinfo {author} {\bibfnamefont {K.~A.}\ \bibnamefont {Ross}}, \bibinfo {author} {\bibfnamefont {J.~P.~C.}\ \bibnamefont {Ruff}}, \bibinfo {author} {\bibfnamefont {B.~D.}\ \bibnamefont {Gaulin}},\ and\ \bibinfo {author} {\bibfnamefont {S.~E.}\ \bibnamefont {Nagler}},\ }\bibfield  {title} {\bibinfo {title} {\textit{SEQUOIA: A Newly Operating Chopper Spectrometer at the SNS}},\ }\href {https://doi.org/10.1088/1742-6596/251/1/012058} {\bibfield  {journal} {\bibinfo  {journal} {Journal of Physics: Conference Series}\ }\textbf {\bibinfo {volume} {251}},\ \bibinfo {pages} {012058} (\bibinfo {year} {2010})}\BibitemShut {NoStop}%
\bibitem [{\citenamefont {Freeman}\ and\ \citenamefont {Watson}(1962)}]{Freeman1962}%
  \BibitemOpen
  \bibfield  {author} {\bibinfo {author} {\bibfnamefont {A.~J.}\ \bibnamefont {Freeman}}\ and\ \bibinfo {author} {\bibfnamefont {R.~E.}\ \bibnamefont {Watson}},\ }\bibfield  {title} {\bibinfo {title} {\textit{Theoretical Investigation of Some Magnetic and Spectroscopic Properties of Rare-Earth Ions}},\ }\href {https://doi.org/10.1103/PhysRev.127.2058} {\bibfield  {journal} {\bibinfo  {journal} {Phys. Rev.}\ }\textbf {\bibinfo {volume} {127}},\ \bibinfo {pages} {2058} (\bibinfo {year} {1962})}\BibitemShut {NoStop}%
\bibitem [{\citenamefont {Gaudet}\ \emph {et~al.}(2018)\citenamefont {Gaudet}, \citenamefont {Hallas}, \citenamefont {Buhariwalla}, \citenamefont {Sala}, \citenamefont {Stone}, \citenamefont {Tachibana}, \citenamefont {Baroudi}, \citenamefont {Cava},\ and\ \citenamefont {Gaulin}}]{Gaudet2018}%
  \BibitemOpen
  \bibfield  {author} {\bibinfo {author} {\bibfnamefont {J.}~\bibnamefont {Gaudet}}, \bibinfo {author} {\bibfnamefont {A.~M.}\ \bibnamefont {Hallas}}, \bibinfo {author} {\bibfnamefont {C.~R.~C.}\ \bibnamefont {Buhariwalla}}, \bibinfo {author} {\bibfnamefont {G.}~\bibnamefont {Sala}}, \bibinfo {author} {\bibfnamefont {M.~B.}\ \bibnamefont {Stone}}, \bibinfo {author} {\bibfnamefont {M.}~\bibnamefont {Tachibana}}, \bibinfo {author} {\bibfnamefont {K.}~\bibnamefont {Baroudi}}, \bibinfo {author} {\bibfnamefont {R.~J.}\ \bibnamefont {Cava}},\ and\ \bibinfo {author} {\bibfnamefont {B.~D.}\ \bibnamefont {Gaulin}},\ }\bibfield  {title} {\bibinfo {title} {\textit{Magnetoelastically Induced Vibronic Bound State in the Spin-Ice Pyrochlore ${\mathrm{Ho}}_{2}{\mathrm{Ti}}_{2}{\mathrm{O}}_{7}$}},\ }\href {https://doi.org/10.1103/PhysRevB.98.014419} {\bibfield  {journal} {\bibinfo  {journal} {Phys. Rev. B}\ }\textbf {\bibinfo {volume} {98}},\ \bibinfo {pages} {014419} (\bibinfo {year} {2018})}\BibitemShut {NoStop}%
\bibitem [{\citenamefont {Ruminy}\ \emph {et~al.}(2017)\citenamefont {Ruminy}, \citenamefont {Chi}, \citenamefont {Calder},\ and\ \citenamefont {Fennell}}]{Ruminy2017}%
  \BibitemOpen
  \bibfield  {author} {\bibinfo {author} {\bibfnamefont {M.}~\bibnamefont {Ruminy}}, \bibinfo {author} {\bibfnamefont {S.}~\bibnamefont {Chi}}, \bibinfo {author} {\bibfnamefont {S.}~\bibnamefont {Calder}},\ and\ \bibinfo {author} {\bibfnamefont {T.}~\bibnamefont {Fennell}},\ }\bibfield  {title} {\bibinfo {title} {\textit{Phonon-Mediated Spin-Flipping Mechanism in the Spin Ices ${\mathrm{Dy}}_{2}{\mathrm{Ti}}_{2}{\mathrm{O}}_{7}$ and ${\mathrm{Ho}}_{2}{\mathrm{Ti}}_{2}{\mathrm{O}}_{7}$}},\ }\href {https://doi.org/10.1103/PhysRevB.95.060414} {\bibfield  {journal} {\bibinfo  {journal} {Phys. Rev. B}\ }\textbf {\bibinfo {volume} {95}},\ \bibinfo {pages} {060414} (\bibinfo {year} {2017})}\BibitemShut {NoStop}%
\bibitem [{\citenamefont {Fennell}\ \emph {et~al.}(2014)\citenamefont {Fennell}, \citenamefont {Kenzelmann}, \citenamefont {Roessli}, \citenamefont {Mutka}, \citenamefont {Ollivier}, \citenamefont {Ruminy}, \citenamefont {Stuhr}, \citenamefont {Zaharko}, \citenamefont {Bovo}, \citenamefont {Cervellino}, \citenamefont {Haas},\ and\ \citenamefont {Cava}}]{Fennell2014}%
  \BibitemOpen
  \bibfield  {author} {\bibinfo {author} {\bibfnamefont {T.}~\bibnamefont {Fennell}}, \bibinfo {author} {\bibfnamefont {M.}~\bibnamefont {Kenzelmann}}, \bibinfo {author} {\bibfnamefont {B.}~\bibnamefont {Roessli}}, \bibinfo {author} {\bibfnamefont {H.}~\bibnamefont {Mutka}}, \bibinfo {author} {\bibfnamefont {J.}~\bibnamefont {Ollivier}}, \bibinfo {author} {\bibfnamefont {M.}~\bibnamefont {Ruminy}}, \bibinfo {author} {\bibfnamefont {U.}~\bibnamefont {Stuhr}}, \bibinfo {author} {\bibfnamefont {O.}~\bibnamefont {Zaharko}}, \bibinfo {author} {\bibfnamefont {L.}~\bibnamefont {Bovo}}, \bibinfo {author} {\bibfnamefont {A.}~\bibnamefont {Cervellino}}, \bibinfo {author} {\bibfnamefont {M.~K.}\ \bibnamefont {Haas}},\ and\ \bibinfo {author} {\bibfnamefont {R.~J.}\ \bibnamefont {Cava}},\ }\bibfield  {title} {\bibinfo {title} {\textit{Magnetoelastic Excitations in the Pyrochlore Spin Liquid ${\mathrm{Tb}}_{2}{\mathrm{Ti}}_{2}{\mathrm{O}}_{7}$}},\ }\href {https://doi.org/10.1103/PhysRevLett.112.017203} {\bibfield
  {journal} {\bibinfo  {journal} {Phys. Rev. Lett.}\ }\textbf {\bibinfo {volume} {112}},\ \bibinfo {pages} {017203} (\bibinfo {year} {2014})}\BibitemShut {NoStop}%
\bibitem [{\citenamefont {Xu}\ \emph {et~al.}(2021)\citenamefont {Xu}, \citenamefont {Man}, \citenamefont {Tang}, \citenamefont {Baidya}, \citenamefont {Zhang}, \citenamefont {Nakatsuji}, \citenamefont {Vanderbilt},\ and\ \citenamefont {Drichko}}]{Xu2021}%
  \BibitemOpen
  \bibfield  {author} {\bibinfo {author} {\bibfnamefont {Y.}~\bibnamefont {Xu}}, \bibinfo {author} {\bibfnamefont {H.}~\bibnamefont {Man}}, \bibinfo {author} {\bibfnamefont {N.}~\bibnamefont {Tang}}, \bibinfo {author} {\bibfnamefont {S.}~\bibnamefont {Baidya}}, \bibinfo {author} {\bibfnamefont {H.}~\bibnamefont {Zhang}}, \bibinfo {author} {\bibfnamefont {S.}~\bibnamefont {Nakatsuji}}, \bibinfo {author} {\bibfnamefont {D.}~\bibnamefont {Vanderbilt}},\ and\ \bibinfo {author} {\bibfnamefont {N.}~\bibnamefont {Drichko}},\ }\bibfield  {title} {\bibinfo {title} {\textit{Importance of Dynamic Lattice Effects for Crystal Field Excitations in the Quantum Spin Ice Candidate ${\mathrm{Pr}}_{2}{\mathrm{Zr}}_{2}{\mathrm{O}}_{7}$}},\ }\href {https://doi.org/10.1103/PhysRevB.104.075125} {\bibfield  {journal} {\bibinfo  {journal} {Phys. Rev. B}\ }\textbf {\bibinfo {volume} {104}},\ \bibinfo {pages} {075125} (\bibinfo {year} {2021})}\BibitemShut {NoStop}%
\bibitem [{\citenamefont {Thalmeier}\ and\ \citenamefont {Fulde}(1982)}]{Thalmeier1982}%
  \BibitemOpen
  \bibfield  {author} {\bibinfo {author} {\bibfnamefont {P.}~\bibnamefont {Thalmeier}}\ and\ \bibinfo {author} {\bibfnamefont {P.}~\bibnamefont {Fulde}},\ }\bibfield  {title} {\bibinfo {title} {\textit{Bound State Between a Crystal-Field Excitation and a Phonon in ${\mathrm{CeAl}}_{2}$}},\ }\href {https://doi.org/10.1103/PhysRevLett.49.1588} {\bibfield  {journal} {\bibinfo  {journal} {Phys. Rev. Lett.}\ }\textbf {\bibinfo {volume} {49}},\ \bibinfo {pages} {1588} (\bibinfo {year} {1982})}\BibitemShut {NoStop}%
\bibitem [{\citenamefont {Thalmeier}(1984)}]{Thalmeier1984}%
  \BibitemOpen
  \bibfield  {author} {\bibinfo {author} {\bibfnamefont {P.}~\bibnamefont {Thalmeier}},\ }\bibfield  {title} {\bibinfo {title} {\textit{Theory of the Bound State Between Phonons and a CEF Excitation in $\mathrm{CeAl}_2$}},\ }\href {https://doi.org/10.1088/0022-3719/17/23/015} {\bibfield  {journal} {\bibinfo  {journal} {Journal of Physics C: Solid State Physics}\ }\textbf {\bibinfo {volume} {17}},\ \bibinfo {pages} {4153} (\bibinfo {year} {1984})}\BibitemShut {NoStop}%
\bibitem [{\citenamefont {Schedler}\ \emph {et~al.}(2003)\citenamefont {Schedler}, \citenamefont {Witte}, \citenamefont {Loewenhaupt},\ and\ \citenamefont {Kulda}}]{Schedler2003}%
  \BibitemOpen
  \bibfield  {author} {\bibinfo {author} {\bibfnamefont {R.}~\bibnamefont {Schedler}}, \bibinfo {author} {\bibfnamefont {U.}~\bibnamefont {Witte}}, \bibinfo {author} {\bibfnamefont {M.}~\bibnamefont {Loewenhaupt}},\ and\ \bibinfo {author} {\bibfnamefont {J.}~\bibnamefont {Kulda}},\ }\bibfield  {title} {\bibinfo {title} {\textit{Coupling Between Crystal Field Transitions and Phonons in the 4f-Electron System $\mathrm{CeCu}_2$}},\ }\href {https://doi.org/https://doi.org/10.1016/S0921-4526(03)00187-X} {\bibfield  {journal} {\bibinfo  {journal} {Physica B: Condensed Matter}\ }\textbf {\bibinfo {volume} {335}},\ \bibinfo {pages} {41} (\bibinfo {year} {2003})},\ \bibinfo {note} {proceedings of the Fourth International Workshop on Polarised Neutrons for Condensed Matter Investigations}\BibitemShut {NoStop}%
\bibitem [{\citenamefont {Loewenhaupt}\ and\ \citenamefont {Witte}(2003)}]{Loewenhaupt2003}%
  \BibitemOpen
  \bibfield  {author} {\bibinfo {author} {\bibfnamefont {M.}~\bibnamefont {Loewenhaupt}}\ and\ \bibinfo {author} {\bibfnamefont {U.}~\bibnamefont {Witte}},\ }\bibfield  {title} {\bibinfo {title} {\textit{Coupling Between Electronic and Lattice Degrees of Freedom in 4f-Electron Systems Investigated by Inelastic Neutron Scattering}},\ }\href {https://doi.org/10.1088/0953-8984/15/5/307} {\bibfield  {journal} {\bibinfo  {journal} {Journal of Physics: Condensed Matter}\ }\textbf {\bibinfo {volume} {15}},\ \bibinfo {pages} {S519} (\bibinfo {year} {2003})}\BibitemShut {NoStop}%
\bibitem [{\citenamefont {Chapon}\ \emph {et~al.}(2006)\citenamefont {Chapon}, \citenamefont {Goremychkin}, \citenamefont {Osborn}, \citenamefont {Rainford},\ and\ \citenamefont {Short}}]{Chapon2006}%
  \BibitemOpen
  \bibfield  {author} {\bibinfo {author} {\bibfnamefont {L.}~\bibnamefont {Chapon}}, \bibinfo {author} {\bibfnamefont {E.}~\bibnamefont {Goremychkin}}, \bibinfo {author} {\bibfnamefont {R.}~\bibnamefont {Osborn}}, \bibinfo {author} {\bibfnamefont {B.}~\bibnamefont {Rainford}},\ and\ \bibinfo {author} {\bibfnamefont {S.}~\bibnamefont {Short}},\ }\bibfield  {title} {\bibinfo {title} {\textit{Magnetic and Structural Instabilities in $\mathrm{CePd}_2\mathrm{Al}_2$ and $\mathrm{LaPd}_2\mathrm{Al}_2$}},\ }\href {https://doi.org/https://doi.org/10.1016/j.physb.2006.01.300} {\bibfield  {journal} {\bibinfo  {journal} {Physica B: Condensed Matter}\ }\textbf {\bibinfo {volume} {378-380}},\ \bibinfo {pages} {819} (\bibinfo {year} {2006})},\ \bibinfo {note} {proceedings of the International Conference on Strongly Correlated Electron Systems}\BibitemShut {NoStop}%
\bibitem [{\citenamefont {Adroja}\ \emph {et~al.}(2012)\citenamefont {Adroja}, \citenamefont {del Moral}, \citenamefont {de~la Fuente}, \citenamefont {Fraile}, \citenamefont {Goremychkin}, \citenamefont {Taylor}, \citenamefont {Hillier},\ and\ \citenamefont {Fernandez-Alonso}}]{Adroja2012}%
  \BibitemOpen
  \bibfield  {author} {\bibinfo {author} {\bibfnamefont {D.~T.}\ \bibnamefont {Adroja}}, \bibinfo {author} {\bibfnamefont {A.}~\bibnamefont {del Moral}}, \bibinfo {author} {\bibfnamefont {C.}~\bibnamefont {de~la Fuente}}, \bibinfo {author} {\bibfnamefont {A.}~\bibnamefont {Fraile}}, \bibinfo {author} {\bibfnamefont {E.~A.}\ \bibnamefont {Goremychkin}}, \bibinfo {author} {\bibfnamefont {J.~W.}\ \bibnamefont {Taylor}}, \bibinfo {author} {\bibfnamefont {A.~D.}\ \bibnamefont {Hillier}},\ and\ \bibinfo {author} {\bibfnamefont {F.}~\bibnamefont {Fernandez-Alonso}},\ }\bibfield  {title} {\bibinfo {title} {\textit{Vibron Quasibound State in the Noncentrosymmetric Tetragonal Heavy-Fermion Compound ${\mathrm{CeCuAl}}_{3}$}},\ }\href {https://doi.org/10.1103/PhysRevLett.108.216402} {\bibfield  {journal} {\bibinfo  {journal} {Phys. Rev. Lett.}\ }\textbf {\bibinfo {volume} {108}},\ \bibinfo {pages} {216402} (\bibinfo {year} {2012})}\BibitemShut {NoStop}%
\bibitem [{\citenamefont {Anand}\ \emph {et~al.}(2021)\citenamefont {Anand}, \citenamefont {Fraile}, \citenamefont {Adroja}, \citenamefont {Sharma}, \citenamefont {Tripathi}, \citenamefont {Ritter}, \citenamefont {de~la Fuente}, \citenamefont {Biswas}, \citenamefont {Sakai}, \citenamefont {del Moral},\ and\ \citenamefont {Strydom}}]{Anand2021}%
  \BibitemOpen
  \bibfield  {author} {\bibinfo {author} {\bibfnamefont {V.~K.}\ \bibnamefont {Anand}}, \bibinfo {author} {\bibfnamefont {A.}~\bibnamefont {Fraile}}, \bibinfo {author} {\bibfnamefont {D.~T.}\ \bibnamefont {Adroja}}, \bibinfo {author} {\bibfnamefont {S.}~\bibnamefont {Sharma}}, \bibinfo {author} {\bibfnamefont {R.}~\bibnamefont {Tripathi}}, \bibinfo {author} {\bibfnamefont {C.}~\bibnamefont {Ritter}}, \bibinfo {author} {\bibfnamefont {C.}~\bibnamefont {de~la Fuente}}, \bibinfo {author} {\bibfnamefont {P.~K.}\ \bibnamefont {Biswas}}, \bibinfo {author} {\bibfnamefont {V.~G.}\ \bibnamefont {Sakai}}, \bibinfo {author} {\bibfnamefont {A.}~\bibnamefont {del Moral}},\ and\ \bibinfo {author} {\bibfnamefont {A.~M.}\ \bibnamefont {Strydom}},\ }\bibfield  {title} {\bibinfo {title} {\textit{Crystal Electric Field and Possible Coupling with Phonons in Kondo Lattice ${\mathrm{CeCuGa}}_{3}$}},\ }\href {https://doi.org/10.1103/PhysRevB.104.174438} {\bibfield  {journal} {\bibinfo  {journal} {Phys. Rev. B}\ }\textbf {\bibinfo
  {volume} {104}},\ \bibinfo {pages} {174438} (\bibinfo {year} {2021})}\BibitemShut {NoStop}%
\bibitem [{\citenamefont {Prather}(1961)}]{Prather1961}%
  \BibitemOpen
  \bibfield  {author} {\bibinfo {author} {\bibfnamefont {J.~L.}\ \bibnamefont {Prather}},\ }\href@noop {} {\emph {\bibinfo {title} {\textit{Atomic Energy Levels in Crystals}}}},\ Vol.~\bibinfo {volume} {19}\ (\bibinfo  {publisher} {US Department of Commerce, National Bureau of Standards, Gaithersburg},\ \bibinfo {year} {1961})\BibitemShut {NoStop}%
\bibitem [{\citenamefont {Hutchings}(1964)}]{Hutchings1964}%
  \BibitemOpen
  \bibfield  {author} {\bibinfo {author} {\bibfnamefont {M.}~\bibnamefont {Hutchings}},\ }\bibfield  {title} {\bibinfo {title} {\textit{Point-Charge Calculations of Energy Levels of Magnetic Ions in Crystalline Electric Fields}}\ }(\bibinfo  {publisher} {Academic Press, New York},\ \bibinfo {year} {1964})\ pp.\ \bibinfo {pages} {227--273}\BibitemShut {NoStop}%
\bibitem [{\citenamefont {Stevens}(1952)}]{Stevens1952}%
  \BibitemOpen
  \bibfield  {author} {\bibinfo {author} {\bibfnamefont {K.~W.~H.}\ \bibnamefont {Stevens}},\ }\bibfield  {title} {\bibinfo {title} {\textit{Matrix Elements and Operator Equivalents Connected with the Magnetic Properties of Rare Earth Ions}},\ }\href {https://doi.org/10.1088/0370-1298/65/3/308} {\bibfield  {journal} {\bibinfo  {journal} {Proceedings of the Physical Society. Section A}\ }\textbf {\bibinfo {volume} {65}},\ \bibinfo {pages} {209} (\bibinfo {year} {1952})}\BibitemShut {NoStop}%
\bibitem [{\citenamefont {Boothroyd}(2014)}]{SPECTRE}%
  \BibitemOpen
  \bibfield  {author} {\bibinfo {author} {\bibfnamefont {A.~T.}\ \bibnamefont {Boothroyd}},\ }\href {https://xray.physics.ox.ac.uk/software.htm} {\bibinfo {title} {\textit{Spectre — A Program for Calculating Spectroscopic Properties of Rare Earth Ions in Crystals}}} (\bibinfo {year} {1990-2014})\BibitemShut {NoStop}%
\end{thebibliography}%
\end{document}